\documentclass[11pt,a4paper]{article}
\pdfoutput=1
\usepackage{jheppub}
\usepackage{amsmath,epsfig}

\usepackage{mathrsfs}
\usepackage{amssymb}
\usepackage{mathtools}
\usepackage{graphics}
\usepackage{axodraw}
\usepackage{slashed}
\usepackage{url}
\usepackage{hyperref}
\usepackage{enumitem}

\usepackage{color}
\usepackage[normalem]{ulem}

\usepackage{multirow}
\usepackage[svgnames,dvipsnames,x11names]{xcolor}
\usepackage[framemethod=default]{mdframed}
\newmdenv[skipabove=7pt,
skipbelow=7pt,
rightline=true,
leftline=true,
topline=true,
bottomline=true,
backgroundcolor=gray!7,
linecolor=gray,
innerleftmargin=5pt,
innerrightmargin=5pt,
innertopmargin=0pt,
innerbottommargin=10pt,
leftmargin=0cm,
rightmargin=0cm,
linewidth=1.5pt]{eBox}

\newcommand{\be}{\begin{equation}}
\newcommand{\ee}{\end{equation}}
\newcommand{\bea}{\begin{eqnarray}}
\newcommand{\eea}{\end{eqnarray}}

\title{Bounds on right-handed neutrino parameters from observable leptogenesis }

\preprint{IFIC/22-20, FTUV-22-0704.1758}

\author[]{ P. ~Hern\'andez,}
\author[]{J. ~L\'opez-Pav\'on,}
\author[]{N. Rius,}
\author[]{and S. Sandner}

\affiliation[]{Instituto de F\'{\i}sica Corpuscular, Universidad de Valencia and CSIC, 
 Edificio Institutos Investigaci\'on, Catedr\'atico Jos\'e Beltr\'an 2, 46980 Spain}
 
 \emailAdd{m.pilar.hernandez@uv.es}
 \emailAdd{jlpavon@ific.uv.es}
\emailAdd{nuria.rius@ific.uv.es}
\emailAdd{stefan.sandner@ific.uv.es}

\abstract{We revisit the  generation of a matter-antimatter asymmetry in the minimal extension of the Standard Model with 
two singlet heavy neutral leptons (HNL) that can explain neutrino masses. 
We derive an accurate analytical approximation to the solution of the complete  linearized set of kinetic equations,  which exposes the non-trivial parameter dependencies in the form of parameterization-independent CP invariants. 
The identification of various washout regimes relevant in different regions of parameter space sheds light on  the relevance of the mass corrections in the interaction rates and clarifies the correlations of baryogenesis with other observables.
In particular, by requiring that the measured baryon asymmetry is reproduced, we derive robust upper or lower bounds on the HNL mixings depending on their masses, and constraints on their flavour structure, as well as on the CP-violating phases of the PMNS mixing matrix, and the amplitude of neutrinoless double-beta decay. We also find certain correlations between low and high scale CP phases. 
Especially emphasizing the testable part of the parameter space we demonstrate that our findings are in very good agreement with numerical results.
The methods developed in this work can help in exploring more complex scenarios. }


\begin{document}

\maketitle


\section{Introduction}
\label{sec:introduction}

Extensions of the Standard Model that can explain the smallness of neutrino masses generically provide a mechanism to explain the matter-antimatter asymmetry in the Universe~\cite{Fukugita:1986hr}. The most minimal of these realizations is arguably the minimal type-I seesaw model~\cite{Minkowski:1977sc,GellMann:1980vs,Yanagida:1979as,Mohapatra:1979ia}, an extension with two Majorana singlet fermions that can couple to the SM via the fermion portal. The massive lepton sector includes light neutrinos, and additional heavy neutral leptons (HNLs) that can be searched for in meson, gauge boson and higgs decays. The possibility to explain the baryon asymmetry of the Universe (BAU) in this model has been studied extensively, and it has been shown to be a robust prediction in a wide range of masses of the heavy states, ranging from sub-GeV up to  
$\sim 10^{15}$ GeV.  The lower limit is set by constraints from cosmology~\cite{Hernandez:2014fha} and big bang nucleosynthesis (see~\cite{Boyarsky:2020dzc} and refs. therein), while the upper limit is set by the requirement of perturbative Yukawa couplings. As dictated by the Sakharov conditions, a matter-antimatter asymmetry can be generated dynamically above the electroweak (EW) phase transition by effective $B+L$-violating sphaleron processes, combined with CP-odd asymmetries created in an out-of-thermal equilibrium process. The type of such process varies depending on the masses of the heavy Majorana singlets. While for heavy masses, the relevant process is out-of-equilibrium decay of these particles at freeze-out~\cite{Fukugita:1986hr,Pilaftsis:2003gt,Abada:2006ea}, in the case of lighter masses, the relevant process is heavy neutrino oscillations at freeze-in~\cite{Akhmedov:1998qx,Asaka:2005pn}. It has been shown recently that there is a description that allows to treat both regimes and interpolates smoothly the region in between~\cite{Klaric:2020phc,Klaric:2021cpi}. A set of  quantum Boltzmann equations need to be solved for some fixed input parameters of the model to obtain a quantitative prediction of the baryon asymmetry. The interaction rates involved in these processes have been computed to a high level of sophistication in~\cite{Ghiglieri:2017gjz}. 

An interesting question is to what extent this scenario can be tested. The answer depends strongly on the scale of the Majorana masses. If these masses are too large to be produced in particle or cosmic accelerators, one could hope to follow the traces left in the form of higher dimensional operators. The leading d=5 being the famous Weinberg operator~\cite{Weinberg:1979sa} that generates light neutrino masses, and can be tested by searching for neutrinoless double-beta decay. Generically also $d \geq 6$ operators are expected~\cite{Broncano:2002rw,Abada:2007ux}, leading potentially to very interesting signals in charged lepton processes, non-unitarity of the leptonic mixing matrix, etc.~\cite{Antusch:2006vwa,Shrock:1980ct,Shrock:1981wq,Langacker:1988ur,Nardi:1994iv,Tommasini:1995ii,Alonso:2012ji,Antusch:2014woa,Fernandez-Martinez:2016lgt}.

A more interesting possibility is, however, that the masses are not so large and these neutrino mass mediators can be produced at colliders, or in rare processes, such as displaced meson decays. This possibility has been studied extensively in recent years and it has been shown that the parameter space that leads to successful baryogenesis can in fact be largely explored~\cite{Canetti:2010aw,Shuve:2014zua,Canetti:2012vf,Canetti:2012kh,Hernandez:2015wna,Abada:2015rta,Hernandez:2016kel,Hambye:2016sby,Drewes:2016jae,Drewes:2016gmt,Ghiglieri:2017csp,Hambye:2017elz,Antusch:2017pkq,Eijima:2018qke,Abada:2018oly}. 

It is well known that there are strong correlations between baryogenesis and  the properties of the HNLs (such as their masses and mixings to the various lepton flavours),  neutrino masses and the amplitude of neutrinoless double-beta decay. 
In particular, upper and lower bounds  on HNLs mixings for successful baryogenesis have been studied numerically in~\cite{Canetti:2010aw,Canetti:2012kh,Drewes:2016jae,Antusch:2017pkq,Eijima:2018qke,Klaric:2021cpi}.
The precise form of these correlations is, however, difficult to reveal from numerical studies. In this paper we address this question analytically, by developing a new perturbative scheme to solve the Boltzmann equations involved in the production of the baryon asymmetry, that takes into account mass effects in the interaction rates, and allows an accurate description of all the washout regimes. A very useful tool in this context is that of CP flavour invariants.
This allows us to accurately rewrite the baryon asymmetry in terms of parameterization-independent CP invariants that can then be easily correlated to other flavour observables. This connection allows us to expose and understand these correlations, and predict the constraints on the baryon asymmetry that could be derived from putative future measurements of HNLs, CP violation in neutrino oscillations and neutrinoless double-beta decay, or alternatively to understand the bounds on HNL parameters from the baryon asymmetry. 
A similar analysis in the context of high-scale leptogenesis led to the celebrated Davidson-Ibarra bound~\cite{Davidson:2002qv}. 

The paper is organized as follows. In sec.~\ref{sec:model} we introduce the model, set our notation and identify the various relevant regimes for the production of the baryon asymmetry and associate each of them to a parameterization-independent CP invariant. We then relate those CP invariants to neutrino masses and HNL parameters in sec.~\ref{sec:cpinv}. In sec.~\ref{sec:ana} we review the Boltzmann equations needed in the computation of the baryon asymmetry, and develop a perturbative method to get an analytical approximate solution to the equations in the various regimes, recovering the expected dependence on the CP invariants.  In sec.~\ref{sec:parambounds}, we  use the analytical results to derive  bounds on the HNL parameters from the baryon asymmetry. In sec.~\ref{sec:num} we present the comparison of the numerical solution to our analytical results and perform a numerical scan of the HNL mixing versus mass testable parameter space for successful baryogenesis.  In sec.~\ref{sec:constraints}, we consider the correlation with other observables such as the flavour of the HNL mixings and neutrinoless double-beta decay.  We conclude in sec.~\ref{sec:conclu}.


\section{The model, Sakharov conditions and CP invariants}
\label{sec:model}

We consider the well-known type-I seesaw model, which includes the SM and $n \geq 2$ additional fermion singlets, $N^i$. The most general renormalizable Lagrangian is
 \begin{eqnarray}
{\cal L} = {\cal L}_{SM}- \sum_{\alpha,i} \bar L^\alpha Y^{\alpha i} \tilde\Phi N^i - \sum_{i,j=1}^n {1\over 2} \bar N^{ic} M_{Rij} N^j+ h.c.\,, \nonumber
\label{eq:lag}
\end{eqnarray}
where $Y$ is a $3\times n$ complex matrix and $M_R$  is a $n\times n$ complex symmetric matrix. $L$ is the fermion doublet and $\tilde\Phi = i\sigma_2 \Phi^*$ is the Higgs doublet.

As long as $n\geq 2$ the model can explain the measured light neutrino masses and mixings, but contains $n$ additional HNLs.  In the limit $M_R\gg Y \langle \Phi \rangle $, the light neutrino masses are well approximated by the well-known seesaw formula:
\be
-m_\nu={v^2}Y M_R^{-1} Y^T\,,
\label{eq:mnu}
\ee
where $\langle \Phi\rangle =v$ and $\sqrt{2} v=246\, \rm{GeV}$, while the masses of the HNLs are the eigenvalues of the matrix $M_R$ up to small corrections.

The HNLs interact with the gauge bosons and the higgs via the mixing: 
\be
\Theta \sim v Y M_R^{-1} \sim {\mathcal O}\left(\sqrt{m_\nu \over M_R}\right)\,.
\label{eq:theta}
\ee
According to this naive scaling, for HNL masses at the electroweak scale, the mixings are very small and  difficult to test. 

It is well known~\cite{Wyler:1982dd,Mohapatra:1986aw,Mohapatra:1986bd,Bernabeu:1987gr,Branco:1988ex,Akhmedov:1995ip,Barr:2003nn,Kersten:2007vk,Gavela:2009cd} that for certain textures of $Y$ and $M_R$, that are consistent with an exact lepton number (LN) symmetry, the naive scaling of eq.~(\ref{eq:theta}) breaks down. Neutrino masses in eq.~(\ref{eq:mnu}) vanish exactly, while $\Theta$ is unsuppressed. 

We will focus on the minimal $n=2$ model for which the symmetric texture is of the form \cite{Gavela:2009cd}
\be
Y=\begin{pmatrix}
y_{e} & 0\\
y_{\mu} & 0 \\
y_{\tau} & 0 
\end{pmatrix},\;\;
M_R=\begin{pmatrix}
0 & \Lambda \\
\Lambda  & 0 
\end{pmatrix}\,,
\label{eq:LNtexture}
\ee
corresponding to a lepton number assignment $L(N_1)= -L(N_2)=1$. The exact lepton number symmetry ensures three massless neutrinos and   degenerate HNLs. Note that the matrix $Y^\dagger Y$ has then a vanishing eigenvalue, which means that one combination of the sterile states does not couple to leptons. 

Obviously, three neutrinos remain exactly massless in the symmetric limit, and beyond this limit they are proportional to the symmetry-breaking entries,  $y'_\alpha$ and $\mu_i$: 
\be
Y=\begin{pmatrix}
y_{e} e^{i \beta_e} & y'_e e^{i \beta_e'}\\
y_{\mu} e^{i \beta_\mu} & y'_{\mu}  e^{i \beta_\mu'}\\\
y_{\tau} e^{i \beta_\tau} & y'_{\tau} e^{i \beta_\tau'}\ 
\end{pmatrix},\;\;
M_R=\begin{pmatrix}
\mu_1 & \Lambda \\
\Lambda  & \mu_2 
\end{pmatrix}\,.
\label{eq:LNVparam2N}
\ee
Note that we use a parameterization where all the complex phases are included in $Y$. 
In appendix~\ref{sec:app}, we will show explicitly that this is the case and, remarkably, that we can also consider $\mu_1=\mu_2$ in all generality.

The breaking of the symmetry induced by the different terms is the same: $\Delta L(\mu_1)=\Delta L(\mu_2)=\Delta L(y'_\alpha)=2$, and therefore it is natural to assume  no large hierarchy between these parameters, in particular $|y'_\alpha/y_\beta|$ and $\mu_i/\Lambda$. On the other hand, while the parameters $y'_\alpha$ and $\mu_2$  contribute to neutrino masses at tree level as:
\be
- \left( m_{\nu}\right)_{\alpha\beta} = \frac{v^2}{\Lambda}\left( Y_{\alpha 1} Y_{\beta 2} + Y_{\alpha 2} Y_{\beta 1} - Y_{\alpha 1} Y_{\beta 1} \frac{\mu_2}{\Lambda} +\mathcal{O}\left(\frac{\mu_i^3}{\Lambda^3}\right)+ \mathcal{O}\left(\frac{\mu_1 y'^2_{\alpha}}{\Lambda}\right)+ \mathcal{O}\left(\frac{\mu_i^2y'_{\alpha}}{\Lambda^2}\right)\right)\,,
\label{eq:mnuLN}
\ee
 the leading $\mu_1$ contribution only shows up at 1-loop. For this reason, $\mu_1$ can be larger than $\mu_2$ without spoiling light neutrino masses.
However, the same parameter can induce a large mass splitting between the HNLs and this is not a favourable regime for low scale leptogenesis. For this reason, we will assume that all symmetry breaking parameters are small compared to the symmetric ones. In particular, we would like to remark that in this symmetry protected scenario, $\mu_1=\mu_2$ can be considered in all generality, as shown in appendix~\ref{sec:app}.

In this paper, we will compute analytically the baryon asymmetry generated in this model by perturbing around the symmetric limit, that is via a series expansion in the small symmetry breaking parameters.

\subsection{Sakharov conditions and regimes}

The necessary Sakharov conditions for the production of the baryon asymmetry are satisfied in this model in the following way. New sources of 
CP violation appear in the couplings $Y$ and $M_R$. Baryon number violation is ensured by sphaleron processes active above the electroweak 
phase transition~\cite{Kuzmin:1985mm}, i.e. $T \geq T_{\rm EW}=131.7$ GeV~\cite{DOnofrio:2014rug}. The out-of-equilibrium condition requires that some of the species are not in thermal equilibrium. In the low mass regime, the asymmetry is generated during the production of the heavy states, $N_i$,  i.e. before they reach full thermal equilibrium. The production of the state $N_i$,   occurs via direct production from inverse decays, $L_\alpha H\leftrightarrow N_i$,  or various $2\rightarrow 2$ scattering processes, with strength $Y_{\alpha i}$. The different mass eigenstates are produced coherently in a state of flavour $\alpha$, $N_\alpha \propto \sum_i Y_{\alpha i} N_i$.  CP asymmetries arise then from the interference of CP violating phases in $Y, M_R$ and the CP conserving oscillation phases $N_\alpha \leftrightarrow N_\beta$~\cite{Akhmedov:1998qx}.

\subsubsection{Time scales and slow modes} 
 
 In an expanding universe, the efficiency of plasma interactions in thermalizing the states involved depends on whether the interaction rates are larger or smaller than the Hubble expansion rate, $H_u(T)$, which in the range of temperatures of our interest is dominated by radiation and given by 
 \begin{eqnarray}
 H_u(T) = {T^2 \over {{M_P^*}}}\,,
 \label{eq:hubble}
 \end{eqnarray}
with 
\bea
M_P^* \equiv  \sqrt{45\over 4 \pi^3 g_*(T)} M_{\rm Planck}\,.
\eea
We assume the number of thermal relativistic degrees of freedom at temperature $T$ to be $g_*(T)= 106.75$ throughout the evolution, that is, we neglect the HNLs contribution. 

The baryon asymmetry is exponentially suppressed if all the relevant processes involved in  its generation are fast compared with the Hubble expansion rate. So, as first noted by Sakharov, the rates for some of these processes must remain below $H_u(T)$. We can distinguish various regimes depending on what modes satisfy this condition at the electroweak phase transition, $T_{\rm EW}$. 

A first relevant scale in the problem is the one related to the vacuum oscillation rate, which is not a thermalization rate, but it is the scale at which CP asymmetries build up:
\begin{eqnarray}
\Gamma_{\rm osc}(T) \propto {M_2^2-M_1^2\over T }\,,
\end{eqnarray}
where $M_i$ are the mass eigenvalues of the heavy states above the EW phase transition.

Secondly, we have the scattering, decay or inverse decay rates. 
At temperatures such that $T\gg M_i$, the HNLs can be assumed relativistic in the corresponding processes. In this case the interaction rate with flavour $\alpha$ is given by:
\begin{eqnarray}
\Gamma_{\alpha}(T) \propto \epsilon_\alpha \Gamma(T), \;\; \Gamma \propto {\rm Tr}[Y Y^\dagger] T, \;\;\; \epsilon_\alpha \equiv {(Y Y^\dagger)_{\alpha\alpha} \over { \rm Tr}[Y Y^\dagger] }\,.
\label{eq:epsilonalpha}
\end{eqnarray}
A flavour hierarchy in the Yukawa couplings can result in a hierarchy in the corresponding interaction rates.  

There are, however, slow modes that do not thermalize with this rate, owing to the approximate zero mode of $Y^\dagger Y$, related to the approximate LN symmetry. 
The thermalization rate of this mode involves oscillations and is of the form
\begin{eqnarray}
\Gamma^{\rm slow}_{\rm osc} \propto P_{\rm osc} \Gamma \leq \Gamma\,,
\label{eq:slowosc}
\end{eqnarray}
where $P_{\rm osc}$ can be thought of as an oscillation probability which is $P_{\rm osc} \rightarrow 1$ when $\Gamma_{\rm osc} \gg \Gamma$, while  $P_{\rm osc} \rightarrow 0$ when $\Gamma_{\rm osc} \ll \Gamma$ (oscillations are damped). Note that this rate may be suppressed even in strong washout, i.e. when $\Gamma \gg H_u$, as long as the  mass difference, $|M_2-M_1|$, is sufficiently small. 

When $M_i/T$ corrections are included in the rates, there is an additional slow mode, related to LN. The corresponding slow rate is (we assume $M\leq T_{EW}$):
\begin{eqnarray}
\Gamma^{\rm slow}_{M} \propto \left({M_i\over T}\right)^2 \Gamma \leq \Gamma\,.
\label{eq:slowM}
\end{eqnarray}
When both slow rates become large compared to $H_u$, and no significantly flavour effects are present, full thermalization is achieved. If this happens before $T_{\rm EW}$, the baryon asymmetry is exponentially suppressed. 

A large fraction of the parameter space of the model, compatible with the light neutrino masses, satisfies
\begin{eqnarray}
\Gamma_{\rm osc}(T_{\rm EW}), \Gamma(T_{\rm EW}) \geq H_u(T_{\rm EW})\,.
\label{eq:ghub}
\end{eqnarray}
This regime is  also the most interesting one, as regards testability prospects, since it corresponds to large mixing of the HNLs. 
In particular most of the accessible  parameter space for future experiments corresponds to the strong washout regime $\Gamma(T_{\rm EW}) \gg H_u(T_{\rm EW})$.

\subsubsection{Washout regimes}
\label{sec:washout}

\begin{figure}[!t]
\centering
\begin{tabular}{cc}
\hspace{-0.5cm} \includegraphics[width=0.5\textwidth]{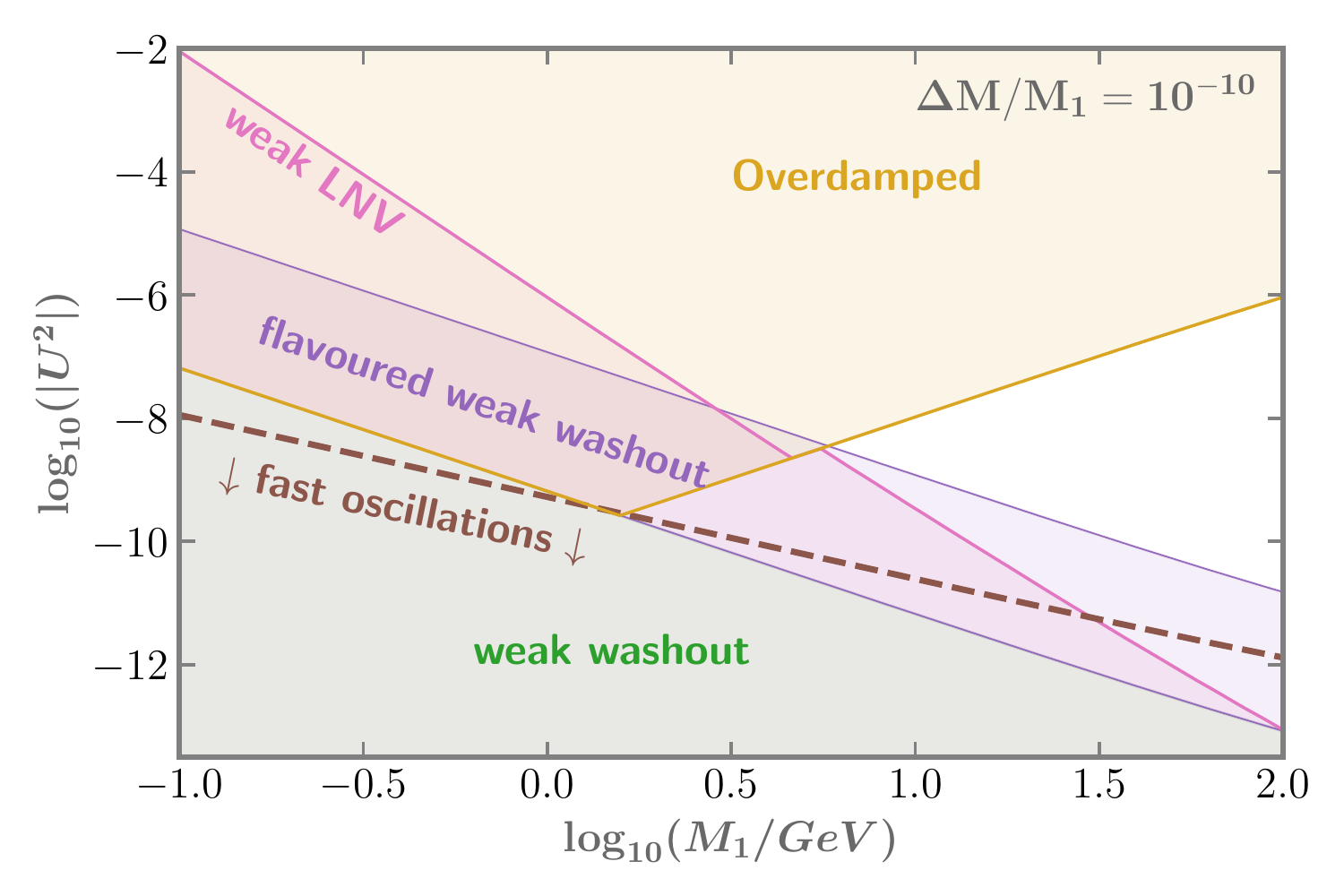} &
\hspace{-0.55cm}  \includegraphics[width=0.5\textwidth]{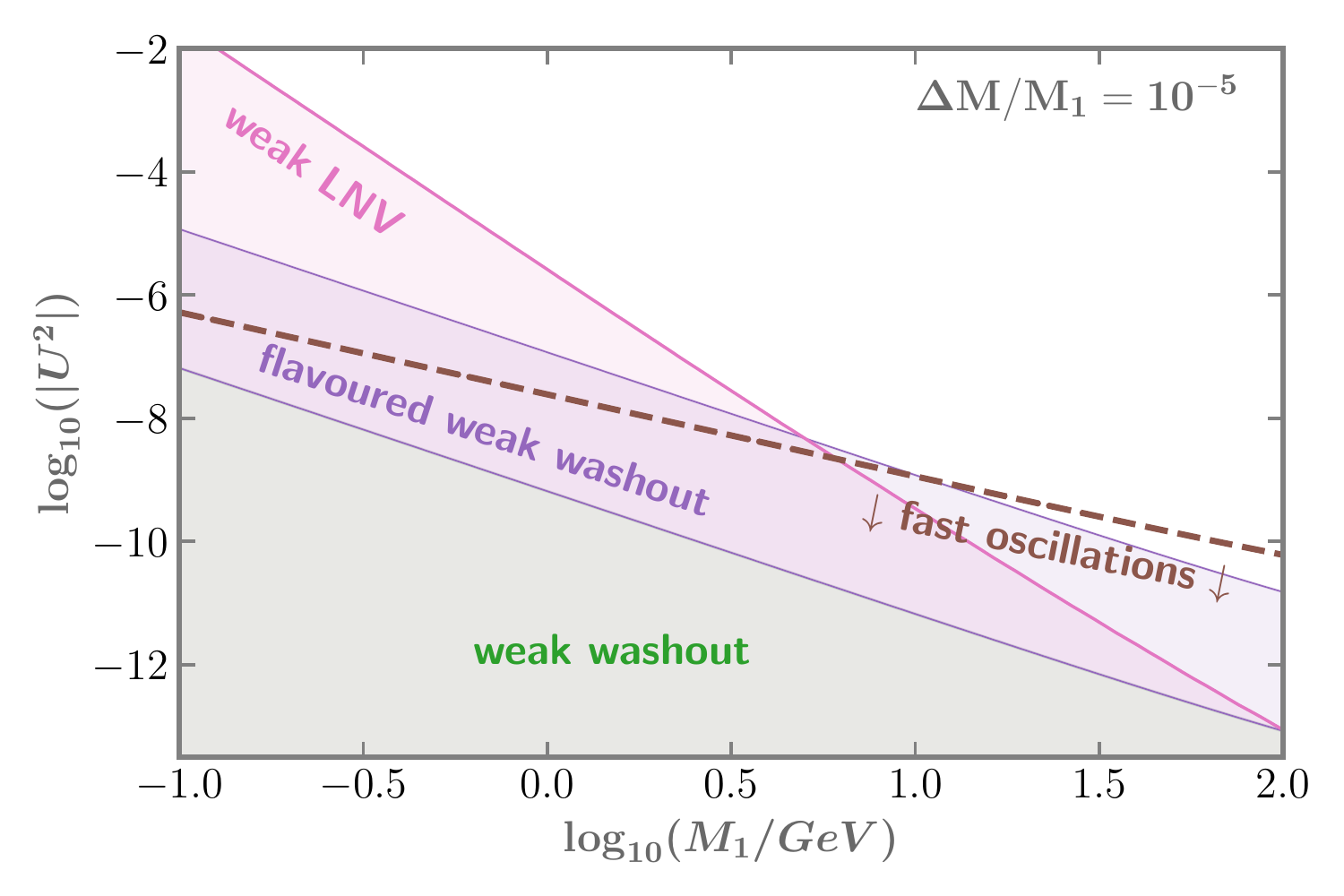}    
\end{tabular}
\vspace{-0.4cm}
\caption{Boundaries of washout regimes and regions described in the text on the plane $|U|^2$ versus M for two choices of $\Delta M/M$.}
\label{fig:regimes}
\end{figure}

Different regimes can be identified depending on the relative strength of the rates that remain below $H_u$ at $T_{EW}$.  
In Fig.~\ref{fig:regimes} we show the different regimes on the plane  mixing of the HNL, $U^2\equiv \sum_\alpha |\Theta_{\alpha 1}|^2\simeq \sum_\alpha |\Theta_{\alpha 2}|^2 $, versus their mass, $M_1\simeq M_2$,  with light neutrino masses properly accounted for (see sec.~\ref{sec:cpinv}), and for two fixed values of the degeneracy $|M_2-M_1|/M_1$. 

\begin{itemize}

\item Weak washout

Defined by the condition
\begin{eqnarray}
\Gamma_\alpha(T_{EW}) < \Gamma(T_{EW}) < H_u(T_{EW})\,.
\label{eq:weak}
\end{eqnarray} 
The thermalization rates of all modes are slow compared to the Hubble rate, so an asymmetry can survive at $T_{\rm EW}$. The condition of eq.~(\ref{eq:weak}) can be translated on the plane $|U^2|$ versus $M_i$, leading to an {\it upper bound} on the mixing, as shown in Fig.~\ref{fig:regimes}. Unfortunately, this regime is beyond reach of future experiments such as \texttt{FCC}~\cite{Blondel:2014bra,FCC:2018byv,FCC:2018evy} and could be reached only in the lowest mass region at SHiP~\cite{SHiP:2015vad,SHiP:2018xqw}, where constraints from BBN are significant. An analytical approximate solution for the baryon asymmetry in this regime was first studied in \cite{Asaka:2005pn}, and including the connection to neutrino masses and other observable parameters in~\cite{Abada:2015rta, Hernandez:2015wna,Hernandez:2016kel}.

\item Flavoured weak washout

When there is a hierarchy  in $\epsilon_\alpha$, eq.~(\ref{eq:epsilonalpha}), we might have
\begin{eqnarray}
\Gamma_\alpha(T_{\rm EW}) < H_u(T_{\rm EW}) < \Gamma(T_{\rm EW})\,,
\label{eq:flavour}
\end{eqnarray}
for some $\alpha=e,\mu,\tau$. The flavour $\alpha$ remains the reservoir of the baryon asymmetry. In Fig.~\ref{fig:regimes},  we show the band corresponding to eq.~(\ref{eq:flavour}). This regime reaches up to two order magnitude larger couplings than in the weak washout. 

\end{itemize}

When $\Gamma_\alpha(T_{\rm EW}) \geq H_u(T_{\rm EW})$ for all $\alpha$, the asymmetry can only survive if any of the slow modes, in eqs.~(\ref{eq:slowosc}) and/or (\ref{eq:slowM}), remain in weak washout. Depending on which one of them does we can distinguish two additional regimes: 

\begin{itemize}
\item Overdamped regime

When
\bea
\epsilon \equiv {\Gamma_{osc} \over \Gamma }\ll 1\,,
\label{eq:ovregime}
\eea
oscillations are damped  by the faster interactions in the plasma and
\begin{eqnarray}
P_{\rm osc} \propto \epsilon^2\,.
\end{eqnarray}
As a result the slow thermalization rate of eq.~(\ref{eq:slowosc}) is suppressed as
\bea
\Gamma^{\rm slow}_{\rm osc} = \epsilon^2 \Gamma\,.
\eea
In strong washout, $\Gamma \gg H_u$,  the overdamped regime is defined by the condition
\begin{eqnarray}
\Gamma^{\rm slow}_{\rm osc}(T_{\rm EW}) < H_u(T_{\rm EW})\,,
\label{eq:ovcondition}
\end{eqnarray}
which implies a {\it lower limit} on the mixing, as is shown in Fig.~\ref{fig:regimes}. 

\item Weak lepton number violating (wLNV) regime

When $M/T$ terms are not negligible but still 
\begin{eqnarray}
\Gamma^{\rm slow}_{M}(T_{\rm EW}) < H_u(T_{\rm EW})\,,
\end{eqnarray}
the asymmetry can survive even when all other rates are larger than $H_u$.  This condition implies an upper limit in the mixing, although significantly less restrictive for small masses than the flavoured weak washout, as shown in Fig.~\ref{fig:regimes}. It is important to stress that this regime is not relevant if 
$M/T$ corrections are neglected: there is effectively an exact lepton number symmetry in this case and no asymmetry is generated in the slow mode direction. 

\end{itemize}

In the unshaded regions of Fig.~\ref{fig:regimes} the Sakharov conditions are not fulfilled at $T_{\rm EW}$, and therefore the asymmetry is exponentially suppressed. Only exponential fine-tunning could reproduce the baryon asymmetry.  

Within the shaded regions the generation of the asymmetry is most important at $T_{\rm osc}$ defined as:
\begin{eqnarray}
\Gamma_{\rm osc}(T_{\rm osc}) = H_u(T_{\rm osc})\,,
\end{eqnarray}
that is when the oscillation rate is the same as the Hubble expansion rate.
As we will see the asymmetry generated depends on the relative strength of $\Gamma_{\rm osc}$ and $\Gamma$ at this temperature. The dashed line on the Fig.~\ref{fig:regimes} separates two regions:
\begin{itemize}
\item Intermediate region (above dashed line)
\begin{eqnarray}
\Gamma_{\rm osc}(T_{\rm osc}) \ll \Gamma(T_{\rm osc})\,.
\label{eq:int}
\end{eqnarray}
\item  Fast oscillation region (below dashed line):  
\begin{eqnarray}
\Gamma_{\rm osc}(T_{osc}) \gg \Gamma(T_{\rm osc})\,.
\label{eq:fastosc}
\end{eqnarray}
\end{itemize}

Analytical approximations for the baryon asymmetry in the fast oscillation region in the limit of $\Gamma_M \rightarrow 0$ have been previously derived in~\cite{Hernandez:2015wna,Hernandez:2016kel,Drewes:2016gmt,Drewes:2017zyw}, while for the overdamped regime semi-analytical solutions in the same limit have been presented in~\cite{Drewes:2016gmt}.

\subsection{ CP--violating flavour invariants and baryogenesis }
\label{subsec:CP_invariants_general}

CP violation is a subtle effect related to the presence of physical complex couplings that generically involve many flavour parameters. The so-called CP flavour invariants~\cite{Jarlskog:1985ht,Jarlskog:1985cw,Bernabeu:1986fc,Branco:2001pq,Jenkins:2007ip,Jenkins:2009dy,Wang:2021wdq,Yu:2021cco} are flavour-basis-independent quantities  that incorporate the involved parameter dependencies that make the complex couplings physical. All CP violating observables such as the baryon asymmetry must be proportional to a combination of such CP flavour invariants. Our goal is to obtain these relations, that will then provide a strong crosscheck of the analytical approximations to the baryon asymmetry derived in sec.~\ref{sec:ana}, which are expected to be proportional to such invariants. Further, this can allow to derive robust connections to other observables.
  
CP flavour invariants are constructed out of the flavour parameters in the model, i.e. the physical parameters in the matrices $Y$ and $M_R$, as well as the charged lepton Yukawa matrix, $Y_l$. If the observable in question can be obtained as a series expansion in these matrices, the relevant CP flavour invariants are polynomials in the matrices, which are invariant under flavour basis transformations and have an imaginary part. 
An exhaustive list of invariants in this model has been found using the Hilbert series in~\cite{Jenkins:2009dy,Wang:2021wdq,Yu:2021cco}. 
Note that the baryon asymmetry is not expected to be proportional to any of those basic invariants, since the dependence on $M$ or $Y$ need not be polynomial, and the thermal plasma provides a reference that distinguishes the charged lepton flavour. However they do contain the building blocks from which the flavoured or unflavoured invariants that appear in leptogenesis can be obtained.
  
  In order to construct the relevant CP invariants to our problem, let us first consider how  $Y$, $Y_l$ and $M_R$ are transformed under a change of flavour basis that leaves the kinetic and gauge interactions invariant:
\bea
Y&\rightarrow V^\dagger Y W,\;\; Y_l \rightarrow V^\dagger Y_l U,\;\; M_R \rightarrow W^TM_R W\,,
\eea
where $U, V,$ and $W$ are respectively generic three or two-dimensional unitary matrices, 

Taking this into account, we can consider the following hermitian combinations 
\bea
h =Y^\dagger Y \rightarrow W^\dagger h W,\;\; \overline{h}=Y^\dagger Y_lY_l^\dagger Y \rightarrow W^\dagger \overline{h} W,\;\;H_M=M_R
^\dagger M_R \rightarrow W^\dagger H_M W\,.
\eea
Combinations that involve $M_R$ only via the hermitian matrix $H_M$ are not sensitive to the Majorana character \footnote{If we consider $M_R$ a spurion that enforces rephasing invariance of the Majorana fields, $M_R$ picks a phase under this transformation while $H_M$ remains invariant.}. They are relevant for the lepton number conserving (LNC) case, i.e. when $M_i/T$ effects in the rates are neglected. 

\subsubsection{LNC invariants}
The simplest invariant built up out of $Y, Y_l$ and $M$, which does not vanish when the Majorana character is irrelevant, is given by~\cite{Jenkins:2009dy}
\be
I_0={\rm Im}\left( {\rm Tr} \left[ h\, H_M \overline{h}\,\right]\right)\,.
\ee
In the basis in which $Y_l$ and $M_R$ are diagonal, with eigenvalues $y_{l_\alpha}$ and $M_i$ respectively, the above quantity can be written as
\bea
I_0 &=&\frac{1}{2}\sum_\alpha y_{l_\alpha}^2 \sum_{i,j}\left(M_j^2-M_i^2\right){\rm Im}\left[ Y_{\alpha j}^*Y_{\alpha i}\left(Y^\dagger Y\right)_{ij}\right]\\
&=& 
\sum_\alpha y^2_{l_\alpha} \sum_{i<j}\left(M_j^2-M_i^2\right) {\rm Im}\left[ Y_{\alpha j}^*Y_{\alpha i}\left(Y^\dagger Y\right)_{ij}\right]
\equiv \sum_\alpha y_{l_\alpha}^2 \, \Delta_\alpha\,.\nonumber
\eea
Note that 
\be
\sum_{\alpha} \Delta_\alpha = {\rm Im}\left( {\rm Tr} \left[ h\, H_M h \,\right]\right)=0\,,
\ee
because  the matrix in the trace is hermitian and therefore its trace is real. 

At the temperatures we are interested in, the plasma can distinguish the charged lepton flavours. The lepton CP asymmetry generated in the neutral lepton sector in flavour $\alpha$ is proportional to the basic quantity $\Delta _\alpha$, and  the net lepton asymmetry is given by a weighted combination of $\Delta_\alpha$, with different weights in different regimes.

{\it  Overdamped regime} 

Since $\Delta_\alpha \propto \Delta M\sim \Gamma_{\rm osc}$, and the coherent oscillation is cut off by $\Gamma^{-1}_\alpha$ we expect
\bea
\Delta^{\rm ov}_{\rm LNC} \propto \sum_\alpha { \Delta_\alpha\over \Gamma_\alpha}.
\eea
Including an extra invariant normalization to match the analytical result in sec.~\ref{subsec:solutions}, the full flavour-dependence of the asymmetry in this regime will be proportional to:
\be
\label{eq:invariant_LNC_ov}
\Delta^{\rm ov}_{\rm LNC}=\frac{1}{\left[{\rm Tr}\left(Y^\dagger Y\right)\right]^2}\sum_{\alpha}\frac{1}{\left(YY^\dagger\right)_{\alpha\alpha}}
\sum_{i<j}\left(M_j^2-M_i^2\right) {\rm Im}\left[ Y_{\alpha j}^*Y_{\alpha i}\left(Y^\dagger Y\right)_{ij}\right]\,.
\ee

{\it Flavoured weak washout}

There must be a weakly coupled flavour, $\alpha$, for the asymmetry to survive. In the \textit{intermediate region}, eq.~(\ref{eq:int}), the net asymmetry is simply the one obtained in flavour $\alpha$: 
\bea
\label{eq:invariant_LNC_int}
\Delta_{\rm LNC}^{\rm int(\alpha)} = \Delta_\alpha\,.
\eea 
In the \textit{fast oscillation region}, eq.~(\ref{eq:fastosc}),  the invariant that controls the production of asymmetry  is not simply proportional to $ \Gamma_{\rm osc}$ since this rate is large. A more general dependence on the masses is expected, but in any case it should be of the form
\bea
\label{eq:invariant_LNC_osc}
\Delta_{\rm LNC}^{\rm osc\, (\alpha)}
= \sum_{i<j} g(M_i,M_j) {\rm Im}\left[ Y_{\alpha j}^*Y_{\alpha i}\left(Y^\dagger Y\right)_{ij}\right]\,,
\eea
where $g(M_i,M_j)$ is an antisymmetric function of the two arguments. The precise form of this function will be fixed after matching to the analytical solution. 

\subsubsection{LNV invariants}

When $M/T$ corrections to the rates cannot be neglected, additional invariants become relevant, that are sensitive to the Majorana character of the HNLs.

The simplest non-vanishing invariant of this type is given by~\cite{Branco:2001pq,Jenkins:2009dy}
\bea
I_1&=&{\rm Im}\left\{ {\rm Tr} \left[ h\, H_M M^*h^* M\,\right]\right\}= \sum_{i<j}\left(M_j^2-M_i^2\right)M_iM_j {\rm Im}\left[\left(h_{ij}\right)^2\right]\nonumber\\
&=& \sum_{\alpha} \sum_{i<j}\left(M_j^2-M_i^2\right)M_iM_j {\rm Im}\left[Y_{\alpha j}Y_{\alpha i}^*\left(Y^\dagger Y\right)_{ij}\right]\equiv \sum_\alpha \, \Delta_\alpha^{M}\,.
\eea
Note that it does not involve the charged lepton Yukawa. 
 
 {\it Overdamped regime}
  
 The asymmetry in the overdamped regime is expected to be proportional to the full invariant up to a normalization:
 \bea
\label{eq:invariant_LNV_ov}
\Delta_{\rm LNV}^{\rm ov}&=& \frac{1}{\left[{\rm Tr}\left(Y^\dagger Y\right)\right]^2}\sum_{\alpha}\Delta_\alpha^M \\
&=& \frac{1}{\left[{\rm Tr}\left(Y^\dagger Y\right)\right]^2}\sum_{\alpha}
\sum_{i<j}\left(M_j^2-M_i^2\right)M_iM_j {\rm Im}\left[Y_{\alpha j}Y_{\alpha i}^*\left(Y^\dagger Y\right)_{ij}\right]\,.\nonumber
\eea
Again, the extra normalization factor is introduced to match the analytical result  to be shown in sec.~\ref{subsec:solutions}.

{\it Flavoured weak washout}

The asymmetry is that obtained in flavour $\alpha$ and the expected invariant is thus given by:
\bea
\label{eq:invariant_LNV_int}
\Delta_{\rm LNV}^{\rm int\, (\alpha)} &=& \frac{\Delta_\alpha^M }{\left[{\rm Tr}\left(Y^\dagger Y\right)\right]^2}\nonumber\\
&=& \frac{1}{\left[{\rm Tr}\left(Y^\dagger Y\right)\right]^2}\sum_{i<j}\left(M_j^2-M_i^2\right)M_iM_j {\rm Im}\left[Y_{\alpha j}Y_{\alpha i}^*\left(Y^\dagger Y\right)_{ij}\right]\,,
\eea
for the \textit{intermediate regime}, where we introduce the same normalization factor as in the previous case, eq.~(\ref{eq:invariant_LNV_ov}).

In the \textit{fast oscillation region} we expect:
\be
\label{eq:invariant_LNV_osc}
\Delta^{\rm osc}_{\rm LNV}=\frac{1}{{\rm Tr}\left(Y^\dagger Y\right)}\sum_{\alpha}\sum_{i<j}{\rm Im}\left[Y_{\alpha j}Y_{\alpha i}^*\left(Y^\dagger Y\right)_{ij}\right] g_M(M_i,M_j)\,,
\ee
where the antisymmetric function $g_M(M_i,M_j)$ will be determined after matching to the analytical solution.


\section{ CP invariants versus neutrino masses}
\label{sec:cpinv}
 
Let us first show the expressions for the CP invariants presented in the previous section considering the parameterization given in eq.~(\ref{eq:LNVparam2N}), and expanding in the small symmetry breaking parameters $y_\beta'$ and  $\mu_2$. 
\bea
\frac{\Delta^{\rm ov}_{\rm LNC}}{M_2^2-M_1^2}&=& -2\sum_\alpha \frac{y_\alpha y'_\alpha \sin\Delta \beta_\alpha}{y^2} \left(\frac{1}{y_\alpha^2}
-\frac{3}{y^2}\right)\,,
\label{eq:CP_Invariants_Yukawa1}\\
\frac{\Delta_{\rm LNC}^{\rm int(\alpha)}}{M_2^2-M_1^2}&=& \frac{ \Delta_{\rm LNC}^{\rm osc\, (\alpha)}}{g(M_1,M_2)}=\frac{1}{2}\sum_{\beta\neq \alpha} \left(  y_\alpha^2  y_\beta y_\beta' \sin\Delta\beta_\beta-y_\alpha y'_\alpha   y_\beta^2\sin\Delta\beta_\alpha\right)\,,
\label{eq:CP_Invariants_Yukawa2}\\
\frac{\Delta_{\rm LNV}^{\rm ov}}{M_1 M_2 (M_2^2-M_1^2)}&=& \frac{1}{y^2}\frac{\Delta^{\rm osc}_{\rm LNV}}{g_M(M_1,M_2)} = -\frac{\sum_\alpha\,y_\alpha y'_\alpha \sin\Delta \beta_\alpha}{y^2}\,,
\label{eq:CP_Invariants_Yukawa3}\\
\frac{\Delta_{\rm LNV}^{\rm int(\alpha)}}{M_1 M_2 (M_2^2-M_1^2)}&=&-\frac{1}{2}\left(\frac{y_\alpha y'_\alpha \sin\Delta \beta_\alpha}{y^2}+\frac{y_\alpha^2}{y^2}\,\frac{\sum_\beta y_\beta y'_\beta \sin\Delta\beta_\beta}{y^2}\right)\,,
\label{eq:CP_Invariants_Yukawa4}
\eea
where $y^2=\sum_\alpha |Y_{\alpha 1}|^2=\sum_\alpha y_\alpha^2$. The CP phases appear in the combinations $\Delta \beta_\alpha = \beta'_\alpha-\beta_\alpha$. This is expected since in the minimal model with two HNLs there are only three physical phases: the Majorana and Dirac phases included in the PMNS matrix and another phase associated to the HNL sector. Recall that in the symmetric limit ($y'_\alpha=\mu_2=0$) there is no CP violation  (see also appendix \ref{sec:app}).
  
On the other hand, the CP invariants can be related to the physical neutrino masses and other observable HNL parameters. Using eq.~(\ref{eq:mnuLN}), the light neutrino mass constraint reads
 
\be
- \left( m_{\nu}\right)_{\alpha\beta} = \frac{v^2}{\Lambda}\left( Y_{\alpha 1} Y_{\beta 2} + Y_{\alpha 2} Y_{\beta 1} - Y_{\alpha 1} Y_{\beta 1} \frac{\mu_2}{\Lambda}\right)=\left(U^*m\,U^\dagger\right)_{\alpha\beta}\,,
\label{eq:mnuU}
\ee 
where $U=U(\theta_{12},\theta_{13},\theta_{23},\delta,\phi)$  is the PMNS matrix\footnote{We use the parameterization of the PDG \cite{ParticleDataGroup:2020ssz}.} describing the light neutrino mixing observed in neutrino oscillation experiments, and $m$ is the diagonal matrix of the light neutrino masses. 
The Yukawa couplings can then be written as a function of the PMNS and neutrino mass parameters~\cite{Gavela:2009cd}. The expressions differ in the normal and inverted hierarchy case.

{\it Normal Hierarchy (NH)}

The Yukawas satisfy
\bea
\label{eq:Yno_PMNS}
Y_{\alpha 1}&=&\frac{e^{-i\theta/2}y}{\sqrt{2}}\left(U^*_{\alpha 3}\sqrt{1+\rho}+U^*_{\alpha 2}\sqrt{1-\rho}\right)\,,\nonumber\\
Y_{\alpha 2}&=&\frac{e^{i\theta/2}y'}{\sqrt{2}}\left(U^*_{\alpha 3}\sqrt{1+\rho}-U^*_{\alpha 2}\sqrt{1-\rho}\right)
+\frac{\Delta M}{4M}\frac{e^{-i\theta/2}y}{\sqrt{2}}\left(U^*_{\alpha 3}\sqrt{1+\rho}+U^*_{\alpha 2}\sqrt{1-\rho}\right)\,,\nonumber\\
\eea
where $y$ is a real free parameter and\footnote{In this parameterization $m_3<0$ ($m_2<0$) for NH (IH)~\cite{Gavela:2009cd}. This negative sign can be reabsorbed with a redefinition of the Majorana phase included in the PMNS matrix $U$.} 
\bea
\rho = \frac{\sqrt{\Delta m^2_{\rm atm}}-\sqrt{\Delta m^2_{\rm sol}}}{\sqrt{\Delta m^2_{\rm atm}}+\sqrt{\Delta m^2_{\rm sol}}},\;\;\;\;\; y' = \frac{M}{2v^2y}\left(\sqrt{\Delta m^2_{\rm atm}}+\sqrt{\Delta m^2_{\rm sol}}\right)\,.
\label{rhoNH}
\eea
Note that besides the phases in the PMNS matrix, there is an additional phase, $\theta$, associated to the HNL sector, that will play a major role in the baryon asymmetry. 

{\it Inverted Hierarchy (IH)}

In this case, we have
\bea
\label{eq:Yio_PMNS}
Y_{\alpha 1}&=&\frac{e^{-i\theta/2}y}{\sqrt{2}}\left(U^*_{\alpha 2}\sqrt{1+\rho}+U^*_{\alpha 1}\sqrt{1-\rho}\right)\,,\nonumber\\
Y_{\alpha 2}&=&\frac{e^{i\theta/2}y'}{\sqrt{2}}\left(U^*_{\alpha 2}\sqrt{1+\rho}-U^*_{\alpha 1}\sqrt{1-\rho}\right)
+\frac{\Delta M}{4M}\frac{e^{-i\theta/2}y}{\sqrt{2}}\left(U^*_{\alpha 2}\sqrt{1+\rho}+U^*_{\alpha 1}\sqrt{1-\rho}\right)\,,\nonumber\\
\eea
where again $y$ is real and arbitrary while
\bea
\rho &=& \frac{\sqrt{\Delta m^2_{\rm atm}}-\sqrt{\Delta m^2_{\rm atm}-\Delta m^2_{\rm sol}}}{\sqrt{\Delta m^2_{\rm atm}}+\sqrt{\Delta m^2_{\rm atm}-\Delta m^2_{\rm sol}}}\,,\nonumber\\
y' &=& \frac{M}{2v^2y}\left(\sqrt{\Delta m^2_{\rm atm}}+\sqrt{\Delta m^2_{\rm atm}-\Delta m^2_{\rm sol}}\right)\,.
\label{rhoIH}
\eea
The parameters of the right handed neutrino Majorana mass matrix are related to the physical HNL masses as  (recall that we assume $\mu_2=\mu_1$)
\bea
\mu_2 =\frac{M_2-M_1}{2}\equiv \frac{\Delta M}{2},\;\;\;\;\; \Lambda = \frac{M_2+M_1}{2}\equiv M\,.
\eea
Note that $y$ essentially gives the magnitude of the $|Y_{\alpha 1}|$ Yukawa couplings, while $y'$ and $\Delta M/M$ sets the scale of $|Y_{\alpha 2}|$. 

The HNL flavour mixing is given by
\be
\label{eq:Hlmixing}
\Theta^*= 
YvM_R^{-1}W^*,
\ee
where $W$ is the unitary matrix which diagonalizes $M_R$, see eq~(\ref{eq:W}). In particular, we obtain
\bea
\label{eq:U2}
U^2&\equiv& \sum_\alpha |\Theta_{\alpha I}|^2=\frac{y^2v^2}{2 M^2} \left[1 \pm\frac{3\Delta M}{4M}\mp 2\,\rho \cos\theta \,\frac{y'}{y}
+\mathcal{O}\left(\frac{y'^2}{y^2}\right)+\mathcal{O}\left(\frac{(\Delta M)^2}{M^2}\right)\right]\nonumber\\
&\approx& \frac{y^2v^2}{2 M^2}\,,
\eea
where the upper (lower) sign corresponds to the first (second) heavy mass eigenstate. 

Using the above expressions, we can rewrite the CP invariants  of eqs.~(\ref{eq:invariant_LNC_ov}), (\ref{eq:invariant_LNC_int}), (\ref{eq:invariant_LNC_osc}), (\ref{eq:invariant_LNV_ov}), (\ref{eq:invariant_LNV_int}) and (\ref{eq:invariant_LNV_osc}) as a function of the physical parameters. In order to illustrate the main dependencies, we will expand over $y'/y$, $\Delta M/M$ and the small light neutrino parameters
\be
r\equiv\frac{\sqrt{\Delta m^2_{\rm sol}}}{\sqrt{\Delta m^2_{\rm atm}}}\sim\theta_{13}\sim|\theta_{23}-\pi/4|\sim 10^{-1}\,.
\ee
At leading order in the expansion parameters we obtain the following simple expressions:

{\it Normal Hierarchy}

\begin{align}
\frac{\Delta^{\rm ov}_{\rm LNC}}{M_2^2-M_1^2}
&\approx -
\frac{v^2 \sqrt{\Delta m^2_{\rm atm}}}{8M^3 U^4} s_\theta\,,
\label{eq:ovlncinvnh}
\\
\frac{\Delta_{\rm LNC}^{\rm int(e)}}{M_2^2-M_1^2}&= \frac{\Delta_{\rm LNC}^{\rm osc\,(e)}}{g(M_1,M_2)}\approx U^2  M^3 \frac{\sqrt{\Delta m^2_{\rm atm}}}{v^4}\,  r\, s_{12}^2s_\theta\,,
\label{eq:intlncinvnh}\\
\frac{\Delta_{\rm LNC}^{\rm int(\mu)}}{M_2^2-M_1^2}&= \frac{\Delta_{\rm LNC}^{\rm osc\,(\mu)}}{g(M_1,M_2)}\approx \frac{U^2 M^3}{2} \frac{\sqrt{\Delta m^2_{\rm atm}}}{v^4} 
\sqrt{r}\,c_{12}\sin(\theta-\phi)\,,\\
\frac{\Delta_{\rm LNC}^{\rm int(\tau)}}{M_2^2-M_1^2}&= \frac{\Delta_{\rm LNC}^{\rm osc\,(\tau)}}{g(M_1,M_2)} \approx -\frac{\Delta_{\rm LNC}^{\rm int(\mu)}}{M_2^2-M_1^2},\\
\frac{\Delta_{\rm LNV}^{\rm ov}}{M_1 M_2 (M_2^2-M_1^2)}&= \frac{v^2}{2 U^2 M^2}\frac{\Delta^{\rm osc}_{\rm LNV}}{g_M(M_1,M_2)} 
\approx -\frac{\sqrt{\Delta m^2_{\rm atm}}}{4M U^2 } s_\theta\,.
\label{eq:osclnvnh}
\end{align}

{\it Inverted Hierarchy}

\bea
\label{eq:CP_invariants_IH_leading_order}
\frac{\Delta^{\rm ov}_{\rm LNC}}{M_2^2-M_1^2} 
&\approx&  \frac{v^2 \sqrt{\Delta m^2_{\rm atm}}}{8M^3 U^4}
\dfrac{\left(1 + 3 c_\phi\sin2\theta_{12}\right) \left(c_\theta s_\phi\sin2\theta_{12} + 
  s_\theta \cos2\theta_{12}\right)}{-1 + c_\phi^2 \sin^22\theta_{12}}\,,
\label{eq:ovlncinvih}\\
\frac{\Delta_{\rm LNC}^{\rm int(e)}}{M_2^2-M_1^2}&=& \frac{\Delta_{\rm LNC}^{\rm osc\,(e)}}{g(M_1,M_2)}\approx 
 \frac{U^2 M^3}{2} \frac{\sqrt{\Delta m^2_{\rm atm}}}{v^4} 
 (\sin2\theta_{12}s_\phi\, c_{\theta}+ \cos2\theta_{12}\,s_{\theta} )\,,\,\,\,\,\,\,\,
\label{eq:intlncinvih}\\
\frac{\Delta_{\rm LNC}^{\rm int(\mu)}}{M_2^2-M_1^2}&=& \frac{\Delta_{\rm LNC}^{\rm osc\,(\mu)}}{g(M_1,M_2)}\approx 
\frac{\Delta_{\rm LNC}^{\rm int(\tau)}}{M_2^2-M_1^2}= \frac{\Delta_{\rm LNC}^{\rm osc\,(\tau)}}{g(M_1,M_2)}\approx -{1\over 2} {\Delta_{\rm LNC}^{\rm int (e)} \over M_2^2 -M_1^2}\,,
\\
\frac{\Delta_{\rm LNV}^{\rm ov}}{M_1 M_2 (M_2^2-M_1^2)}&=& \frac{v^2}{2 U^2 M^2}\frac{\Delta^{\rm osc}_{\rm LNV}}{g_M(M_1,M_2)} 
\approx -\frac{\sqrt{\Delta m^2_{\rm atm}}}{8M U^2} r^2 s_\theta\,.
\label{eq:osclnvih}
\eea

All the CP invariants depend on the ``high scale'' phase $\theta$ and, remarkably, $\Delta_{\rm LNV}^{\rm osc}$ and $\Delta_{\rm LNV}^{\rm ov}$ only depend on this phase. Indeed, it can be easily checked that this is a general result, satisfied to all orders in the expansion. All the other invariants are also functions of the PMNS CP phases $\phi$ (Majorana) and $\delta$ (Dirac). Even if this dependence can be subleading (as it is always the case for $\delta$ due to the suppression in $\theta_{13}$), the corrections may be relevant for values of the parameters that suppress the leading order. 

Our results can be mapped to the Casas-Ibarra parameterization following the prescription given in appendix~\ref{sec:CasasIbarra}.


\section{Baryon asymmetry: kinetic equations and analytical approximations}
\label{sec:ana}

\subsection{Kinetic equations}
The quantum kinetic equations that describe the generation of the baryon asymmetry have been studied in detail before (see for instance~\cite{Ghiglieri:2017gjz} for the complete derivation of the kinetic equations). 
We use the same equations as derived in~\cite{Hernandez:2016kel}, but adding the LNV corrections to the rates that have been computed in~\cite{Ghiglieri:2017gjz}. 
We have checked that they are equivalent to those in~\cite{Ghiglieri:2017gjz}, but neglecting the hypercharge chemical potential, which is a small effect. 
We consider only the momentum-averaged approximation, which reproduces the full momentum computation up to $\mathcal{O}(1)$ effects in the BAU~\cite{Ghiglieri:2018wbs, Asaka:2011wq}.

We work in the basis where $M={\rm diag}(M_1,M_2)$, with $M_2 > M_1 > 0$. 
We define the normalized heavy neutrino density matrices for the two helicities:
\begin{eqnarray}
r_N = {\rho_N \over \rho_F}, \;\;\;\; r_{\bar N} = {\rho_{\bar N}\over \rho_F}\,,
\end{eqnarray}
where  $\rho_F(z) =( \exp z + 1)^{-1}$ with $z=k/T$ is the Fermi-Dirac distribution.
The evolution of these matrices as a function of the scale factor $x  = a = T^{-1}$ is dictated by the equations:
\begin{eqnarray}
x H_u {\text{d} r_N\over \text{d} x} &=& -i [\langle H\rangle, r_N]  -{\langle\gamma^{(0)}_N\rangle\over 2} \{Y^\dagger Y, r_N-1\}-  x^2{\langle s^{(0)}_N\rangle\over 2} \{M Y^T Y^* M, r_N-1\}\nonumber\\
&+&  \langle \gamma_N^{(1)} \rangle Y^\dagger \mu Y  -  x^2 \langle s_N^{(1)} \rangle M Y^T \mu Y^* M \nonumber\\
&-& {\langle \gamma_N^{(2)}\rangle \over 2}  \big\{Y^\dagger \mu Y,r_N\big\} + x^2 \frac{\langle s_{N}^{(2)} \rangle}{2} \{M Y^T \mu Y^* M, r_N\} \,,\nonumber\\ 
x H_u {\text{d} r_{\bar N}\over \text{d} x} &=& -i [\langle H^*\rangle, r_{\bar N}]  -{\langle\gamma^{(0)}_N\rangle\over 2} \{Y^T Y^*, r_{\bar N}-1\} -x^2 {\langle s^{(0)}_N\rangle\over 2} \{M Y^\dagger Y M, r_{\bar N}-1\} 
\nonumber\\
&-&  \langle \gamma_N^{(1)} \rangle Y^T \mu Y^*   + x^2 \langle s_N^{(1)} \rangle M Y^\dagger \mu Y M \nonumber\\   
&+&  {\langle \gamma_N^{(2)}\rangle \over 2}  \big\{Y^T \mu Y^*,r_{\bar N}\big\} - x^2 \frac{\langle s_{N}^{(2)} \rangle}{2} \{M Y^{\dagger} \mu Y M, r_{\bar{N}}\} \,,\nonumber\\  
x H_u {\text{d} {\mu}_{B/3-L_\alpha}\over \text{d} x} & = & {\int_k \rho_F\over \int_k \rho'_F}  \left[{\langle \gamma_N^{(0)}\rangle \over 2} (Y r_N Y^\dagger- Y^* r_{\bar N} Y^T) - x^2 {\langle s_N^{(0)}\rangle \over 2} (Y^* M r_N M Y^T- Y M r_{\bar N} M Y^\dagger) \right.\nonumber\\
&-&\left.\mu_\alpha \left(\langle\gamma_N^{(1)}\rangle YY^\dagger +x^2 \langle s_N^{(1)}\rangle Y M^2Y^\dagger   \right)
 +   {\langle\gamma_N^{(2)}\rangle\over 2} \mu_\alpha (Y r_N Y^\dagger+Y^* r_{\bar N} Y^T)    \right. \nonumber\\
&+&\left. x^2 \frac{\langle s_{N}^{(2)}\rangle}{2} \mu_{\alpha}\left(YMr_{\bar{N}}MY^{\dagger} + Y^*Mr_{N}MY^{T} \right) \right]_{\alpha\alpha}\,,
\label{eq:rhonrhonbarav}
\end{eqnarray}
where $H_u(T)$ is the Hubble parameter of eq.~(\ref{eq:hubble}) and $\rho_{F}'=\text{d}\rho_{F}/\text{d}z$. In these equations, the matrix $\mu\equiv {\rm diag}(\mu_\alpha)$ and $\mu_\alpha$  is the lepton chemical potential in flavour $\alpha$. $\mu_{B/3-L_\alpha}$ is related to the approximately conserved charge densities as:
\bea 
\label{eq:n_to_mu}
n_{B/3-L_\alpha} \equiv -2 \mu_{B/3 -L_\alpha} \int_k \rho'_F = {1\over 6} \mu_{B/3 -L_\alpha} T^3\,.
\eea
The relation between the two is 
\bea
\mu_{\alpha} &=& - \sum_\beta C_{\alpha\beta} \mu_{B/3-L_\beta}\,,
\eea
where the matrix $C$ is given by~\cite{Abada:2018oly}
\bea
C = -\frac{1}{711}\left(
\begin{array}{ccc}
257 & 20 & 20  \\
 20 & 257 & 20 \\
 20 & 20 & 257 \\
\end{array}
\right)\,.
\label{eq:Cmatrix}
\eea
The Hamiltonian term is given by\footnote{We neglect mass effects in the thermal mass \cite{Antusch:2017pkq}, since we have checked that in the parameter space considered they are negligible. They can become relevant for smaller mass splittings that those considered.}
\begin{eqnarray}
H \equiv {M^2 \over 2 k_0} + V_N(k), \;\;\; V_N(k) \equiv {T^2 \over 8 k_0} Y^\dagger Y\,.
\end{eqnarray}
The LNC rates including  $1\leftrightarrow 2$ and $2\leftrightarrow 2$ processes have been expanded  to linear order in the leptonic chemical potential:
\begin{eqnarray}
\gamma_N(k,\mu_\alpha) \simeq \gamma_N^{(0)}+ \gamma_N^{(2)} \mu_\alpha\,,
\end{eqnarray}
while 
\bea
\label{eq:gamma1_definition}
\gamma_N^{(1)} \equiv \gamma_N^{(2)} - {\rho'_F\over \rho_F} \gamma_N^{(0)}\,.
\eea
The $s_N$ rates are expanded analogously. 
All the rates are momentum averaged:
\begin{eqnarray}
\label{eq:average_definition}
\langle (...)\rangle \equiv {\int_z (...) \rho_F(z)\over \int_z \rho_F(z)}\,.
\end{eqnarray}
Lastly, we define the factor
\begin{eqnarray}
{\int_k \rho_F\over \int_k \rho'_F} = -{9 \xi(3)\over \pi^2} \equiv -\kappa\,.
\end{eqnarray}

\begin{figure}[!t]
\centering
\begin{tabular}{cc}
\hspace{-0.5cm} \includegraphics[width=0.5\textwidth]{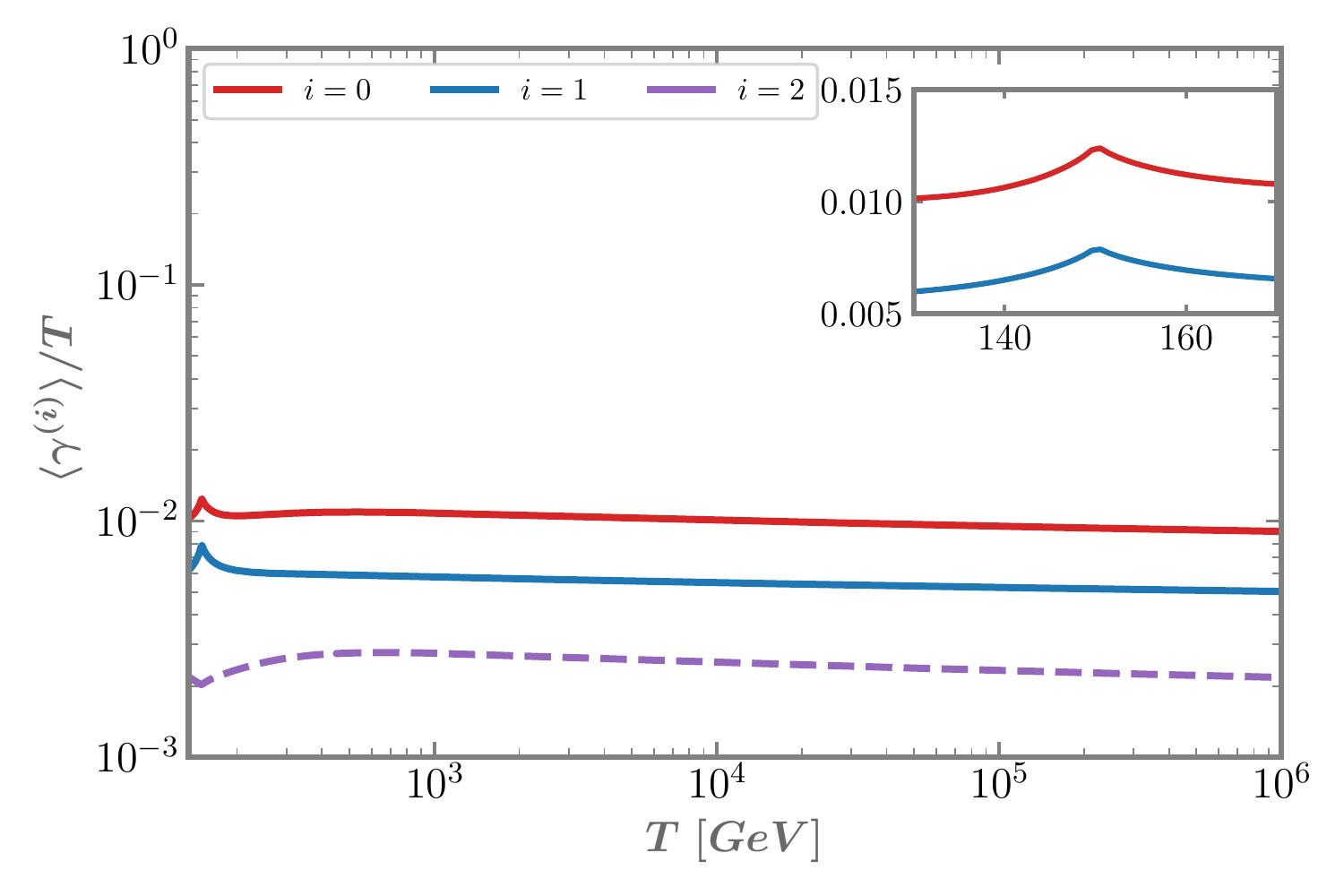} & \hspace{-0.55cm}  \includegraphics[width=0.5\textwidth]{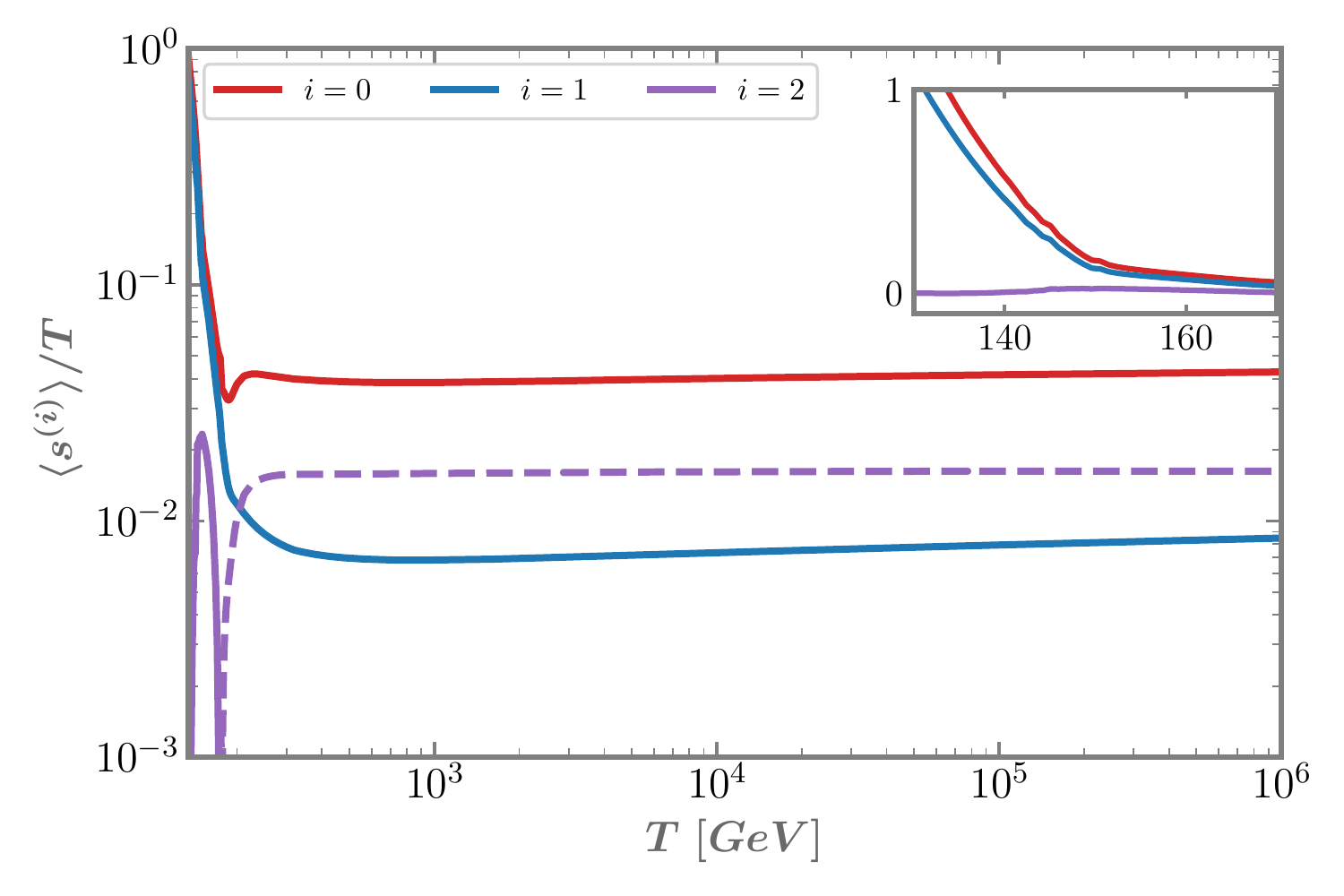}    \\
\end{tabular}
\vspace{-0.4cm}
\caption{ Temperature dependence of the ratios $\langle \gamma_N^{(n)}\rangle/T$ and $\langle s_N^{(n)}\rangle/T$. A dashed line indicates a negative contribution of a partial rate.}
\label{fig:rates}
\end{figure}

In table~\ref{tab:rates} we show the results for $\langle \gamma_N^{(n)}\rangle/T$ and $\langle s_N^{(n)}\rangle/T$ for $T=10^{6}~\text{GeV}$\footnote{Averaging over the Boltzmann distribution instead gives results that vary at the $\%$ level.}. 
\begin{table}[!t]
\begin{center}
\begin{tabular}{ccc}
$n$ & $\langle \gamma^{(n)}_N(T)\rangle/T $ & $\langle s^{(n)}_N(T)\rangle/T$ \\
\hline
\hline
0 &  0.0091 & 0.0434  \\
1 &  0.0051 & 0.0086 \\
2 & -0.0022 & -0.0165\\
\hline
\hline
\end{tabular}
\caption{Coefficients in the momentum averaged rates at $T= 10^6$ GeV.}
\label{tab:rates}
\end{center}
\end{table}
Their dependence with the temperature is shown in Fig.~\ref{fig:rates}. 
At large temperatures both quantities go to a constant.
When approaching the electroweak phase transition the LNV rates grow very significantly.

Note that in the above equations the terms proportional to $\langle \gamma_N^{(2)}\rangle$ and  $\langle s_N^{(2)}\rangle$ are non-linear. 
The equations are evolved from some small initial time, $x_{\rm ini}\sim 0$, where $r_N$ and $r_{\bar N}$ as well as the $\mu_{B/3-L_\alpha}$ vanish, up to the electroweak phase transition $x_{EW}$.\footnote{The effects associated to a non-zero initial abundance of HNLs has been studied in~\cite{Asaka:2017rdj}.}

\subsubsection{Interaction rates beyond the relativistic regime}

The relativistic approximation has been used in deriving the interaction rates of the HNL with the plasma in the kinetic equations. This is a good approximation when $M/T \ll 1$, but in the regime  $M/T_{\text{EW}} \lesssim1$, which can be tested at FCC, non-relativistic corrections  become important. In the absence of a full calculation of these corrections,  we adopt an educated guess\footnote{We thank M. Laine for this suggestion.}: the LNC rates are modified as 
\begin{align}
\gamma_N \to \frac{\mathcal{E} + k }{2 \mathcal{E}}   \gamma_N  \,,
\end{align}
while the LNV ones as 
\begin{align}
 s_N  {M^2\over T^2} \to  2\frac{k^2}{T^2} \frac{\mathcal{E} - k }{\mathcal{E}} s_N\,.
\end{align}
Here $k$ is the momentum and $\mathcal{E} = \sqrt{M^2 + k^2}$ is the particle energy. 

\begin{figure}[!t]
\centering
\begin{tabular}{cc}
\hspace{-0.5cm} \includegraphics[width=0.5\textwidth]{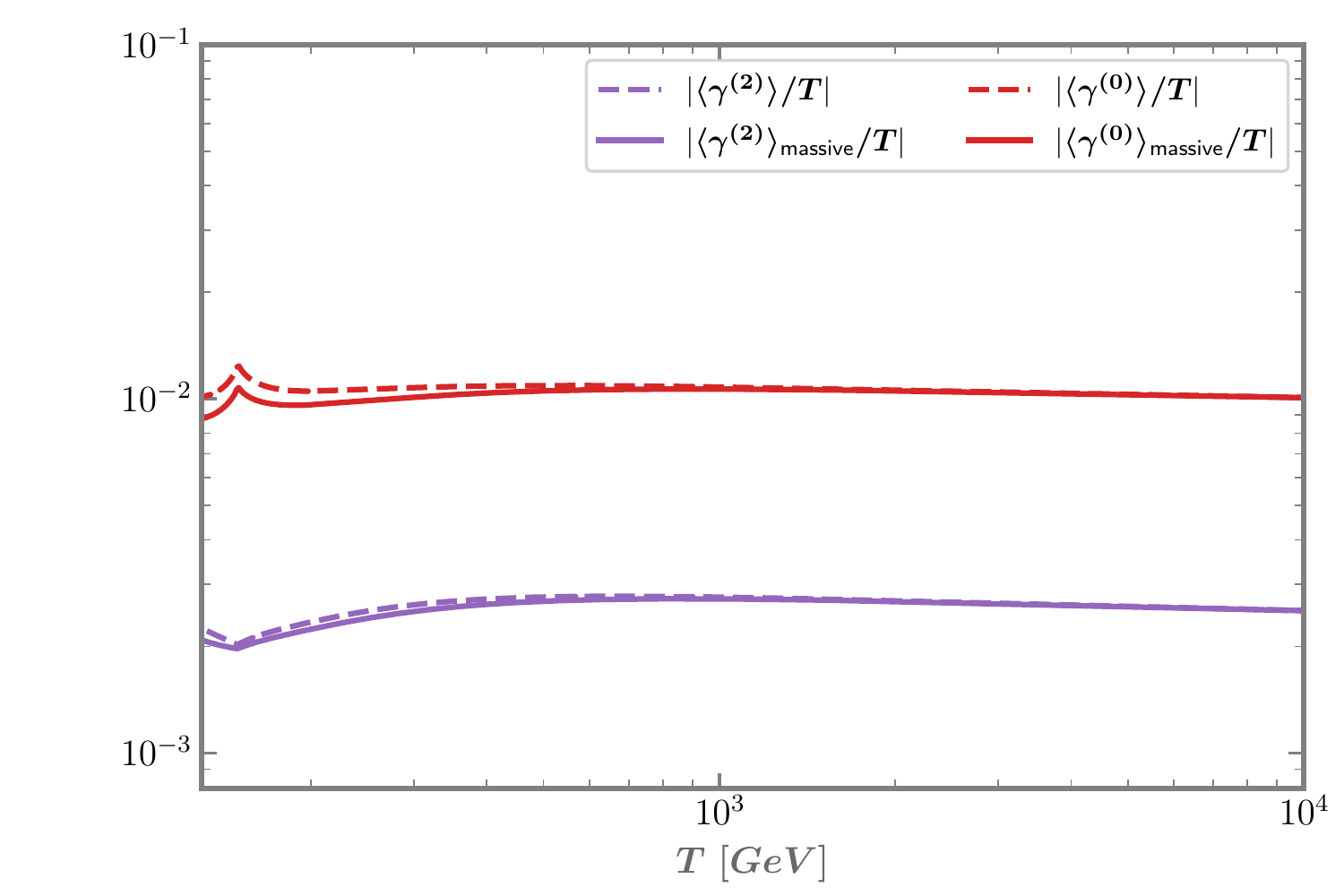} & \hspace{-0.55cm}  \includegraphics[width=0.5\textwidth]{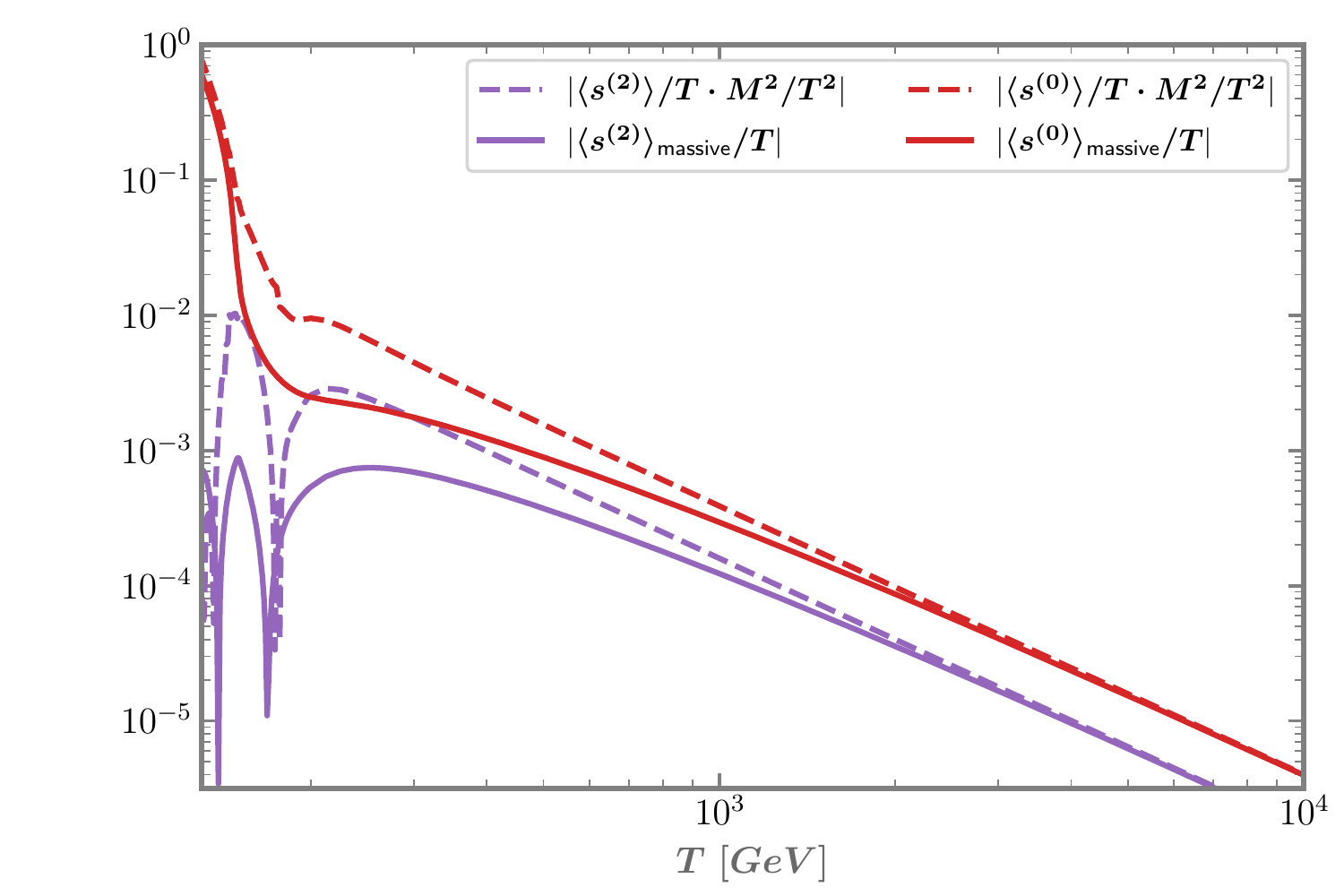}   
\end{tabular}
\vspace{-0.4cm}
\caption{ Effective interactions rates entering the kinetic equation including (excluding) non-relativistic corrections in bold (dashed). The effect in the LNC case (left) is negligible whereas in the LNV case (right) the effective rate contribution can change by $\mathcal{O}(1-10)$. We omit  the $\gamma^1$ and $s^1$ rates as they can be derived from the shown results via eq.~\eqref{eq:gamma1_definition}.}
\label{fig:rates_non_rel}
\end{figure}

The momentum and temperature dependent rates are taken from ref.~\cite{Ghiglieri:2017gjz}, 
and we average the new rates over the Fermi-Dirac distribution as indicated in eq.~\eqref{eq:average_definition}, but including the non-relativistic corrections.
Note that for the LNV rates we include explicitly the mass dependence, which means that the mass matrix is factorized by a common mass $M \simeq M_1 \simeq M_2$.
Hence, in the terms involving the LNV rates in the kinetic equations the mass matrix $M$ has to be replaced by the unit matrix.
The effects of the mass correction in the LNC case is at most of order percent but in the LNV case it can lead to $\mathcal{O}(1 -10)$ suppressions.
We show in Fig.~\ref{fig:rates_non_rel} the effect of the non-relativistic corrections on the rate for a mass of $M=100~\text{GeV}$ which will be the upper bound of our numerical search, and roughly the upper bound for direct searches at \texttt{FCC}. 

\subsection{Perturbation and adiabatic approximation}

In order to obtain an analytical approximation to these equations, we simplify them neglecting the non-linear terms, and also simplifying the matrix $C$ to take a diagonal form,  $C\rightarrow {\rm diag}(-1/2,-1/2,-1/2)$. 

Defining the 11-dimensional vector
\bea 
r(x) &\equiv& \left([r_N]_{11}, [r_N]_{22}, {\rm Re}([r_N]_{12}), {\rm Im}([r_N]_{12}),[r_{\bar N}]_{11}, [r_{\bar N}]_{22}, {\rm Re}([r_{\bar N}]_{12}), {\rm Im}([r_{\bar N}]_{12}),\right.\nonumber\\
& & \left. \mu_{B/3-L_e},\mu_{B/3-L_\mu},\mu_{B/3-L_\tau}\right)\,,
\eea
we can write the linearized differential equations in the compact form
\bea
{\text{d} r(x)\over \text{d} x} = A(x) r(x) + h(x)\,. 
\label{eq:rdot}
\eea

The goal is to find an analytical solution to these equations perturbing around the symmetric textures for $Y$ and $M$ in eq.~(\ref{eq:LNtexture}), and in the $M/T$ corrections in the rates. 
Hence, we can write 
\bea
A(x) &=& A^{(0)}(x) + A^{(1)}(x) +{\mathcal O}(y_\alpha',  (xM)^2)^2\,, \nonumber\\
 h(x) &=& h^{(0)}(x)+  h^{(1)}(x)+{\mathcal O}(y_\alpha',  (xM)^2)^2\,.
\eea
The leading order $r^{(0)}(x)$ solution satisfies:
\bea
{\text{d} r^{(0)}(x)\over \text{d} x} = A^{(0)}(x) r^{(0)}(x) + h^{(0)}(x)\,.
\label{eq:lo}
\eea
This equation is still hard to solve analytically, because $A^{(0)}(x)$ cannot be diagonalized by an $x$-independent change of basis. However, an adiabatic approximation can be employed when there is a large hierarchy between $\Gamma_{\rm osc}$ and $\Gamma$ independently of which is larger. 

At fixed $x$, we can diagonalize the matrix $A^{(0)}$:
 \begin{eqnarray}
 A^{(0)}(x) = V(x) \lambda(x) V(x)^{-1}\,,
 \end{eqnarray}
where $V(x)$ is the matrix of the eigenvectors in columns and $\lambda$ is the diagonal matrix containing the corresponding eigenvalues. If we neglect the $x$-variation of $V(x)$, the solution is the adiabatic approximation:
\begin{eqnarray}
r_a(x) = V(x) e^{\Lambda(x)} \int^x_0  e^{-\Lambda(z)} V^{-1}(z) h^{(0)}(z) \text{d}z\,,
\end{eqnarray}
with
\begin{eqnarray}
\Lambda(x) \equiv \int_0^x \lambda(z) \text{d}z\,.
\end{eqnarray}
This solution satisfies the equation:
\begin{eqnarray}
\dot r_a(x) = (A^{(0)}(x) +\dot V(x) V^{-1}(x) ) r_a(x) +h^{(0)}(x)\,.
\end{eqnarray}
In the overdamped regime,  $\dot V V^{-1}$ can be expanded in $\epsilon$,  eq.~(\ref{eq:ovregime}), and is found to be  ${\mathcal O}(\epsilon)$. Therefore we can include it as a perturbation up to corrections of higher order in $\epsilon$. Adding the correction:
\begin{eqnarray}
\delta r_a(x) = -V(x) e^{\Lambda(x)} \int^x_0  e^{-\Lambda(z)} V^{-1}(z) \dot V(z) V(z)^{-1} r_a(z) \text{d}z\,,
\end{eqnarray}
it is easy to show that the solution of eq.~(\ref{eq:lo}) is 
\begin{eqnarray}
r^{(0)}(x) = r_a(x)+ \delta r_a(x)\,,
\label{eq:r0}
\end{eqnarray}
up to ${\mathcal O}(\epsilon^2)$\,.

In the fast oscillation regime, $\Gamma_{\rm osc} \gg \Gamma$, we can instead expand in $\epsilon^{-1}$ and we find $\dot V V^{-1} = {\mathcal O}(\epsilon^{-1})$, so  the adiabatic solution can be obtained as in eqs.~(\ref{eq:r0}) and (\ref{eq:r1}), up to corrections ${\mathcal O}(\epsilon^{-2})$. 

We can now include the first order perturbation, $A^{(1)}$ and $h^{(1)}$, in the small parameters. The first order correction satisfies the equation:  
\begin{eqnarray}
{\text{d} r^{(1)}(x)\over \text{d} x} = A^{(0)}(x) r^{(1)}(x) + A^{(1)} r^{(0)}(x)+ h^{(1)}(x)\,,
\end{eqnarray}
which again can be solved in the adiabatic approximation. Defining
\begin{eqnarray}
{\tilde r}_a(x)\equiv V(x) e^{\Lambda(x)} \int^x_0  e^{-\Lambda(z)} V(z)^{-1} \left[A^{(1)}(z) r^{(0)}(z) + h^{(1)}(z) \right] \text{d}z\,,
\end{eqnarray}
and
\begin{eqnarray}
{\delta  \tilde r}_a\equiv - V(x) e^{\Lambda(x)} \int^x_0  e^{-\Lambda(z)} V(z)^{-1} \dot V(z) V(z)^{-1}  {\tilde r}_a(z) \text{d}z\,,
\end{eqnarray}
the first order correction to the solution is
\begin{eqnarray}
r^{(1)}(x) = {\tilde r}_a(x) + \delta {\tilde r}_a(x)\,,
\label{eq:r1}
\end{eqnarray}
up to ${\mathcal O}(\epsilon^2)$.

In the LNV case, we need to perturb simultaneously in ${\mathcal O}(y_\alpha')$ and ${\mathcal O}(M/T)^2$ corrections. In this case, it is necessary to go to second order. The corresponding expressions are straightforward. 

\subsubsection{Thermalization rates}
The thermalization rates are related to the real part of the eigenvalues of the matrix $A$. 
All the eigenvalues of the matrix $A(x)$ have negative real parts. The solution at $x\rightarrow \infty$ can then be shown to be the thermal equilibrium one: $r=(1,1,0,0,1,1,0,0,0,0,0)$, a limit which is approached exponentially. 

The approach to the asymptotic limit is controlled by the eigenvalues of $A^{(0)}$ in the adiabatic approximation. More precisely 
\begin{eqnarray}
\propto e^{- \Lambda_i(x)} \equiv \exp\left({-\int_0^x \text{d}z  |{\rm Re}(\lambda_{i}(z))|}\right)\,,
\end{eqnarray}
 with $\lambda_{i}$ the eigenvalues of $A^{(0)}$. 

We normalize $x$ such that at $T = T_{\text{EW}}$ we have $x_{\text{EW}}=1$, and define the dimensionless combinations
\begin{eqnarray}
\label{eq:defs}
\Delta \equiv {c_H\over 2}{ |M_2^2-M_1^2| M_P^*\over T_{\rm EW}^3}, ~\gamma_i \equiv {\langle \gamma^{(i)}\rangle\over T}  {M_P^*\over T_{\rm EW}},  ~s_i \equiv {\langle s^{(i)}\rangle\over T}  {M_P^*\over T_{\rm EW}}, ~\omega\equiv {c_H\over 8} {M_P^*\over T_{\rm EW}}\,,
\end{eqnarray}
with
\bea
c_H\equiv  {\pi^2 \over 18 \zeta(3)}\,.
\eea

The largest real part corresponds to the strong rate, that we can identify with $\Gamma$:
\begin{eqnarray}
\label{eq:Lambdamax}
\Lambda_{\rm max}(x) =\int_0^x \text{d} z~ {\rm Max}(|{\rm Re}(\lambda(z))|)  = {1\over 2} y^2 \gamma_0 x \equiv \int_0^x \text{d}z ~{\Gamma\over z H_u}\,.
\end{eqnarray}

Similarly, we can identify the slow rates described in sec.~\ref{sec:model} with  those associated to the eigenvalues of $A^{(0)}$ with the smallest real parts. In order for the corresponding mode not to thermalize before the EW transition it is necessary that
\bea
\Lambda_i(x_{\rm EW}) \leq 1\,.
\label{eq:noneq}
\eea

In the overdamped regime, we find modes that are suppressed by $\epsilon$: 
\begin{eqnarray}
\label{eq:eps}
{\rm Min}(|{\rm Re}(\lambda(x))|) = \epsilon^2  {\gamma_0^3 y^2 \over \gamma_0^2 + 4 \omega^2 }, ~~\epsilon(x) =  {x^2 \Delta\over   y^2 \gamma_0 }\,,
\end{eqnarray}
therefore
\begin{eqnarray}
\Lambda_{\rm ov}(x)= {x^5 \Delta^2\over 5 y^2} {\gamma_0 \over \gamma_0^2 + 4 \omega^2 }\equiv\int_0^x \text{d}z ~{\Gamma_{\rm osc}^{\rm slow} \over z H_u}\,.
\label{eq:betaov}
\end{eqnarray}
The boundary of the overdamped region is defined by 
\begin{eqnarray}
\Lambda_{\rm ov}(x_{\rm EW}) = 1\,.
\label{eq:ovboundary}
\end{eqnarray}
In the flavoured weak washout region, a slow mode remains in flavour $\alpha$ provided there is a hierarchy in the yukawas $y_\alpha/y \ll 1$. The slow rate of the flavoured weak washout regime is identified from the  corresponding eigenvalue
\begin{eqnarray}
\Lambda_\alpha(x)  \simeq {1\over 2} y_\alpha^2 \kappa \gamma_1 x \equiv \int_0^x \text{d} z {\Gamma_\alpha(z)\over z H_u(z)}\,.
\label{eq:gammaalpha}
\end{eqnarray}
The boundary of the weak flavour washout region is therefore
\begin{eqnarray}
\Lambda_{\alpha}(x_{\rm EW}) = 1\,.
\label{eq:wLNV}
\end{eqnarray}

On the other hand, one of the eigenvalues of $A^{(0)}$  is always  zero. This mode is associated with LN. It remains decoupled in the LNC limit (when $M/T\rightarrow 0$ in the rates), but it is weakly coupled when  $M/T$ terms are included. This mode is different in the overdamped regime or outside. 
In the overdamped regime we find 
\begin{eqnarray}
\Lambda^{\rm ov}_{\rm M}(x) =  {1\over 3} {M^2\over T_{\rm EW}^2} x^3 s_0 y^2 \equiv \int_0^x \text{d} z {\Gamma ^{\rm slow}_M(z)\over z H_u} \,,
\label{eq:Gamma_slow_M_ov}
\end{eqnarray}
while in the intermediate or fast oscillations we find
\begin{eqnarray}
\Lambda^{\rm int}_{\rm M}(x)= {1\over 3} \frac{M^2}{T_{\rm EW}^2} x^3 \frac{\gamma_1 s_0 + \gamma_0 s_1}{3\gamma_0  + \gamma_1 \kappa} \kappa y^2\,.
\label{eq:betaMint}
\end{eqnarray}
The boundary of the wLNV region is defined by the condition
\begin{eqnarray}
\Lambda_M(x_{\rm EW}) = 1\,,
\end{eqnarray}
and, as shown in Fig.~\ref{fig:regimes}, it is slighly different in the overdamped or intermediate regimes.

Finally the oscillation rate, which controls the generation of the asymmetry, is related instead to  the CP conserving phases corresponding to ${\rm Im}(\lambda_i)$
\begin{eqnarray}
\propto e^{-i \Lambda_{\rm osc}(x)} \equiv e^{-i \int_0^x dz  |{\rm Im}(\lambda_{i}(z))|}\,.
\end{eqnarray}
 It is found to be
\begin{eqnarray}
\Lambda_{\rm osc}(x)= {x^3\over 3} \Delta \equiv \int_0^x d z~  {\Gamma_{osc}\over z H_u}\,.
\end{eqnarray} 
The oscillation rate and the Hubble expansion are equal at $x_{\rm osc}$, which is  defined therefore by the condition
\begin{eqnarray}
\label{eq:x_osc}
\Lambda_{\rm osc}(x_{\rm osc}) = 1\,.
\end{eqnarray}

\subsubsection{Projection method}

In the intermediate regime, the asymmetry is basically built up at early times, when the evolution is in the overdamped regime, but it exits this regime before $x_{\rm EW}$. In this case, the adiabatic solution is not valid at the crossover between regimes. On the other hand, in these cases a quasi-stationary solution is found. A good approximation can be obtained from the solution in the overdamped regime evolved up to some threshold, $x_{\rm th}$ and then projecting it on the slow mode(s) direction(s). 

Let us denote by $v_i\,(w_i)$ the right (left) eigenvectors of $A$. They satisfy the orthonormality relation $w_i^\dagger v_j =\delta_{ij}$. Let us assume that at some time $x_{\rm th}$ a strong washout regime is reached with all modes strongly coupled with the plasma except one, with associated right (left) eigenvector $v_0\,(w_0)$ and corresponding to an approximate zero mode. 
Let us assume, as will be the case later, that these eigenvectors do not depend on $x$, and $w_0^\dagger\cdot h=0$, then:
\begin{eqnarray}
w_0^\dagger \cdot {\text{d} r(x)\over \text{d} x} = {\text{d} w_0^\dagger \cdot r(x)\over \text{d} x} \simeq 0\,.
\label{eq:proj}
\end{eqnarray}
Writing $r(x)$ in the basis of right eigenvectors: 
\begin{eqnarray}
r(x) = \sum_i a_i(x) v_i\,,
\end{eqnarray}
and substituting in eq.~(\ref{eq:proj}) implies that $a_0(x)$ is constant. Since all the other directions should have achieved thermalization, the large time quasi-stationary solution is therefore
\begin{eqnarray}
r(x) \simeq \left(w_0^\dagger\cdot r(x_{\rm th}) \right) v_0\,.
\end{eqnarray}
This result assumes $\Lambda_0(x) \simeq 0$.  At later times, this might not be a good approximation.  The time evolution in this case is well described by
\begin{eqnarray}
r(x) \simeq \left(w_0^\dagger\cdot r(x_{\rm th}) \right) v_0~ e^{-(\Lambda_0(x)-\Lambda_0(x_{\rm th}))}\,,
\end{eqnarray}
so that when $\Lambda_0(x) \gg 1$, the asymmetry is exponentially suppressed.

In some cases, we have two weakly coupled modes, with eigenvectors, $w_0, w_1$. In this case, a good approximation is 
\begin{eqnarray}
r(x) \simeq \left(w_0^\dagger\cdot r(x_{\rm th}) \right) v_0  e^{-(\Lambda_0(x)-\Lambda_0(x_{\rm th}))}+ \left(w_1^\dagger\cdot r(x_{\rm th}) \right) v_1 e^{-(\Lambda_1(x)-\Lambda_1(x_{\rm th}))}\,.
\label{eq:twozm}
\end{eqnarray}

\subsection{Solutions}
\label{subsec:solutions}

\begin{figure}[!t]
\centering
\hspace{-0.5cm} 
\includegraphics[width=0.97\textwidth]{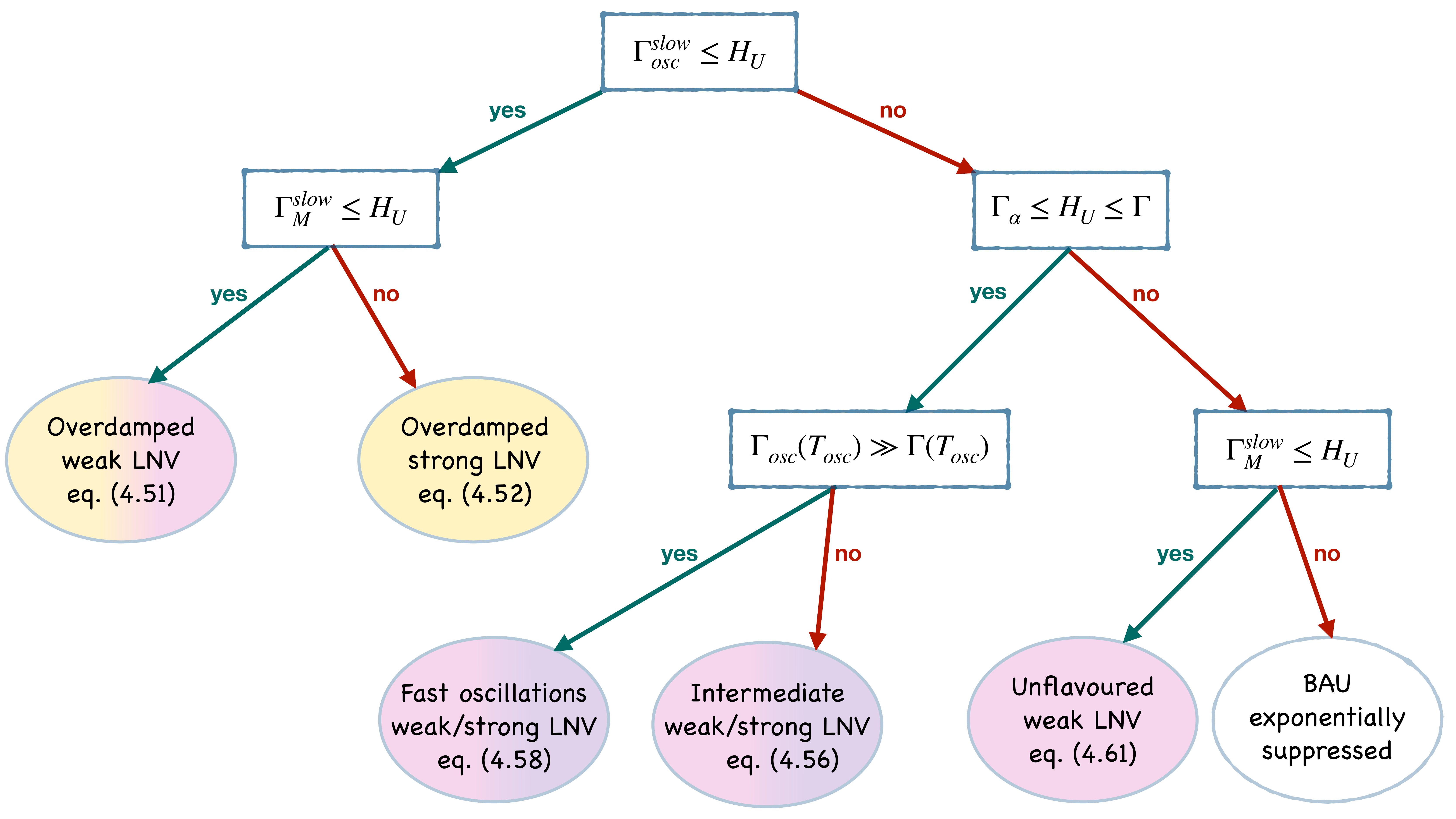} 
\vspace{-0.4cm}
\caption{ 
Chart summarizing the analytical solution in each  washout regime.
The interaction rates and the Hubble expansion rate need to be evaluated at $T_{\rm EW}$.
}
\label{fig:regimes_ana_sol_graph}
\end{figure}

We are interested in the strong washout regime $\Lambda_{\rm max}(x_{\rm EW})\gg 1$ since most of the \texttt{SHiP} and \texttt{FCC} accessible regions are in this regime. Simultaneously, at least one mode must remain weakly coupled at $x_{\rm{EW}}$.
The different alternatives and the corresponding analytical solutions are summarized in the flow chart~\ref{fig:regimes_ana_sol_graph}.
All analytical results are expressed in terms of the CP invariants as derived in section~\ref{sec:cpinv}.
In terms of the parameters of eq.~(\ref{eq:LNVparam2N}) they are given in eqs.~(\ref{eq:CP_Invariants_Yukawa1})-(\ref{eq:CP_Invariants_Yukawa4}).
Their relation to physical observable quantities is given in eqs.~(\ref{eq:ovlncinvnh})-(\ref{eq:osclnvnh}) (eqs.~(\ref{eq:ovlncinvih})-(\ref{eq:osclnvih})) for NH (IH).

\subsubsection{Overdamped regime}

The overdamped regime is defined by the condition
\begin{eqnarray}
\Lambda_{\rm ov}(x_{\rm EW}) \leq 1\,,
\end{eqnarray}
where $\Lambda_{\rm ov}$ is defined in eq.~(\ref{eq:betaov}). 
This condition can be satisfied in the region of interest for sufficiently small $\Delta M/M$, as shown in Fig.~\ref{fig:regimes}.

There are LNC and LNV contributions to the asymmetry. The former is ${\mathcal O}(y_\alpha')$ and can be obtained from the adiabatic solution in 
 eq.~(\ref{eq:r0}) and eq.~(\ref{eq:r1}), which is a good approximation at all times. 
 When $M/T$ terms are included in the rates, there are additional ${\mathcal O}(y_\alpha' (M/T)^2)$ contributions to the asymmetry. These LNV contribuions depend on whether the rate $\Lambda_M(x_{\rm EW})$ is smaller or larger than one. The former case ($\Lambda_M(x_{\rm EW})<1$) corresponds to the weak LNV regime (wLNV), while the latter ($\Lambda_M(x_{\rm EW})>1$) is the strong LNV regime (sLNV). Let us denote by $x^{\rm ov}_{\rm M}$ as the time at which $\Lambda_M(x^{\rm ov}_{\rm M})=1$. Using eq.~(\ref{eq:Gamma_slow_M_ov}), we find
 \begin{eqnarray}
x^{\rm ov}_{\rm M}= \left( {3 T_{\rm EW}^2 \over  M^2 s_0  y^2}\right)^{1/3}\,.
\label{eq:xMov}
\end{eqnarray}

Within the wLNV regime, i.e. $x^{\rm ov}_{\rm M} \geq 1$, we get
\begin{eqnarray}
 \label{eq:overdamped_wLNV_sol}
 \left(\sum_\alpha\mu_{B/3-L_\alpha}\right)^{\rm ov-wLNV}&\simeq& -{4 \kappa \Delta x^2 \over   6\gamma_0+\kappa \gamma_1} {\gamma_0^2\over \gamma_0^2+ 4 \omega^2}\sum_\alpha {y_\alpha y'_\alpha \sin\Delta\beta_\alpha\over y^2}  \left({1 \over y_\alpha^2} -{3 \over y^2}\right) \nonumber\\
 &+ &{48\over 5} {\kappa s_0  \Delta x^5\over 6 \gamma_0+\kappa \gamma_1} {\gamma^2_0  \over \gamma_0^2+ 4 \omega^2}  {M^2\over T^2_{\rm EW}}   \sum_\alpha {y_\alpha y'_\alpha \sin\Delta\beta_\alpha\over y^2}\,,
 \end{eqnarray}
 that can be written in terms of the CP invariants
 \begin{eqnarray}
 \label{eq:overdamped_wLNV_sol_inv}
 \left(\sum_\alpha\mu_{B/3-L_\alpha}\right)^{\rm ov-wLNV}&\simeq&  {\kappa x^2 \over   6\gamma_0+\kappa \gamma_1}{\gamma_0^2\over \gamma_0^2+ 4 \omega^2} \frac{c_H M_P^*}{T_{EW}^3}\left( \Delta^{\rm ov}_{\rm LNC}-{24\over 5} {s_0 x^3\over T_{\rm EW}^2}\Delta^{\rm ov}_{\rm LNV} \right)\,. \,\,\,\,\,\,\,\,\,\,
 \end{eqnarray}

When $x^{\rm ov}_{\rm M} \leq 1$, the asymmetry stops growing at $x^{\rm ov}_{\rm M}$ and a quasi-stationary solution is found , as long as $\Lambda_{\rm ov}(x) \leq 1$. The asymmetry can be obtained by the projection method, that is projecting the wLNV solution at $x^{\rm ov}_{\rm M}$ on the slow mode direction. The result is
 \begin{eqnarray}
  \label{eq:overdamped_sLNV_sol_inv}
  \left(\sum_\alpha\mu_{B/3-L_\alpha}\right)^{\rm ov-sLNV}&\simeq & - {24\over 5} {\kappa s_0^2 (x^{\rm ov}_{\rm M})^5\over 6 \gamma_0 s_0 + \kappa \gamma_0 s_1+\kappa \gamma_1 s_0} {\gamma^2_0  \over \gamma_0^2+ 4 \omega^2}  \frac{c_H M_P^*}{T_{EW}^5}\Delta_{\rm LNV}^{\rm ov}\,.
\end{eqnarray}

Note that only the LNV invariant appears in the sLNV regime: the LNC contributions do not generate any asymmetry in the direction of the  slow mode in this regime as it is connected to LN.

 \subsubsection{ Intermediate regime} 
 
In the intermediate regime,  $\epsilon(x_{\rm osc}) \ll 1$, but at some point, $x_0$, before the EW phase transition, the slow oscillation modes thermalize roughly when 
 $\Lambda_{\rm ov}(x_0) =1$, which according to eq.~(\ref{eq:betaov}) corresponds to 
 \begin{eqnarray}
x_0= \left(5 {(\gamma_0^{2} + 4 \omega^2)  y^2\over  \gamma_0 \Delta^2}\right)^{1/5}\,.
\label{eq:x0}
\end{eqnarray}
A good approximation for the asymmetry in this case is obtained by evolving the overdamped solution until $x_0$ and approximating the asymmetry by projecting on the slow mode(s). The latter can be that of the flavoured weak regime, i.e. flavour direction $\alpha$, and/or the slow mode in the weak LNV regime. The latter enters strong washout at $x_{\rm M}^{\rm int}$, defined by $\Lambda_M(x_{\rm M}^{\rm int})=1$ (see eq.~(\ref{eq:betaMint})):
\be
 x_{\rm M}^{\rm{int}} = \left( 3 \frac{T_{\rm EW}^2}{M^2} \frac{3\gamma_0 + \gamma_1 \kappa }{\kappa y^2(\gamma_1 s_0 + \gamma_0 s_1)}\right)^{1/3}\,.
\ee
In the parameter range of interest we always have $x_0 \ll x_{\rm M}^{\rm int}$.

{\it Flavoured weak washout}

A good approximation is obtained from the overdamped solution evolved up to $x=x_0$ and projected on the two slow modes.  In the relevant part of the parameter space, the LNV slow mode might get strong before $x_{\rm EW}$, so we need to include the time evolution of this contribution according to eq.~(\ref{eq:twozm}), such that
\begin{eqnarray}
 \label{eq:intermediate_fw_sol}
 \left(\sum_\alpha\mu_{B/3-L_\alpha}\right)^{\rm fw-int}&\simeq &{2 \over 3} { x_0^3 \Delta \over y^4}\left({2\kappa \gamma_0 \over  2\gamma_0+\kappa \gamma_1} e^{-\delta\Lambda_M^{\rm int}(x)} -\kappa \right) {\gamma^2_0  \over \gamma_0^2+ 4 \omega^2} \nonumber \\
  &\times& \sum_{\beta\neq\alpha} { y_\beta^2 y_\alpha y_\alpha' \sin\Delta\beta_\alpha  - y_\alpha^2 y_\beta y_\beta' \sin\Delta\beta_\beta }\,,
 \end{eqnarray}
where $\delta \Lambda_{\rm M}^{\rm int}(x) \equiv \Lambda_{\rm M}^{\rm int}(x) -\Lambda_{\rm M}^{\rm int}(x_0)$, with $\Lambda_{\rm M}^{\rm int}$ given by eq.~(\ref{eq:betaMint}).  
In terms of the CP invariants this can be expressed as 
\begin{eqnarray}
 \label{eq:intermediate_fw_sol_inv}
  \left(\sum_\alpha\mu_{B/3-L_\alpha}\right)^{\rm fw-int}&\simeq &-{2 \over 3} { x_0^3 \over y^4}\left({2\kappa \gamma_0 \over  2\gamma_0+\kappa \gamma_1} e^{-\delta\Lambda_M^{\rm int}(x)} -\kappa \right) {\gamma^2_0  \over \gamma_0^2+ 4 \omega^2}  \frac{c_H M_P^*}{T_{EW}^3}  \Delta^{\rm int(\alpha)}_{\rm LNC}\,. \,\,\,\,\,\,\,\,\,\,\,\,\,\,
\end{eqnarray}
The LNV contribution in this regime is very small and has been neglected for simplicity.

{\it Unflavoured wLNV}
 
When the only slow mode is the LNV one we get instead
\begin{eqnarray}
\label{eq:lnv_int_sol}
\left( \sum_\alpha\mu_{B/3-L_\alpha}\right)^{\rm wLNV-int} &\simeq& {24 \over 5}\kappa{ s_0\Delta x_0^5\over 3 \gamma_0 +\kappa \gamma_1} {\gamma_0^{2}\over \gamma_0^{2}+4 \omega^2 }  {M^2\over T^2_{\rm EW}}   \sum_\alpha{ y_\alpha y_\alpha' \sin\Delta\beta_\alpha\over y^2}\nonumber\\
&=&
-24 { \kappa s_0 \gamma_0 \over 3 \gamma_0 +\kappa \gamma_1}   {1 \over T_{\rm EW}^2} \Delta_{\rm LNV}^{\rm osc}\,.
\end{eqnarray}
Note that there is no contribution from the LNC invariants. 
This is because the LNC contribution projected on the LNV slow-mode direction vanishes. 

\subsubsection{Fast oscillation regime}

Contrary to the intermediate regime, the fast oscillation regime is characterized by $\epsilon(x_{\rm osc}) \gg 1$.
Again we can have two weakly coupled modes at $T_{\rm{EW}}$, which are the same as discussed for the intermediate regime.

{\it Flavoured weak washout}

With the adiabatic approximation we find
\begin{eqnarray}
 \label{eq:lnc_osc_sol}
 \left( \sum_\alpha\mu_{B/3-L_\alpha} \right)^{\rm fw-osc} &\simeq &- \left( \gamma_0^2 \kappa -  \frac{2 \gamma_0^3 \kappa}{2\gamma_0 + \gamma_1 \kappa} e^{-\delta\Lambda_M^{\rm int}(x)}  \right)
{\rm Im}\left(J_{200}(\Delta, -\Delta, x_0) \right)  \nonumber\\
& \times& \sum_{\beta\neq \alpha}  y_\alpha y'_\alpha \sin\Delta\beta_\alpha y_\beta^2- y_\alpha^2  y_\beta y_\beta' \sin\Delta\beta_\beta \nonumber\\
 &=&
\left( \gamma_0^2 \kappa -  \frac{2 \gamma_0^3 \kappa}{2\gamma_0 + \gamma_1 \kappa} e^{-\delta\Lambda_M^{\rm int}(x)}  \right) \Delta_{\rm LNC}^{\rm osc(\alpha)}\,,
\end{eqnarray}
which is valid once the system only possess one flavoured weak mode $\alpha$, i.e. $\Lambda_\alpha < 1$ and $\Lambda_\beta > 1$ for the other flavoures $\beta$.
The mass function entering the CP invariant is found to be $g(M_1,M_2)=  {\rm Im}\left(J_{200}(\Delta, -\Delta, x_0) \right)$, which is defined by
 \begin{eqnarray}
 J_{2nm}(\Delta, -\Delta, x) \equiv \int_0^x du ~ u^n ~e^{i {\Delta u^3\over 3}} \int_0^u dz ~ z^m~e^{-i {\Delta z^3\over 3}}\,.
 \end{eqnarray}
The asymptotic solution of the integral is
 \begin{eqnarray}
{\rm Im} J_{200}(\Delta,-\Delta,\infty) = -{2^{4/3}\over 3^{1/3}} {\pi^{3/2}\over \Gamma[-1/6] }{{\rm sign}(\Delta) \over |\Delta|^{2/3}}\,.
 \end{eqnarray}
 This result is parametrically the same as the intermediate regime result at $x_0 = x_{\rm osc}$, see eq.~(\ref{eq:intermediate_fw_sol_inv}). The two solutions therefore match appropriately.
The LNV contribution in this regime is very small and has been neglected for simplicity.

 {\it Unflavoured wLNV} 
 
A good approximation in this case can be obtained from the result in the weak washout regime and projecting it on the zero mode at the thermalization time $\Lambda_{\rm max}(x_{\rm th})=1$ (see eq.~\eqref{eq:Lambdamax}):
 \begin{eqnarray}
 \label{eq:lnv_osc_sol}
 \left( \sum_\alpha\mu_{B/3-L_\alpha}\right)^{\rm wLNV-osc}_{LNV} &\simeq& 12 { \kappa \gamma_0^2 s_0 \over 3 \gamma_0 +   \kappa \gamma_1} {M^2\over T^2_{\rm EW}}  {x_{\rm th}\over \Delta}y^2 \sum_\alpha{ y_\alpha y_\alpha' \sin\Delta\beta_\alpha}\nonumber\\
&\simeq& -24 { \kappa s_0 \gamma_0 \over 3 \gamma_0 +\kappa \gamma_1}  {1\over T_{\rm EW}^2} \Delta_{\rm LNV}^{\rm osc}\,.
\end{eqnarray}
Note that, remarkably, this result matches the one obtained in the corresponding intermediate region, see eq.~(\ref{eq:lnv_int_sol}).

\subsection{Relating to the baryon asymmetry}

To relate the chemical potentials to the baryon asymmetry we go beyond the widely used instantaneous sphaleron freeze-out approximation and use a smooth transition between $T\in [T_{\text{C}}, T_{\text{EW}}] =[160 \, \text{GeV}, 131.7 \,\text{GeV}]$, following the method of ref.~\cite{Eijima:2017cxr} (see also~\cite{Ghiglieri:2017csp} for other approach to the treatment of the sphaleron rate).

\begin{figure}[!t]
\centering
\begin{tabular}{cc}
\hspace{-0.5cm} \includegraphics[width=0.5\textwidth]{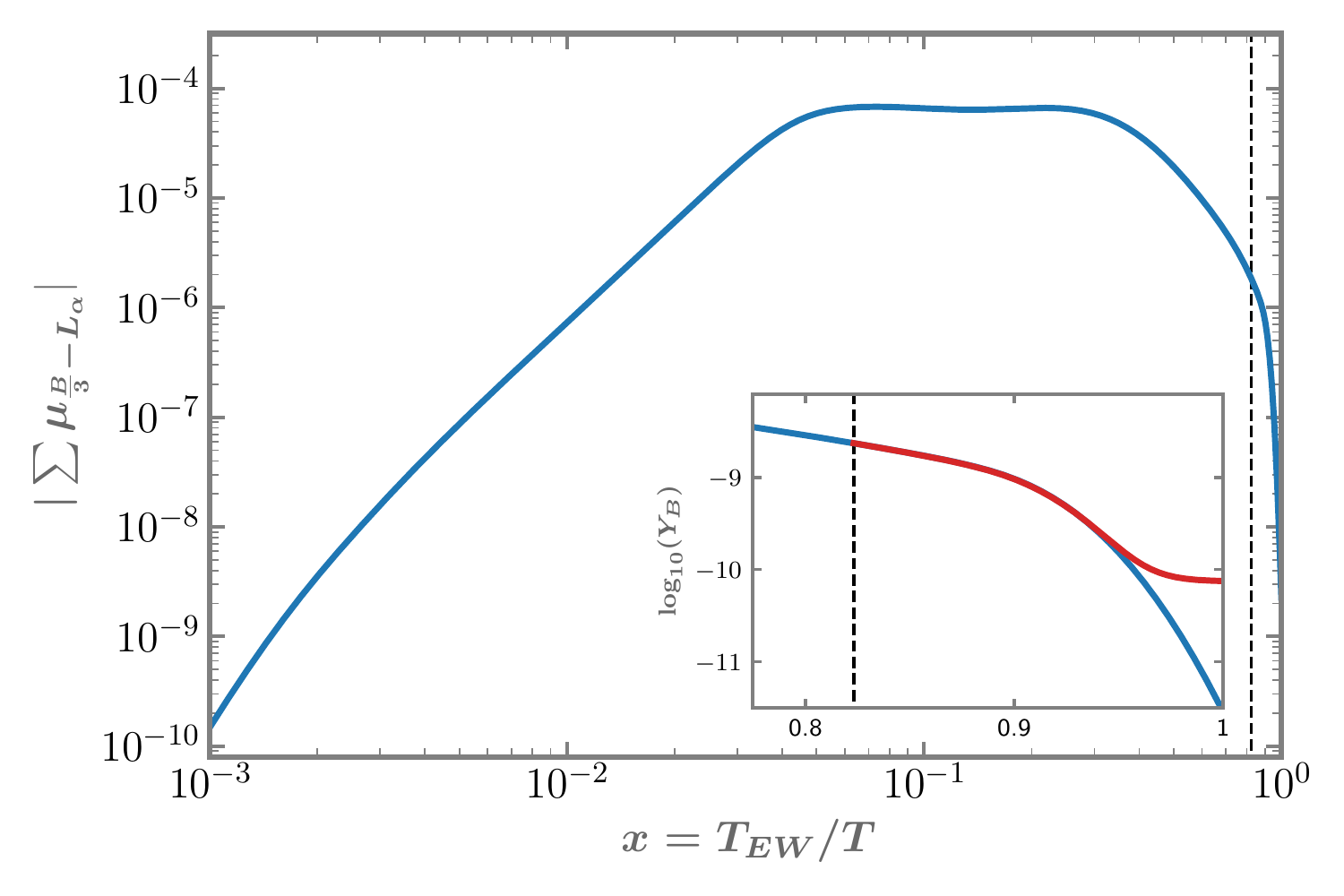} & \hspace{-0.55cm}  \includegraphics[width=0.5\textwidth]{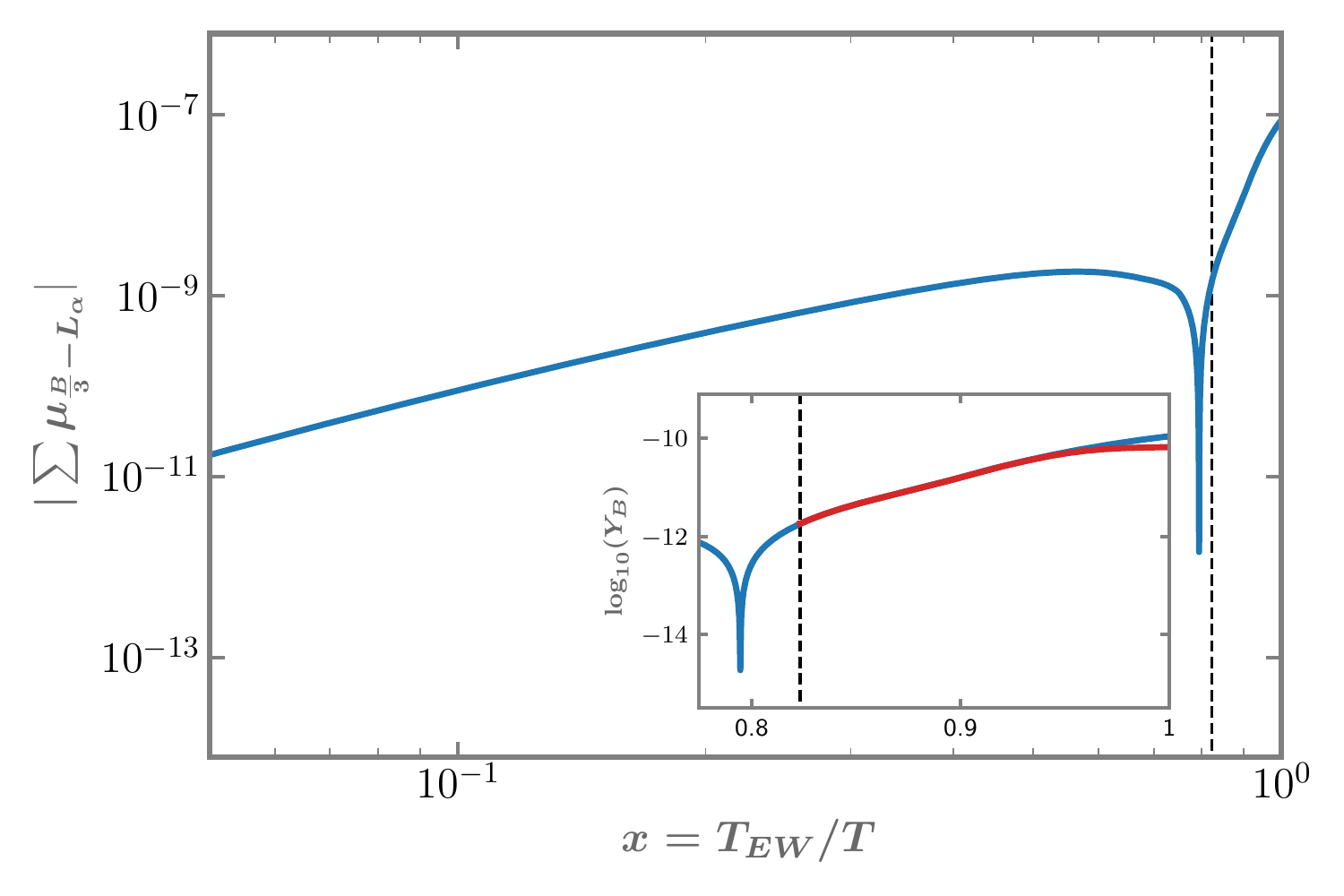}   \\
\end{tabular}
\vspace{-0.4cm}
\caption{ Comparison of the smooth sphaleron freeze-out (red) to the instantaneous approximation (blue). Deviations can be as large as $\mathcal{O}(10)$ if all modes leave the weak coupling regime close to $T_{\text{EW}}$ (left) and at most a factor of two if at least one mode remains weakly coupled (right). }
\label{fig:sphaleron_fo_example}
\end{figure}

We have seen that this effect is not relevant in most of the parameter space, but it is very relevant when all flavours enter the strong washout close to $x_{\rm EW}$. In this case, the smooth sphaleron freeze-out has two important effects: i) it counteracts the effect of the significant growth of the LNV rates in the range $[T_{\text{C}}, T_{\text{EW}}]$ and ii) it reduces the washout of the asymmetry below $T_{\rm C}$. In these situations the prediction of the BAU can be changed by $\mathcal{O}(10)$, see Fig.~\ref{fig:sphaleron_fo_example}.

Following~\cite{Eijima:2017cxr}, the smooth sphaleron freeze-out is implemented as follows. We introduce an additional differential equation  for the baryon number in the range $T < T_{\text{C}}$ 
\begin{align}
x H_u \frac{\text{d}}{\text{d}x}Y_B = - \Gamma_B (Y_B - Y_B^{\rm eq})\,,
\end{align}
where 
\begin{align}
\Gamma_B = 3^2 \frac{869 + 333 (\sqrt{2} \langle \Phi \rangle/ T)^2}{792 + 306 (\sqrt{2} \langle \Phi \rangle/ T)^2}  \frac{\Gamma_{\text{diff}}}{T^3}\,,
\end{align}
and the temperature dependent higgs vev below $T_{\rm C}$ is $\langle \Phi \rangle^2 = v^2  ( 1 -  T/T_{\rm C} ) $.
The critical temperature $T_{\rm C}$ and the Chern-Simons diffusion rate 
\begin{align}
\Gamma_{\text{diff}}^{T < T_{\rm C}} = \Gamma_{\text{diff}} = \exp\left(-147.7 + \frac{0.83T}{ {\rm GeV} }\right) T^4\,,
\end{align} 
are obtained from a lattice calculation~\cite{DOnofrio:2014rug}. 

On the other hand, in the instantaneous freeze-out  the sphalerons are in full equilibrium up to $T_{\text{EW}}$, and the relation between the baryon asymmetry and the chemical potential is given by~\cite{Khlebnikov:1996vj,Burnier:2005hp}
\begin{equation}
\label{eq:Y_B_eq_def}
Y_B^{\rm eq} \simeq 3.6 \times 10^{-3}  \chi(T) \sum_{\alpha} \mu_{\frac{B}{3} - L_{\alpha}}
\,,\;\text{with}\;\;
\chi(T) 	\simeq \frac{4(27 (\sqrt{2} \langle \Phi \rangle/T)^2 + 77) }{333 ( (\sqrt{2} \langle \Phi \rangle/T)^2+ 869)}\,,
\end{equation}
where the factor in the equilibrium relation arises from the relation of the chemical potential to the particle number density in a comoving volume, see eq.~\eqref{eq:n_to_mu}, normalized to a constant entropy density $s=(2\pi^2)/45 g_{s*}T^3$. 
For $T=T_{\text{EW}}$ we obtain 
\begin{eqnarray}
Y_B^{\rm eq} \simeq 1.26\times 10^{-3} \sum_{\alpha} \mu_{\frac{B}{3} - L_{\alpha}}\,.
\label{eq:insfo}
\end{eqnarray}
The experimentally measured value of the asymmetry is~\cite{Planck:2018vyg}
\begin{align}
Y_{B}^{\mathrm{exp}} = (8.66 \pm 0.05) \times 10^{-11}\,.
\end{align}

As long as one mode remains weakly coupled at $x_{\rm EW}$ the gradual sphaleron freeze-out differs from the instantaneous decoupling approximation at most by a factor of two if the asymmetry is dominated by the contribution of eq.~\eqref{eq:overdamped_wLNV_sol_inv}, see Fig.~\ref{fig:sphaleron_fo_example}, and by a few percent
if any other weakly coupled mode dominates the asymmetry generation. However deviations can be as large as $\mathcal{O}(10)$ if all modes leave the weak coupling regime at $x_{\text{EW}}$, since in this case the washout of the asymmetry is exponential and therefore very sensitive to the details of the sphaleron freeze-out.


\section{Parameter constraints from the baryon asymmetry}
\label{sec:parambounds}

From the analytical results of the previous sections we can easily derive the constraints imposed by successful baryon asymmetry on the masses and mixings of the HNLs. For these estimates we use the instantaneous sphaleron freeze-out approximation of eq.~(\ref{eq:insfo}) and evaluate the rates at $T= 150\,\text{GeV}$ for fixed $M_{1} = 1\,\text{GeV}$.  
In the next section we will compare the constraints derived here with the results from the full numerical analysis. In appendix~\ref{app:lnc_limit}, we consider the bounds for the pure LNC case, that is neglecting $M/T$ corrections in the rates. 

\subsection{Overdamped regime}

The overdamped regime is defined by $\Lambda_{\rm ov}(x_{\rm EW}) \leq 1$ which translates into
\begin{eqnarray}
(U^2)_{\rm ov} \geq 8 \times 10^9 \left({\Delta M\over M} {M\over 1{\rm GeV}}\right)^2\,.
\label{eq:dmovmboundov}
\end{eqnarray} 
On the other hand, the dynamics heavily depends on whether LNV rates are weak or strong.
Using eq.~\eqref{eq:Gamma_slow_M_ov} we find that the wLNV regime requieres mixings
\be
\left(U^2 \right)_{\rm wLNV} \leq 1\times10^{-6} \left( \frac{1\,\rm{GeV}}{M}\right)^{4}\,,
\ee
while for larger mixings LNV rates are strong.
We consider both cases separately.

\subsubsection{wLNV regime}
The analytical solution in this regime is given by eq.~(\ref{eq:overdamped_wLNV_sol_inv}), in terms of the CP invariants. Using eq.~(\ref{eq:ovlncinvnh}) (eq.~(\ref{eq:ovlncinvih})) for NH (IH), and the relation between the $B-L$ chemical potentials and the final baryon asymmetry as given by eq.~(\ref{eq:insfo}), the asymmetry within the wLNV can be expressed as
\begin{eqnarray}
\left( Y_B \right)_{\rm ov}^{\rm wLNV} \simeq 2\times 10^{-1}  \frac{\Delta M}{M} \left(\frac{10^{-7}}{U^2}\right) \frac{1\,\text{GeV}}{M} \left( \left(\frac{M}{1\,\text{GeV}}\right)^4 f_{\rm LNV}^{\rm H}  - \left(\frac{10^{-7}}{U^2}\right) f_{\rm LNC}^{\rm H} \right)\,.
\label{eq:Yb_ov_wLNV}
\end{eqnarray} 
The functions $f^{\rm H}_{\rm LNC/LNV}$ isolate the angular dependence of the CP invariants, associated to both the LNC and LNV contributions, on the PMNS angles and phases, as well as the {\it high scale} phase $\theta$. They are {\it naturally} expected to be $\mathcal{O}(1)$ quantities.
The superscript $\rm H$ corresponds to the unknown neutrino hierarchy. At leading order in the expansion parameters, $r$, $\theta_{13}$ and $\theta_{23}$,  c.f. eq.~(\ref{eq:ovlncinvnh}) and eq.~(\ref{eq:ovlncinvih}), we find
\begin{eqnarray}
f_{\rm LNC}^{\rm IH} = \frac{(1+3c_\phi \sin2\theta_{12})(c_\theta s_\phi \sin2\theta_{12} + s_\theta \cos2\theta_{12})}{1 - c_\phi^2 \sin^22\theta_{12}}\,,
\label{eq:f_lnc_ih}
\end{eqnarray} 
and
\be
f_{\rm LNC}^{\rm NH} = f_{\rm LNV}^{\rm NH} = 2/r^2 f_{\rm LNV}^{\rm IH} =  s_\theta\,.
\label{eq:f_lnc_lnv_H}
\ee
For a fixed set of $(\theta, \phi)$, the asymmetry within the wLNV regime can have different signs depending on the particular value of the HNL masses.
This is explained by the dominance of the LNC contribution (second term in eq.~(\ref{eq:Yb_ov_wLNV})) or the LNV one (first term), since both contributions to the final asymmetry estimation have opposite sign. 

Solving for the mixing $U^2$ in eq.~\eqref{eq:Yb_ov_wLNV} to match the observed BAU we find
\begin{eqnarray}
\nonumber
\left( U^2 \right)_{\rm ov}^{\rm wLNV} &=& 1.3 \times 10^{2}  f_{\rm LNV}^{\rm H} \frac{\Delta M}{M} \left(\frac{M}{1\,\text{GeV}}\right)^3 + 7.2 \times 10^{6} \sqrt{\frac{\Delta M}{M}} \sqrt{\frac{1\,\rm{GeV}}{M}} \,\\
&\times& \sqrt{  3.5\times 10^{-10} \left(f_{\rm LNV}^{\rm H}\right)^2 \left( \frac{\Delta M}{M} \right) \left( \frac{M}{1\,\text{GeV}} \right)^{7} - 4.3 \times 10^{-19} f_{\rm LNC}^{\rm H} }\,.
\label{eq:U2_ov_wLNV_max_fixed_DMM}
\end{eqnarray} 
 The square root in eq.~\eqref{eq:U2_ov_wLNV_max_fixed_DMM} must be real and this results in a mass threshold of
 \be
M_* \simeq 5 \times 10^{-2} \left(\frac{\Delta M}{M} \frac{|f_{\rm LNV}^{\rm H}|^2}{|f_{\rm LNC}^{\rm H}|}\right)^{-1/7} \,\rm{GeV}\,.
\ee
For $M\leq M_*$, the LNC contribution dominates and the positivity requirement of $\left( Y_B \right)_{\rm ov}^{\rm wLNV}$ selects $f_{\rm LNC}^{\rm H} < 0$. For $M\geq M_*$, when LNV dominates instead, matching the BAU requires $f_{\rm LNV}^{\rm H} > 0$.

Maximizing the functions $f^H$ in absolute value over the unknown phases $(\theta, \delta, \phi)$, an {\it upper} bound on the HNL mixing, for fixed $\Delta M/M$ and $M$, can be derived. 
For NH we find\footnote{We note that the next-to-leading order contribution can enhance $f_{\rm LNC}^{\rm NH}$ by a factor of $1.8$ and, therefore, we include it in the numerical evaluations.}:
\begin{eqnarray}
\nonumber
\left. \left( U^2 \right)_{\rm ov}^{\rm wLNV}\right|_{\rm NH} &\leq& \mp 1.3 \times 10^{2}  \frac{\Delta M}{M} \left(\frac{M}{1{\rm GeV}}\right)^3 + 7.2 \times 10^{6} \sqrt{\frac{\Delta M}{M}} \sqrt{\frac{1{\rm GeV}}{M}} \,\\
&\times& \sqrt{  3.5\times 10^{-10} \left( \frac{\Delta M}{M} \right) \left( \frac{M}{1\,{\rm GeV}} \right)^{7} \pm 7.7 \times 10^{-19}  }\,,
\label{eq:U2_ov_wLNV_max_fixed_DMM_NH}
\end{eqnarray} 
where the upper (lower) sign corresponds to $M< M_*$ ($M> M_*$).
For low values of $M$ the bound is saturated for $\theta=3\pi/2$, while in the large mass limit this occurs for $\theta=\pi/2$.

In the IH case $M_*$ is always inside the strong LNV regime.
Therefore, the LNV contribution can be neglected for all the range of masses and the bound can be simplified to
\be
\left.\left( U^2 \right)_{\rm ov}^{\rm wLNV}\right|_{\rm IH}\lesssim 15 \times 10^{-3} \sqrt{\frac{\Delta M}{M}} \sqrt{\frac{1\,\rm GeV}{M}}\,,
\label{eq:U2_ov_wLNV_max_LNC_fixed_DMM}
\ee  
which is saturated for $(\theta, \phi) = (3\pi/2,0)$.

\begin{figure}[!t]
\centering
\begin{tabular}{cc}
\hspace{-0.5cm}  \includegraphics[width=0.5\textwidth]{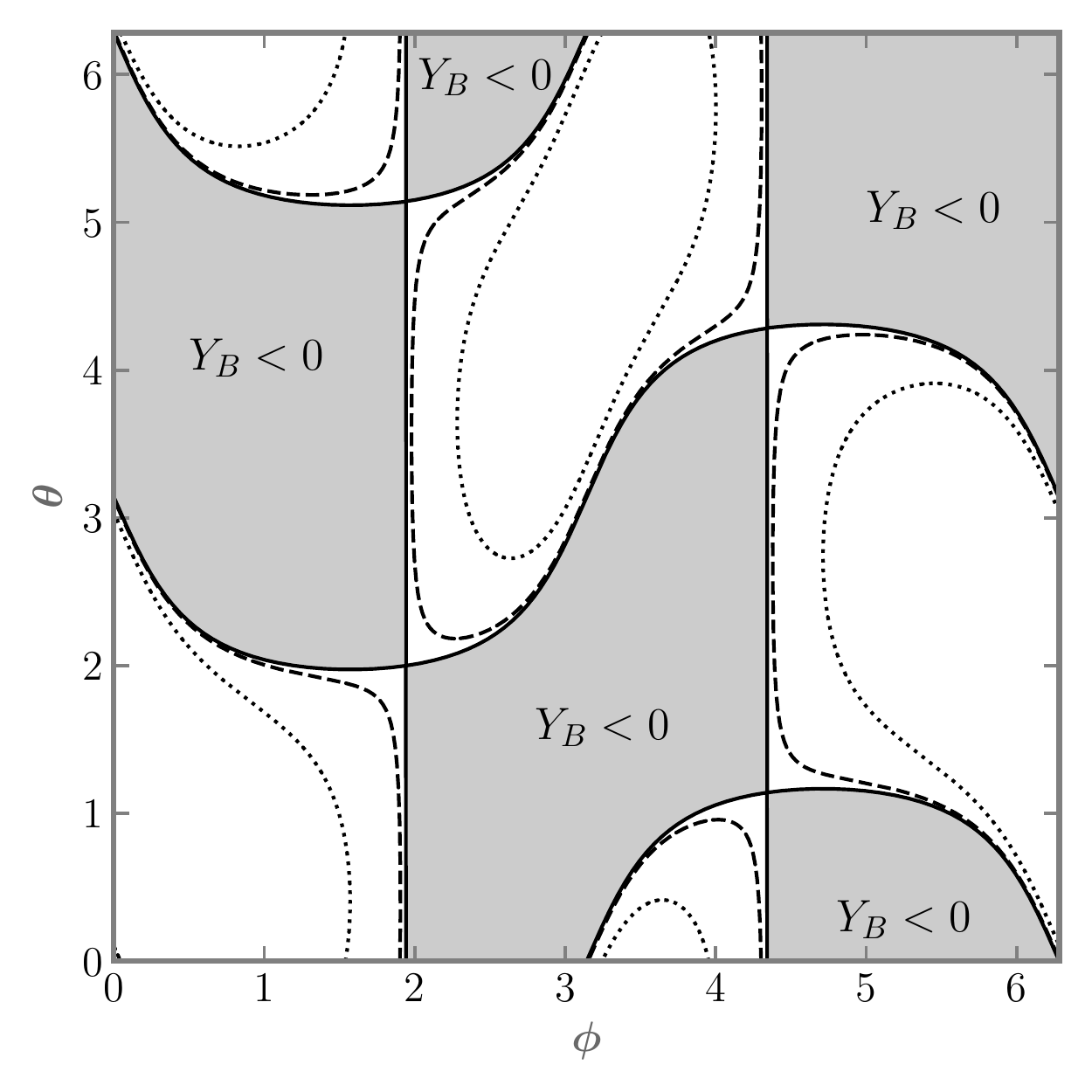} &
\hspace{-0.55cm}  \includegraphics[width=0.5\textwidth]{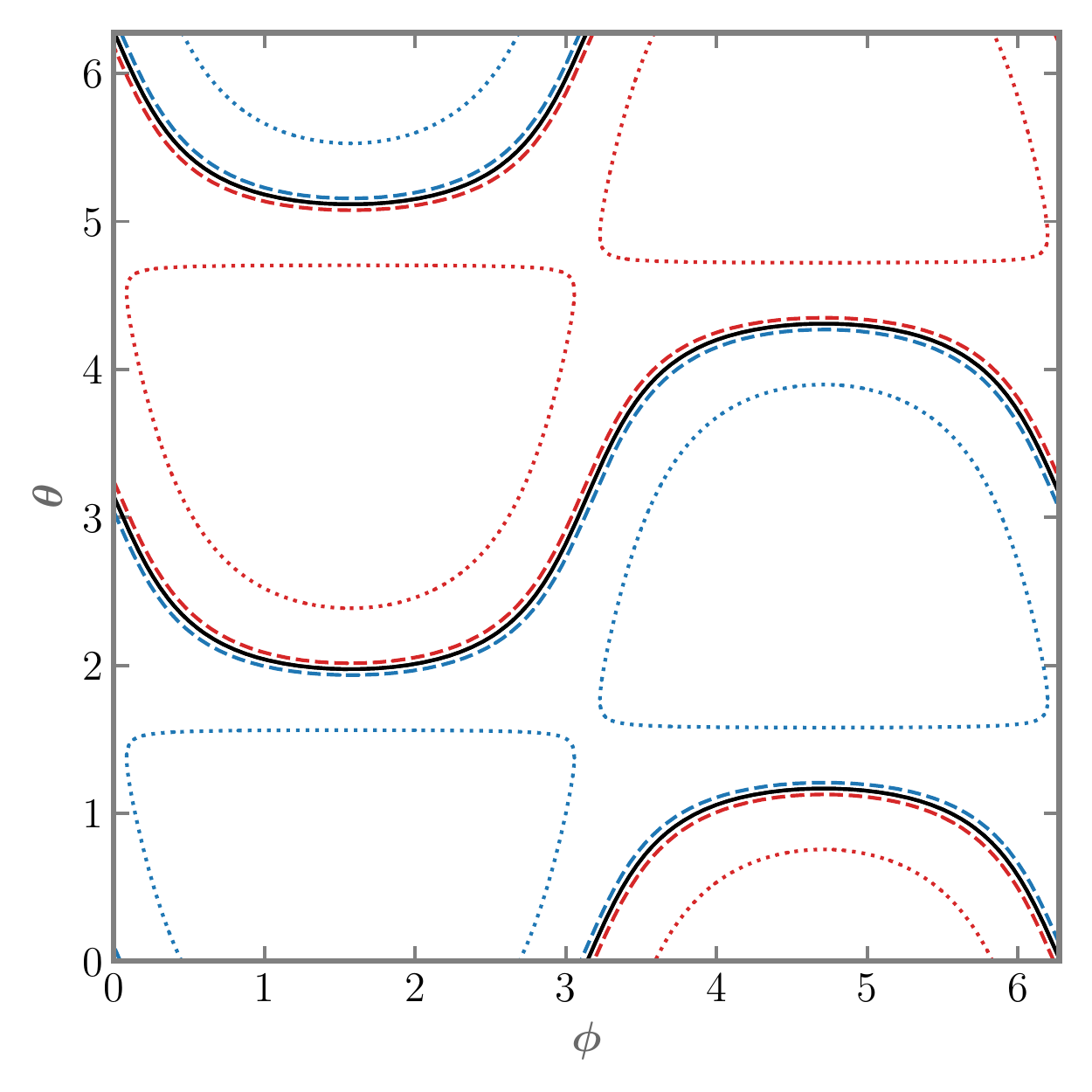} 
\end{tabular}
\vspace{-0.4cm}
\caption{ 
\textit{Left}: Contour lines corresponding to $f_{\rm LNC}^{\rm IH}  = (0, 0.1, 1)$ (solid, dashed, dotted). Grey shaded regions lead to a negative baryon asymmetry.
\textit{Right}: Contour lines in red corresponding to $\tilde{f}^{\mu,\tau}_{\rm IH}  = (0.01, 0.1)$ (dashed, dotted) and in blue to $-\tilde{f}^{\mu,\tau}_{\rm IH} = \tilde{f}^{e}_{\rm IH}/2=(0.01,0.1)$. The black solid line represents $\tilde{f}^\alpha_{\rm IH} = 0$.
}
\label{fig:f_contours}
\end{figure}

Values of the mixing much smaller than the upper bound necessarily require a suppression from $f^H$ to match the BAU. 
For NH this is controlled by only one parameter, $\theta$\footnote{Higher order corrections in the expansion must be considered if $f^H$ is less than $10\%$ of its maximum value. }.
In contrast, in the IH case the required suppression of the BAU depends on $(\theta, \phi)$ and involves a strong correlation between these two phases as shown on the left panel in Fig.~\ref{fig:f_contours}.
Any numerical scan (e.g. Markov Chain Monte Carlo, Bayesian Nested Sampling, etc.) that treat both phases as independent parameters may have difficulties in finding the required correlation. The analytical result is therefore a necessary guide to optimize the scan of parameter space.

Finally, note that the upper limit on the HNL mixing is proportional to $\Delta M/M$. However, the overdamped regime leads to an upper bound on $\Delta M/M$, see eq.~\eqref{eq:dmovmboundov}. 
Therefore, saturating the bound on $\Delta M/M$ from eq.~\eqref{eq:dmovmboundov} and substituting the resulting expression into eq.~\eqref{eq:U2_ov_wLNV_max_fixed_DMM}, leads to the maximal attainable mixing compatible with the BAU in this regime. The resulting expression is not particularly illuminating and cannot be solved analytically for $U^2$.
However, a reasonably good approximation is obtained neglecting the LNV contributions for both hierarchies. We obtain the compact expression
\be
U^2 \lesssim 5\,(17) \times 10^{-7} \left(\frac{1\,\rm{GeV}}{M}\right)^{4/3}  \,\,\,\,\, \rm{NH\,(IH)}\,.
\label{eq:U2_ov_wLNV_max_LNC}
\ee
We remark that this is an absolute upper bound valid in the wLNV, i.e. for $M \lesssim \mathcal{O}(1\,\rm{GeV})$.

\subsubsection{sLNV regime}
The analytical solution in this regime is given in eq.~(\ref{eq:overdamped_sLNV_sol_inv}). 
Using the expression for the CP invariants given by eq.~(\ref{eq:ovlncinvnh}) (eq.~(\ref{eq:ovlncinvih})) for NH (IH), and the relation between the $B-L$ chemical potentials and the final baryon asymmetry as given by eq.~(\ref{eq:insfo}), we obtain
\be
\left( Y_B \right)_{\rm ov}^{\rm sLNV} \simeq 9 \times 10^3 \frac{\Delta M}{M} \left(\frac{10^{-7}}{U^2}\right)^{8/3} \left(\frac{1\,\rm{GeV}}{M}\right)^{11/3}f_{\rm{LNV}}^{\rm H} \,.
\label{eq:Yb_ov_sLNV}
\ee
The angular function $f^H_{LNV}$ is defined in eq.~(\ref{eq:f_lnc_lnv_H}) and has its maximum at $\theta = \pi/2$, and this leads to the upper bound
\be
\left( U^2 \right)_{\rm ov}^{\rm sLNV} \leq 15\,(3) \times 10^{-4} \left(\frac{\Delta M}{M}\right)^{3/8}  \left(\frac{1\,\rm{GeV}}{M}\right)^{11/8} \,\,\,\,\, \rm{NH\,(IH)}\,.
\label{eq:U2_ov_sLNV_max_fixed_DMM}
\ee
Including the upper bound on $\Delta M/M$ such that the overdamped condition of eq.~\eqref{eq:dmovmboundov} is fulfilled we arrive at
\be
U^2 \lesssim  16\, (2.3) \times 10^{-7} \left(\frac{1\,\rm{GeV}}{M}\right)^{28/13}   \,\,\,\,\, \rm{NH\,(IH)}\,.
\label{eq:U2_ov_sLNV_max}
\ee
This should be seen as an absolute upper bound on the mixing for HNLs with masses $M \gtrsim \mathcal{O}(1\,\rm{GeV})$, if the asymmetry is to be explained with the asymptotic overdamped mode.
A more conservative estimate, which will still be satisfied if we allow for some suppression due to strong washout, is given by the maximal asymmetry which can be generated before the LNV rates become strong. 
Namely, the asymmetry within the overdamped wLNV regime at the point $x^{\rm ov}_M$, eq.~(\ref{eq:xMov}). 
Evaluating eq.~(\ref{eq:overdamped_wLNV_sol_inv}) at $x = x^{\rm ov}_{\rm M}$, using eq.~(\ref{eq:ovlncinvnh}) (eq.~(\ref{eq:ovlncinvih})) for NH (IH) and the translation of the $B-L$ chemical potentials to the final baryon asymmetry, we obtain a conservative estimate which coincides with eq.~(\ref{eq:U2_ov_sLNV_max_fixed_DMM}) for NH. For IH, this conservative estimate is a factor $\times 4$ larger than the corresponding result in eq.~(\ref{eq:U2_ov_sLNV_max_fixed_DMM}). Similarly, introducing the maximum $\Delta M/M$ that satisfies the overdamped condition, the corresponding conservative bound is that of eq.~(\ref{eq:U2_ov_sLNV_max}) for NH, while for IH it is a factor $\times 6$ larger than eq.~(\ref{eq:U2_ov_sLNV_max}).

\subsection{Intermediate regime}

In the parameter space outside the overdamped region, i.e. for mixings that do not satisfy eq.~(\ref{eq:dmovmboundov}), the analytical estimate depends on whether we are in the intermediate or fast oscillation regime.
They are separated by the line
\be
\left( U^2\right)_{\rm osc/int} \simeq 10^{-6} \left( \frac{\Delta M}{M} \right)^{1/3} \left( \frac{1\,\rm GeV}{M} \right)^{4/3}\,,
\label{eq:U2_int_osc}
\ee
corresponding to $\epsilon(x_{\rm osc}) =1$, where $x_{\rm osc}$ is given by eq.~\eqref{eq:x_osc} and $\epsilon(x)$ by eq.~\eqref{eq:eps}.
For larger mixings we are in the intermediate regime and for smaller in the fast oscillation regime.

We have seen that the asymmetry in this regime requires that either at least one flavour $\alpha$ remains weakly coupled, i.e. $\Lambda_\alpha(x_{\rm EW})\leq 1$, and/or the LNV mode does, i.e. $\Lambda_M(x_{\rm EW})\leq 1$.
Again we need to distinguish these cases. 

\subsubsection{Flavoured weak washout}

Using eq.~(\ref{eq:gammaalpha}), the {\it necessary} (but not sufficient) condition to have (at least) one flavour $\alpha$ that remains weak at $x_{\rm EW}$ and at least one strongly coupled is given by 
\begin{eqnarray}
 10^{-9} \left({1{\rm GeV}\over M}\right)^2{1\over {\rm Max}(\epsilon_\alpha)} \leq (U^2)_{\rm fw}\leq   10^{-9} \left({1{\rm GeV}\over M}\right)^2{1\over {\rm Min}(\epsilon_\alpha)}\,,
\label{eq:boundfwo}
\end{eqnarray}
where $\epsilon_\alpha \equiv y_\alpha^2/y^2$, which depends only on the PMNS parameters and in particular the unknown CP phases, $(\delta, \phi)$.   While the maximum of $\epsilon_\alpha$ is ${\mathcal O}(1)$, the minimum is obtained for a given flavour in each hierarchy \footnote{For IH there are particular solutions for $(\delta,\phi)$ which can lead to $\text{Min}(\epsilon_\mu) \simeq 5 \times 10^{-4}$.}
\begin{eqnarray}
{\rm Min}(\epsilon_\tau)_{\rm IH} \simeq {\rm Min}(\epsilon_e)_{\rm NH} = 5\times 10^{-3}  \,.
\end{eqnarray}
The range of phases that lead to a small $\epsilon_\alpha$ are shown in Fig.~\ref{fig:flavourselect}.

If a flavour remains slow until $x_{\rm EW}$, but the LNV mode becomes strong earlier, the asymmetry is well approximated by eq.~(\ref{eq:intermediate_fw_sol_inv}).
Including the CP invariants from eq.~\eqref{eq:intlncinvnh} (eq.~\eqref{eq:intlncinvih}) for NH (IH), the final asymmetry is well approximated by
\begin{eqnarray}
\left( Y_B\right)_{\rm fw-int} \simeq 9.5 \times 10^{-9} \eta~ {\tilde f}^\alpha_{\rm NH/IH} \left({\Delta M\over M}\right)^{-1/5} \left({1 \rm GeV \over M}\right)^{1/5} \left({10^{-9}\over U^2}\right)^{2/5}\,,
\label{eq:ybfwint}
\end{eqnarray}
where $\eta$ is a constant factor that depends on whether the LNV becomes strong or not before $x_{\rm EW}$. $\eta$ is a constant factor equal to $1$ in the weak LNV limit ($x_{M}^{\rm int} \geq x_{\rm EW}$) and
\be
\eta = \frac{\gamma_1\kappa}{2\gamma_0 + \gamma_1 \kappa} \simeq 4\,,
\label{eq:d_factor}
\ee
in the strong LNV case ($x_{M}^{\rm int} \leq x_{\rm EW}$).
The angular functions are given by
\begin{eqnarray}
\tilde f^e_{\rm NH} = r s_{12}^2 s_\theta,\;\;\; \tilde f^{\mu,\tau}_{\rm IH} = -\tilde f^{e}_{\rm IH}/2= -{1\over 4}(\sin 2\theta_{12} s_\phi c_\theta + \cos 2 \theta_{12} s_\theta)\,.
\end{eqnarray}

Maximizing the factors of ${\tilde f}^\alpha_{\rm NH/IH}$ over then unknown CP phases, and requiring that the asymmetry is the observed one, leads to the following upper bound
\be
\left( U^2\right)_{\rm int} \leq 1(40) \times 10^{-6} \eta \left( \frac{\Delta M}{M} \right)^{-1/2} \left( \frac{1\,\rm GeV}{M}\right)^{1/2}\,.
\label{eq:boundU2int}
\ee
This upper bound on $U^2$ set by the BAU is  less stringent than the one impossed by the required weak flavour condition of eq.~\eqref{eq:boundfwo}. Therefore, the latter sets the upper bound, which means that the asymmetry can always be explained inside the region defined by eq.~\eqref{eq:boundfwo}. On the other hand, since the upper limit on $U^2$ driven by eq.~\eqref{eq:boundfwo} is more stringent than that in eq.~(\ref{eq:boundU2int}), a significant suppression from ${\tilde f}^\alpha_{\rm NH/IH}$ is needed to match the BAU in this region. 
For NH this is mostly controlled by $s_\theta$, while for IH involves a non-trivial correlation between the two phases $(\theta,\phi)$ as shown on the right panel of Fig.~\ref{fig:f_contours}.
Matching the asymmetry involves therefore an interplay of a minimization in the flavour hierarchy $\epsilon_\alpha$ and the angular function ${\tilde f}^\alpha_{\rm NH/IH}$. While for NH a significant suppression of $\epsilon_\alpha$ is only possible for the electron flavour, in the IH case a similar suppression can be achieved for all three flavours. Note, however, that ${\tilde f}^e_{\rm IH}$ has the opposite sign to ${\tilde f}^{\mu/\tau}_{\rm IH}$. 

\subsubsection{Unflavoured weak LNV}

For $U^2$ exceeding the weak flavour region given by eq.~(\ref{eq:boundfwo}), an asymmetry is only achievable if the LNV mode is weak. Using eq.~\eqref{eq:betaMint}, this requires 
\begin{eqnarray}
(U^2)_{\rm wLNV}\leq 4 \times 10^{-6} \left({M\over 1{\rm GeV}}\right)^{-4}\,.
\label{eq:WLNV}
\end{eqnarray}
According to the analytical result obtained for this regime, given by eq.~\eqref{eq:lnv_int_sol}, and using eq.~\eqref{eq:intlncinvnh} (eq.~\eqref{eq:intlncinvih}) for NH (IH), it is easy to check that the corresponding asymmetry is independent of the mixing $U^2$. Maximizing over the unknown CP phases, we have found that the maximum asymmetry achievable in this regime is much smaller than the observed BAU for the relevant range of HNL masses. Therefore this regime fails in reproducing the BAU.

\subsection{Fast oscillation regime}

In the fast oscillation regime the analytical approximations are valid for mixings smaller than the one given in eq.~\eqref{eq:U2_int_osc}.
As in the intermediate regime, two qualitatively different regimes need to be considered: if the flavour $\alpha$ remains weak until $x_{\rm EW}$, or if it is the LNV mode the one remaining weak.
In the latter case the analytical approximation matches exactly the one of the intermediate regime and thus the same conclusion as in the previous subsection applies: the BAU can not be explained. 
However, with flavour effects, which are possible in the range defined by eq.~\eqref{eq:boundfwo}, the asymmetry can be expressed by using eq.~\eqref{eq:lnc_osc_sol}, and eq.~(\ref{eq:intlncinvnh}) (eq.~(\ref{eq:intlncinvih})) for NH (IH), as
\begin{eqnarray}
\left( Y_B\right)_{\rm fw-osc} = -4.3 \times 10^{-12} \eta {\tilde f}^\alpha_{\rm NH/IH} \left({U^2\over 10^{-9}}\right) \left({\Delta M\over M}\right)^{-2/3} \left({M\over 1{\rm GeV}}\right)^{5/3}\,,
\label{eq:Yb_fw_osc}
\end{eqnarray}
with the same constant factor $\eta$ and angular function ${\tilde f}^\alpha_{\rm NH/IH}$ as in the intermediate regime.
Successful baryogenesis then implies a lower limit on $U^2$ given by
\begin{eqnarray}
(U^2)_{\rm osc} &\geq& 18\,(3.7) \times 10^{-8} \eta \left({\Delta M\over M}\right)^{2/3} \left({1\,{\rm GeV}\over M}\right)^{5/3} {\rm NH\,(IH)}\,.
\label{eq:U2_bound_fast_osc}
\end{eqnarray}
When this lower limit becomes larger than the upper limit of flavoured weak washout, eq.~(\ref{eq:boundfwo}), which happens at large $\Delta M/M$, no solution is possible. 
Thus, these two conditions can be used to set an upper bound on $\Delta M/M$ for which the BAU can be reproduced within the fast oscillation regime
\begin{eqnarray}
{\Delta M\over M} \leq 4.2 \times 10^{-3\,(4)} \left({1{\rm GeV} \over M} \right)^{1/2} {1 \over {\rm Min}(\epsilon_\alpha)^{3/2}} ~ {\rm NH\,(IH)}\,.
\label{eq:dmbound}
\end{eqnarray}


\section{Numerical results: comparison with analytical approximations and parameter scan}
\label{sec:num}

As we have seen, the generation of a baryonic asymmetry via right handed neutrino oscillations generally involves various time scales which may be very different. The stiffness of a (linear) numerical system such as eq.~(\ref{eq:rdot}) is dictated by the ratio of the largest to smallest non-zero eigenvalue  of $A$, $\mathrm{max}(|\lambda|)/\mathrm{min}(|\lambda|)$. 
If this happens to be much bigger than unity the system is affected by a \textit{stiff} behaviour.
The standard method to overcome the problem is to use variable-order implicit methods.
We find, in agreement with ref.~\cite{Eijima:2018qke}, that the \texttt{FORTRAN77 ODEPACK}  implementation of the \texttt{LSODA} algorithm efficiently solves the full non-linear set of kinetic equations.
Furthermore, significant speed up can be achieved in the fast oscillating regime, $\Gamma_{\rm osc}/\Gamma \gg 1$, by switching to an incoherent evolution. 
We average out the oscillations once they reach a frequency of $10^5$ or $10^3$ oscillations are completed. 
With these optimizations the solver integrates within seconds, and therefore an extensive scan of the parameter space is possible. 
The software used, \href{https://github.com/stefanmarinus/amiqs}{\tt amiqs}~\cite{pilar_hernandez_2022_6866454}, is made publicly available.

\subsection{Analytical results versus numerical solutions}

The derived analytical solutions presented in sec.~\ref{subsec:solutions} represent asymptotic solutions for $\sum_\alpha \mu_{B/3-L_\alpha}$. 
For the intermediate and fast oscillation regimes we only give the large time asymptotic result. 
Although the full time dependence can also be obtained, the expressions are too lengthy and not particularly illuminating. 
To verify the accuracy of the analytical solutions we confront them with i) the numerical solution within the same approximations used in the analytical derivation (i.e.  linearization of the full system, constant rates ($\gamma_i, s_i$), and a diagonal $C$ matrix), and ii) the full non-linear numerical solution.
In order to easily select the different regimes and for clarity we make use of the parameterization  in eq.~\eqref{eq:LNVparam2N}, i.e. we do not include the light neutrino mass constraints here.
Including them does not change anything qualitatively, but different regimes become non-linearly connected to the input parameters.

Considering the CP invariants given by eqs.~(\ref{eq:CP_Invariants_Yukawa1})-(\ref{eq:CP_Invariants_Yukawa4}), it is evident 
that unequal $y_\alpha$ are necessary to generate a non-zero asymmetry within the LNC limit.
In contrast LNV contributions are non-zero in a flavour democratic scenario with equal $y_\alpha$. Such choice actually isolates the pure LNV contribution. In the general case of unequal $y_\alpha$ both, LNC and LNV contributions, contribute to the final asymmetry. Also, recall that outside the overdamped regime flavour effects are necessary to explain the BAU, see section~\ref{sec:parambounds}. 
In Tab.~\ref{tab:ana_vs_num_ini}, we present various choices of the input parameters considering unequal $y_\alpha$ that we use to test the agreement of our analytical expressions to the numerical result. 
\begin{table}[!t]
\begin{center}
\begin{tabular}{c| c c c c c c c c}
Scenario & $\log_{10}(M)$ & $\log_{10}({\Delta M\over 2})$ & $\log_{10}(y_{e})$ & $\log_{10}(y_{\mu})$ & $\log_{10}(y_{\tau})$ &  $\Delta \beta_e$ & $\Delta \beta_\mu$ & $\Delta \beta_\tau$ \\
\hline
\hline
(a) & $ 0 $ & $-10$ & $-5$ & $-5.1$  & $-5.2$ & $0$ & $\pi/2$ & $\pi/2$  \\
\hline
(b) & $ 1 $ & $-10$ & $-5$ & $-5.1$  & $-5.2$ & $0$  & $\pi/2$ & $\pi/2$ \\
\hline
(c) & $ 1 $ & $-5$ & $-5$ & $-5.1$ & $-8.2$ & $0$ & $\pi/2$ & $\pi/2$  \\
\hline
(d) & $ 1.5 $ & $-1$ & $-8$ & $-5.4$  & $-5.5$ & $\pi/2$ & $\pi/2$ & $0$  \\
\hline
\hline
\end{tabular}
\caption{Input parameters for the comparison between the analytical and numerical solutions shown in Fig.~\ref{fig:comparison_ana_num}. 
The perturbative $y'$ parameters are always taken to be the same: $y_e' =10^{-9}, y_\mu'=10^{-9.1}, y_\tau' =10^{-9.2}$.
}
\label{tab:ana_vs_num_ini}
\end{center}
\end{table}
\begin{figure}[!t]
\centering
\begin{tabular}{cc}
\hspace{-0.5cm} \includegraphics[width=0.5\textwidth]{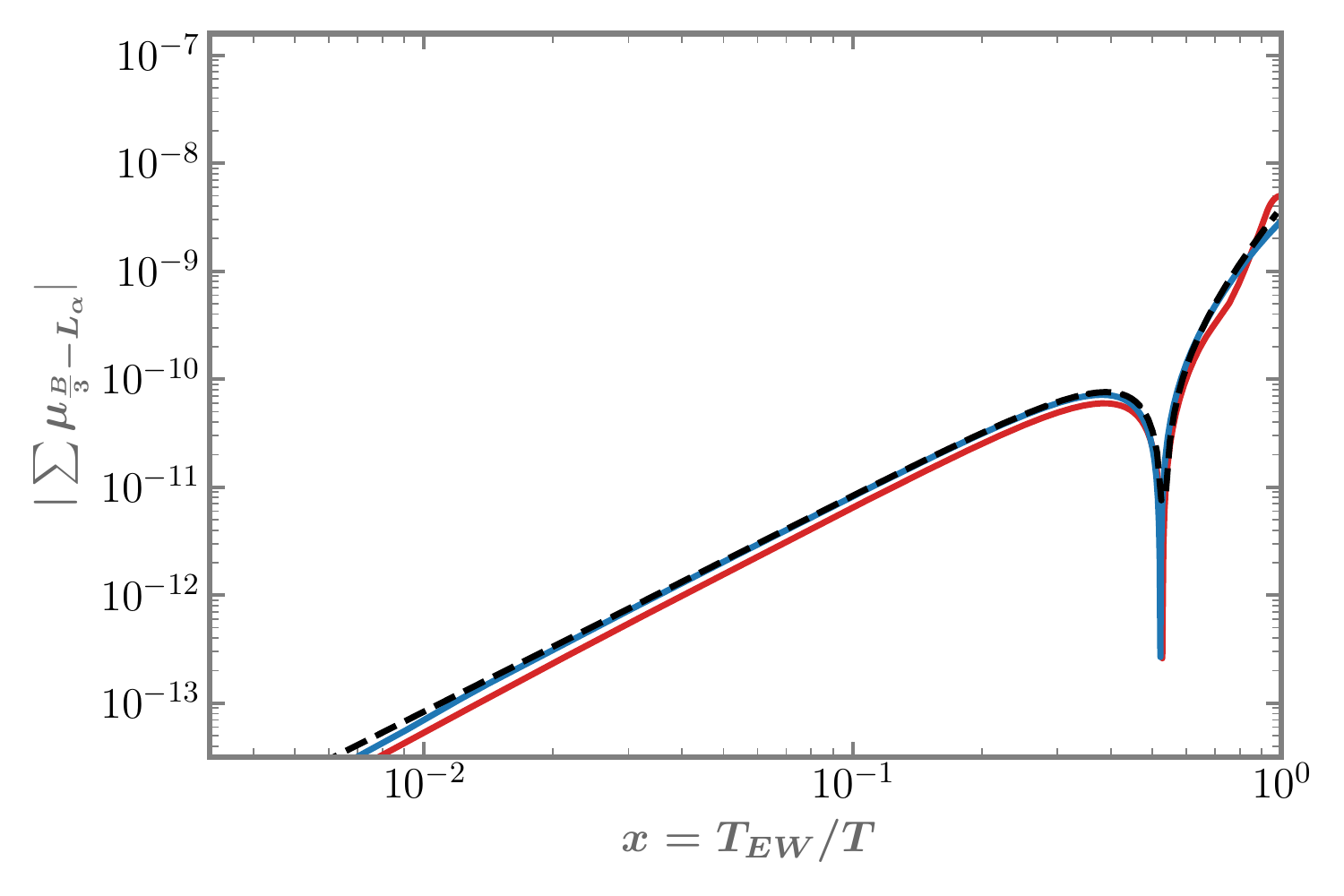} & \hspace{-0.55cm}  \includegraphics[width=0.5\textwidth]{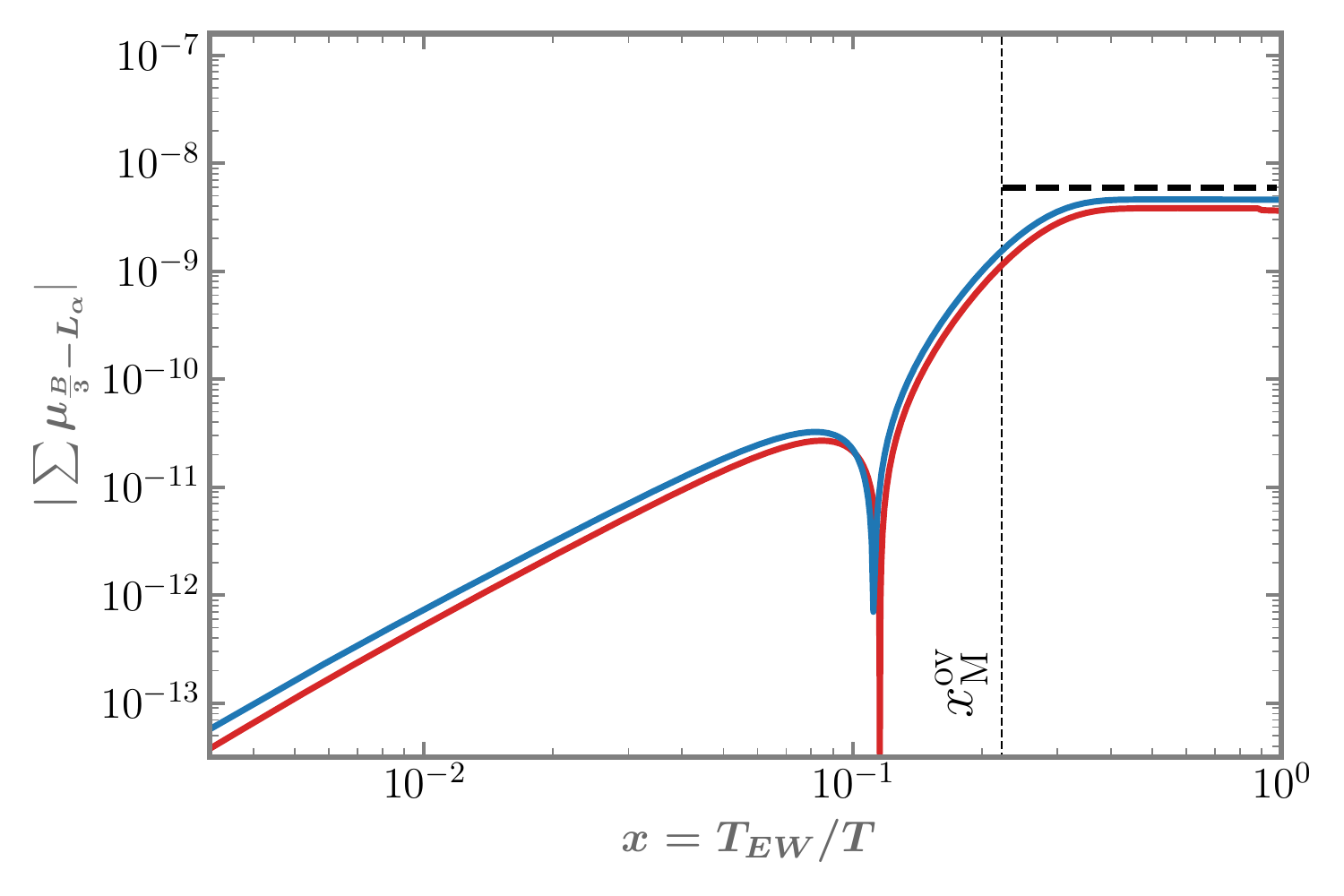}   \\
\hspace{-0.5cm} \includegraphics[width=0.5\textwidth]{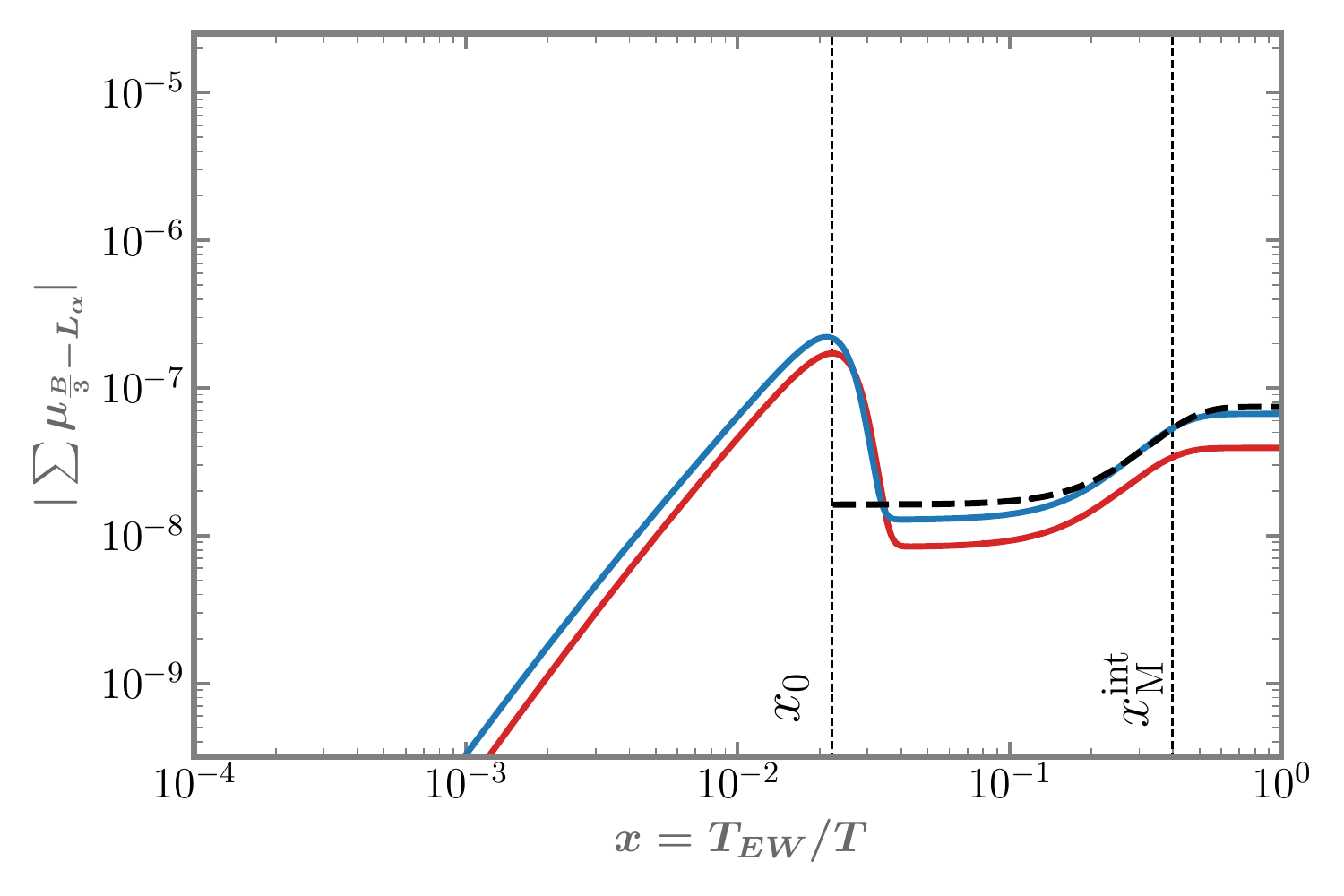} & \hspace{-0.55cm}  \includegraphics[width=0.5\textwidth]{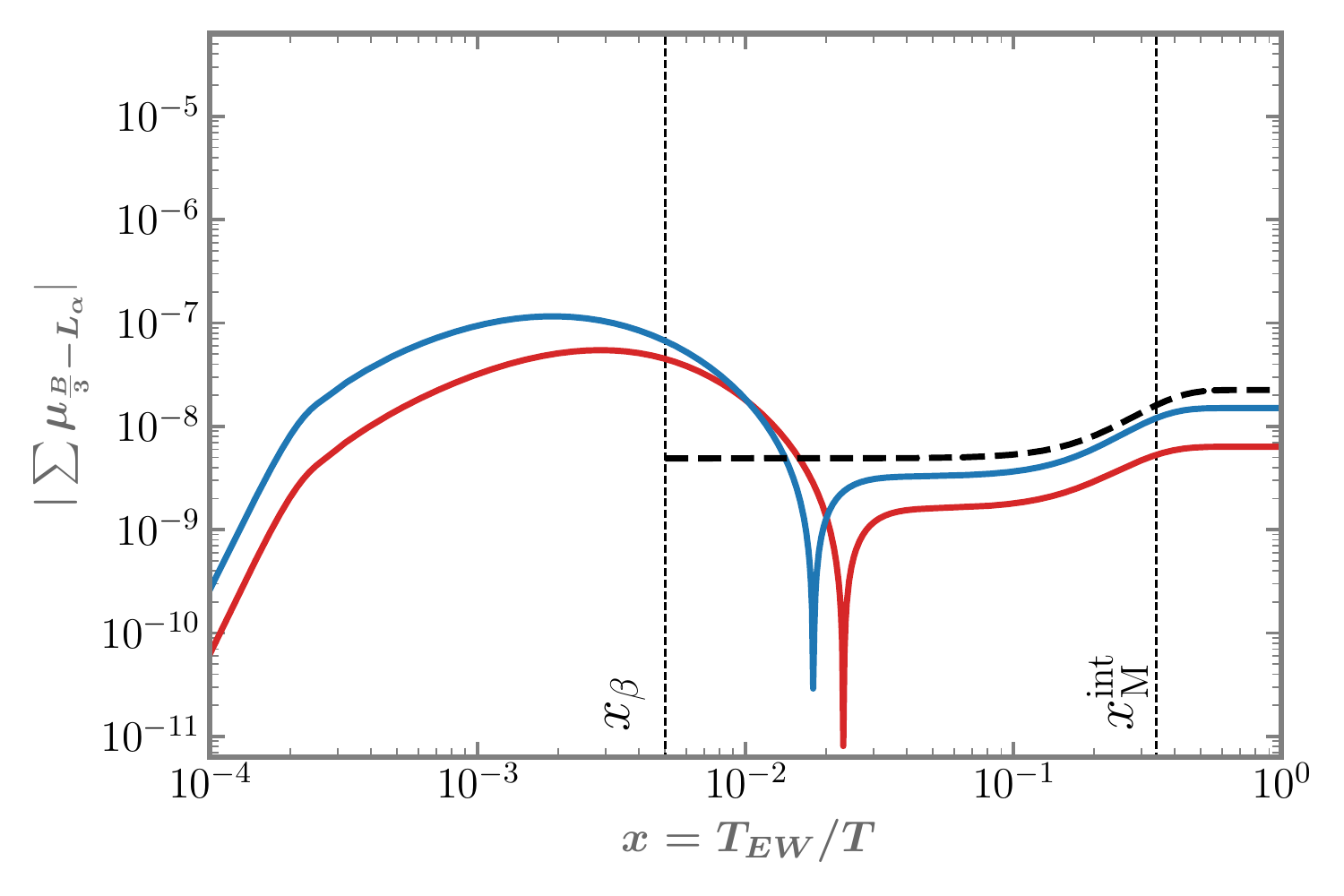}   \\
\end{tabular}
\vspace{-0.4cm}
\caption{ Comparison of the asymptotic analytical result (black dashed) to i) the numerical result with the same settings (blue) and ii) the full non-linear numerical solution (red) in the four scenarios (a)-(d) as described in the main text.
In the top left we show the scenario (a), in the top right the scenario (b), in the bottom left the scenario (c) and in the bottom right the scenario (d). 
The vertical dashed lines indicate projection times used for the analytical derivation.
}
\label{fig:comparison_ana_num}
\end{figure}
Our choice of parameters allow us to exemplify the different regimes that are relevant in different regions of the parameter space, namely
\begin{enumerate}[label=(\alph*)]
\item{Overdamped regime with weak LNV as given by eq.~\eqref{eq:overdamped_wLNV_sol_inv},}
\item{Overdamped regime with strong LNV as given by eq.~\eqref{eq:overdamped_sLNV_sol_inv},}
\item{Intermediate regime with slow flavour $\alpha$ and strong LNV as given by eq.~\eqref{eq:intermediate_fw_sol_inv},}
\item{Fast oscillation regime with slow flavour $\alpha$ and strong LNV as given by eq.~\eqref{eq:lnc_osc_sol}.}
\end{enumerate}
Our results are shown in Fig.~\ref{fig:comparison_ana_num}. 
The comparison of the analytical result, indicated by the dashed line, with the numerical solution obtained in the same approximations used in the analytical analysis, shown in blue, is very good in all cases. 
The exact numerical result (red) including non-linear terms, the $C$ matrix of eq.~(\ref{eq:Cmatrix}) and temperature dependent rates differ within a factor of two at most with the analytical estimate. This is mainly due to the difference in the rates considered.

\subsection{Parameter scan of testable baryogenesis}

We have performed a numerical scan of the parameter space compatible with successful baryogenesis for HNL masses in the range $0.1 \leq M\leq 100$ GeV. 
In this range, the best testability options will be provided by \texttt{SHiP}~\cite{SHiP:2015vad} and \texttt{FCC} running at the Z-peak~\cite{Blondel:2014bra}. Our main goal is to study the correlation between the BAU and different observables, such as the masses and mixings of the HNLs, and therefore we have restricted the scan to the part of the parameter space that can be probed by these future experiments.   

 We use a Bayesian estimation from the log-likelihood
\begin{align}
\log(\mathcal{L}) = -\frac{1}{2} \left(\frac{Y_{B}(T_{\text{EW}}) - Y_{B}^{\mathrm{exp}}}{\sigma_{Y_{B}^{\mathrm{exp}}}} \right)\,,
\end{align}
which we implement in the nested sampling algorithm \href{https://github.com/JohannesBuchner/UltraNest}{\tt UltraNest}~\cite{2021JOSS....6.3001B}.

The result of a bayesian estimation is always dependent on the concrete choice of the prior distribution.
Being restricted to the minimal scenario with two HNLs, the parameter space which can explain the light neutrino data is spanned by $6$ independent variables: three phases $(\delta, \phi, \theta)$, two parameters fixing the heavy neutrino mass scale ($M, \Delta M$) and one parameter which essentially fixes the Yukawa scale, $y$. We agnostically choose flat priors linear in the three phases and logarithmic in $M_1, \Delta M/M_1$ and $y$, see Tab.~\ref{tab:priors}. 
\begin{table}[!b]
\begin{center}
\begin{tabular}{c c c c c c}
$\log_{10}(M_1)$ & $\log_{10}(\Delta M/M_1)$ & $\log_{10}(y)$ & $\theta$ & $\delta$ & $\phi$ \\
\hline
\hline
$[-1,2]$ & $[-14,-1]$ & $[-8,-4]$ & $[0,2\pi]$ & $[0,2\pi]$ & $[0,2\pi]$   \\
\hline
\hline
\end{tabular}
\caption{Priors for the nested sampling.}
\label{tab:priors}
\end{center}
\end{table}
Additionally, since we are mainly interested in the testability of this mechanism within \texttt{SHiP} and \texttt{FCC}, the sampler is programmed to automatically reject points which fall outside the sensitivity reach or are already experimentally excluded, thereby augmenting the speed of parameter space volume shrinking towards a higher likelihood.
A further constraint on the parameter space arises from imposing that the symmetry breaking parameter $y'/y < 0.1$, see eq.~\eqref{rhoNH} (eq.~\ref{rhoIH}) for NH (IH). The lower bound on $\Delta M/M$ is somewhat arbitrary since the evolution is overdamped in the region of the parameter space that can be probed by \texttt{SHiP} (\texttt{FCC}) already for $\Delta M/M \sim 10^{-10} (10^{-12})$.
Even though the analytical results seem to indicate that asymmetries vanish in the limit of $\Delta M\rightarrow 0$, at higher order in $y'$ there are additional CP invariants~\cite{Drewes:2022kap} that may be relevant in this limit~\cite{Antusch:2017pkq}. This case will be considered elsewhere.

Let us first analyze the case in which $\Delta M/M$ is fixed to different values, i.e. $\Delta M/M = 10^{-10}\,,10^{-5},\,10^{-2}$, before we turn to discuss the global scan varying $\Delta M/M$.
This separates different regimes (overdamped, intermediate, fast oscillations) to be relevant in different parts of the parameter space.

{\it Highly degenerate HNLs with $\Delta M/M = 10^{-10}$}

\begin{figure}[!t]
\centering
\begin{tabular}{cc}
\hspace{-0.55cm} \includegraphics[width=0.5\textwidth]{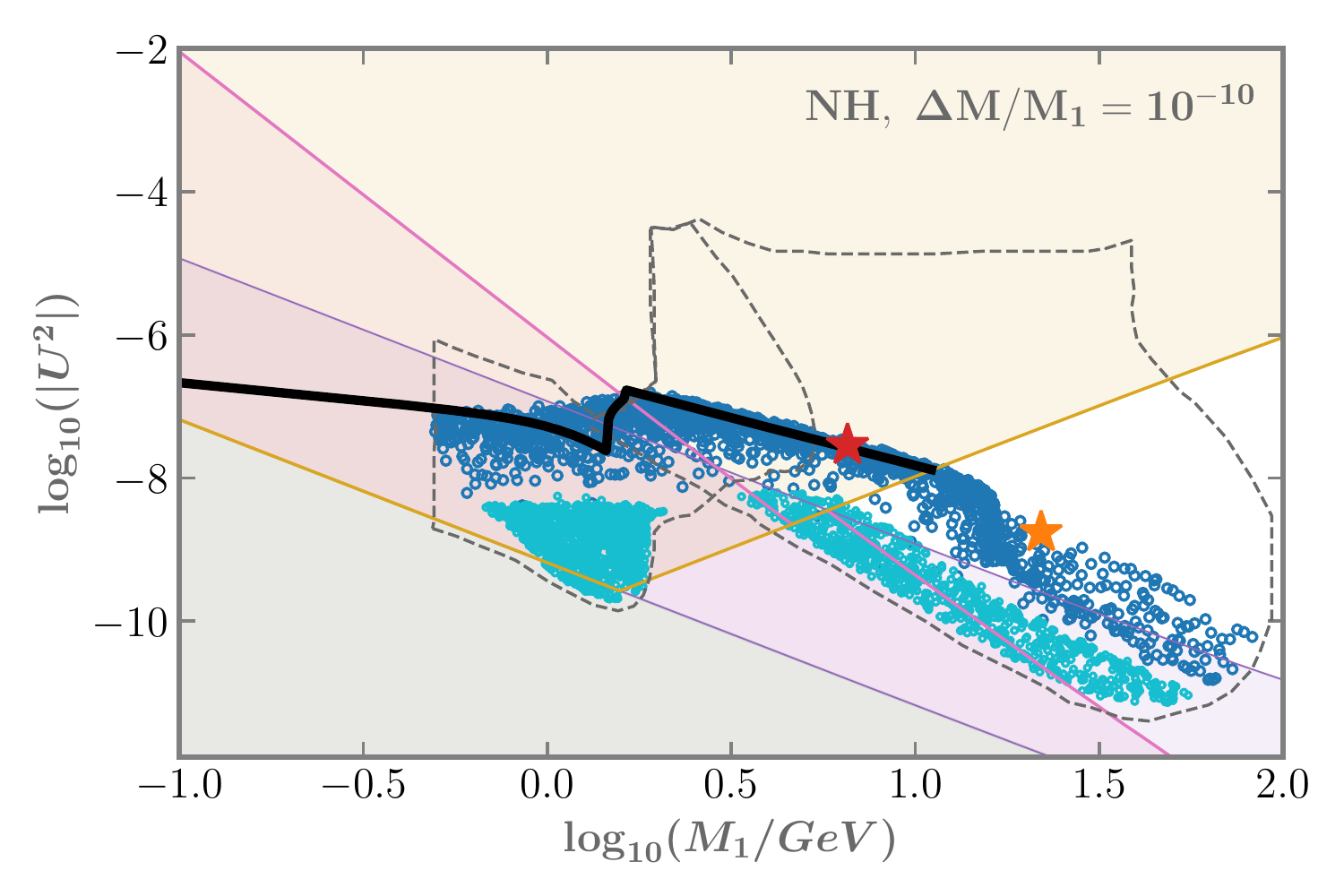} & \hspace{-0.55cm}  \includegraphics[width=0.5\textwidth]{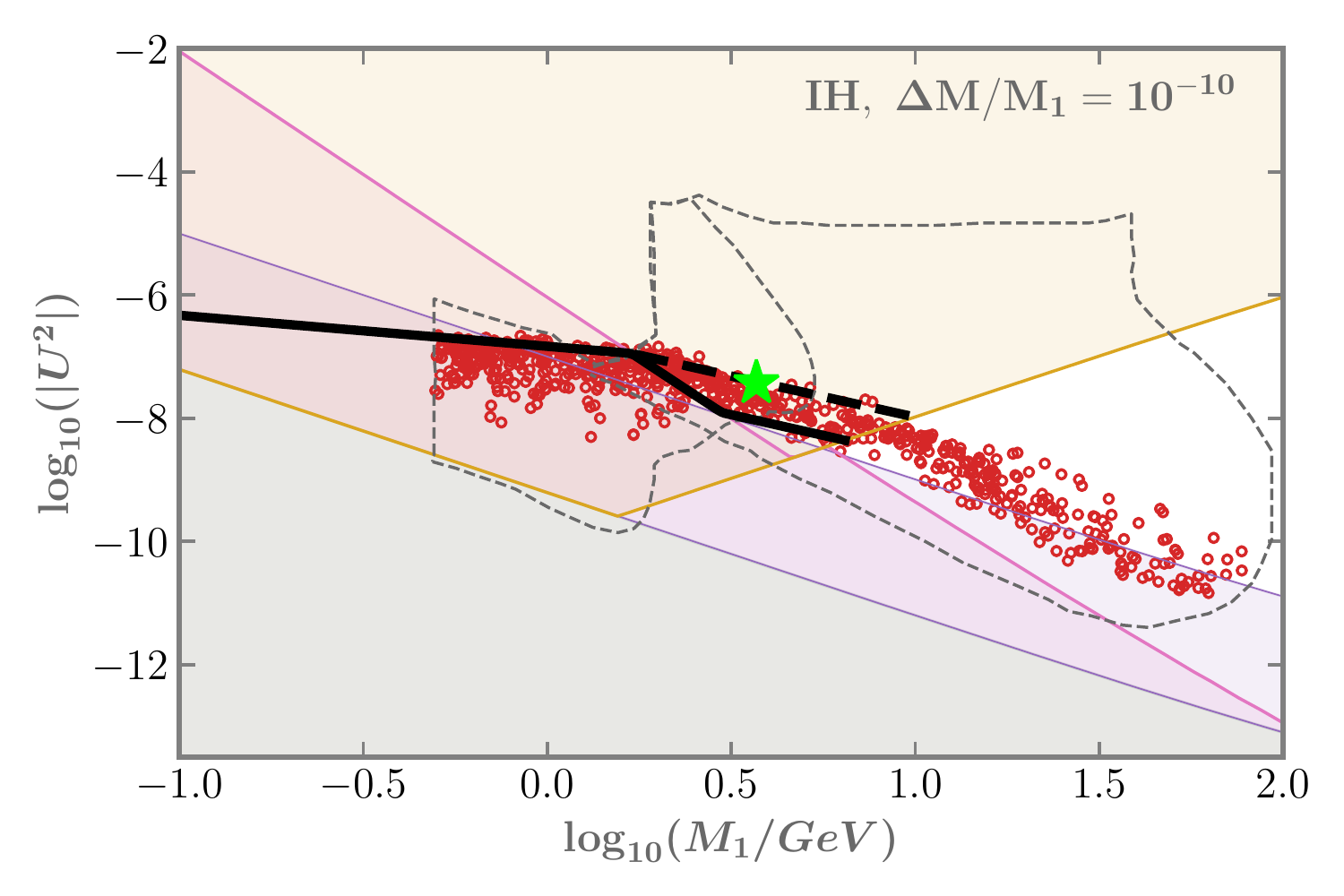}    
\end{tabular}
\vspace{-0.4cm}
\caption{ 
Result of the numerical scan for $\Delta M/M = 10^{-10}$ shown in blue (red) for NH (IH) with standard priors in the phases. For NH, we 
include in lighter blue the result obtained using priors for the phases that are flat in a logarithmic scale. 
The black lines represent the analytical upper bound on the mixing, while the dashed for IH is the conservative bound described in the text. The stars indicate benchmark points, see main text. Color coding for the shaded regions as in Fig.~\ref{fig:regimes}.
}
\label{fig:scan_fixed_dmm10}
\end{figure}

For mass degeneracies of $\Delta M/M \lesssim 10^{-8} (10^{-9})$ the overdamped regime starts to apply in part of the parameter space covered by \texttt{SHiP} and \texttt{FCC}. In this case, successful BAU does not require flavour effects and 
HNL mixings beyond the constrained flavoured weak washout region as defined in eq.~\eqref{eq:boundfwo} are possible.
However, the mixing is not unrestricted because the BAU imposes an upper bound (if light neutrino masses are accounted for), which depends on whether the LNV rates are weak, c.f. eqs.~\eqref{eq:U2_ov_wLNV_max_fixed_DMM} and  \eqref{eq:U2_ov_wLNV_max_LNC_fixed_DMM}, or strong, c.f. eq.~\eqref{eq:U2_ov_sLNV_max_fixed_DMM}.
These upper bounds are represented as black lines in Fig.~\ref{fig:scan_fixed_dmm10}.
The dashed line in the IH scenario shows the conservative bound 
resulting from the maximal achievable asymmetry within the wLNV regime, as explained in sec.~\ref{sec:parambounds}, which is partially washed out. 
Note that in the NH scenario both estimates are identical.
In order to have a more quantitative understanding, it is useful to analyze representative benchmark points.
We choose three benchmark points which account for different properties:
\newpage
\begin{enumerate}[label=(\alph*)]
\item{Red star: saturating the upper bound on the mixing.}
\item{Green star: saturating the conservative upper bound.}
\item{Orange star: point within the region in which the BAU is reached via exponential fine-tuning.}
\end{enumerate}
The corresponding evolution of the baryon asymmetry is depicted in Fig.~\ref{fig:bp_comparison}, and shows the expected behavior in accordance with the analytical understanding. 

Recall that points saturating the upper bound on the mixing are achieved via the \textit{natural} value of the angular part of the CP invariants $f_{\rm LNC}^{\rm H} \simeq f_{\rm LNV}^{\rm H} \simeq \mathcal{O}(1)$, see eqs.~\eqref{eq:f_lnc_ih}-\eqref{eq:f_lnc_lnv_H}.
For smaller mixings, suppressed angular functions are needed and this implies a non-trivial correlation between the CP phases, see Fig.~\ref{fig:f_contours}. 
Our bayesian analysis, with flat priors in all three phases, was not able to resolve the necessary pattern and hence the density of points decreases with the distance to the upper bound.
As a proof of principle, we made an additional scan for NH with logarithmic priors in all phases within the range $[-5, -2]$. Since for NH the angular function depends mostly on $\theta$, the logarithmic flat prior in this parameter should help. 
Indeed, this separate analysis finds points compatible with the BAU up to the sensitivity limit of \texttt{SHiP} and \texttt{FCC}.
This result demonstrates the well known fact that the posterior result is strongly dependent on the prior assumptions, as well as the difficulty of exploring  such large parameter space without an analytical understanding.

{\it Mildly degenerate HNLs with  $\Delta M/M = 10^{-5}$ }

\begin{figure}[!t]
\centering
\begin{tabular}{cc}
\hspace{-0.5cm} \includegraphics[width=0.5\textwidth]{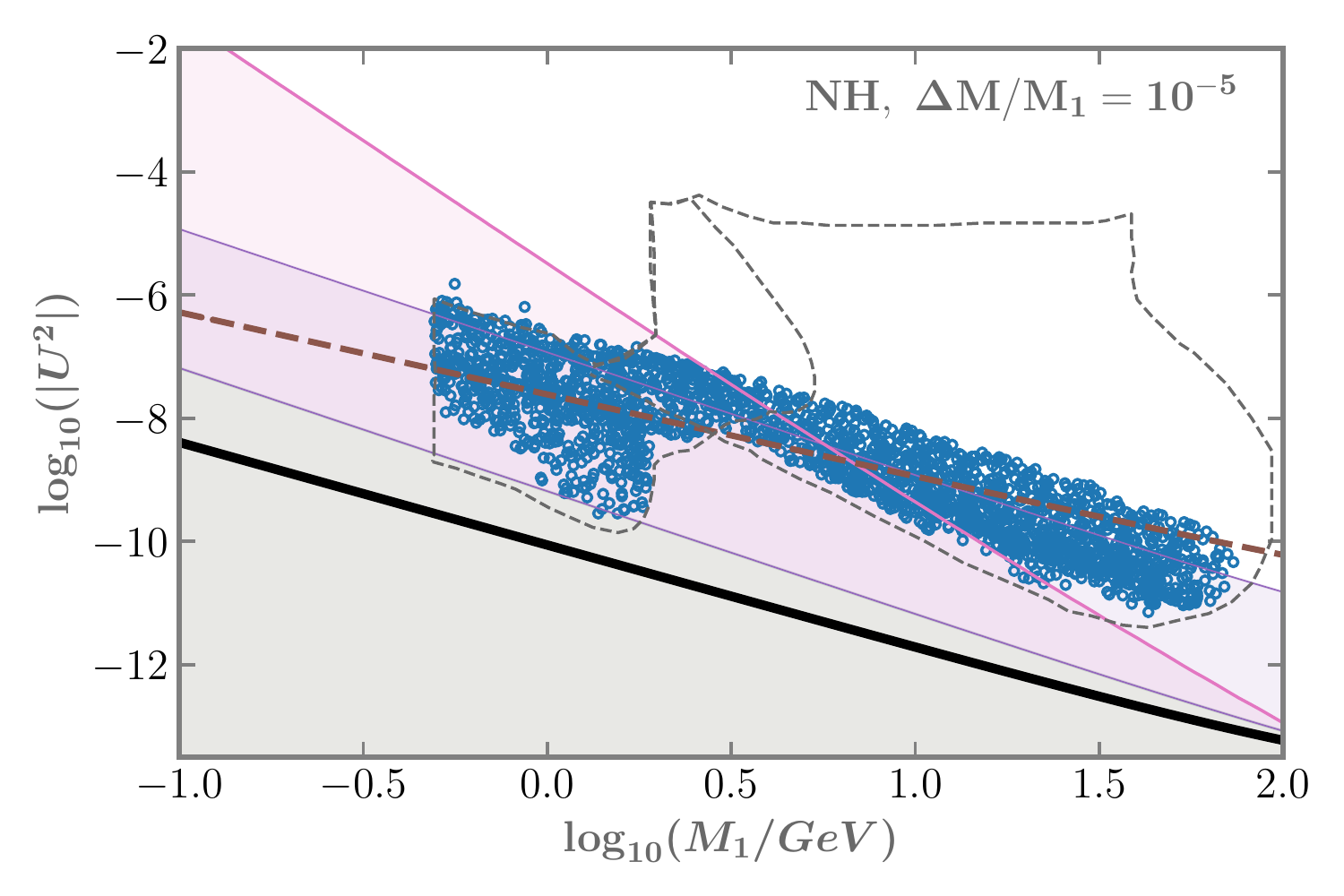} &
\hspace{-0.55cm}  
\includegraphics[width=0.5\textwidth]{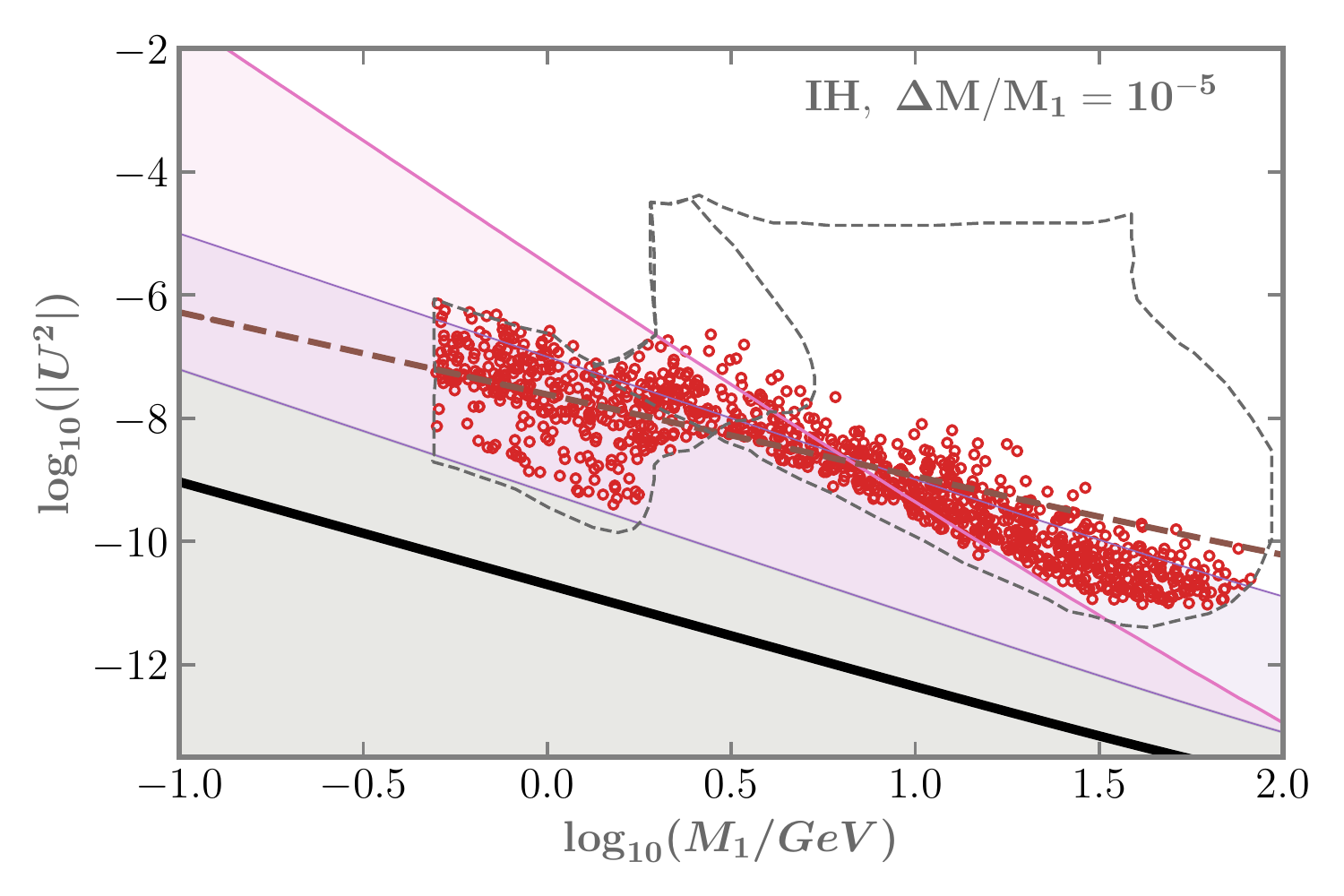}   
\end{tabular}
\vspace{-0.4cm}
\caption{ 
Result of the numerical scan for $\Delta M/M = 10^{-5}$ shown in blue (red) for NH (IH).
The mixing is only bounded by the requirement of having a slow flavour $\alpha$ at $T_{\rm EW}$.
Color coding for the shaded regions as in Fig.~\ref{fig:regimes}.
}
\label{fig:scan_fixed_dmm5}
\end{figure}

For mildly degenerate HNLs two different regimes become relevant, i.e. the intermediate and fast oscillation regime.
They are separated by the line defined by eq.~\eqref{eq:U2_int_osc}.
In both cases the HNL mixing is only bounded from above via the requirement of having a weak flavour at $T_{\rm EW}$, see eq.~\eqref{eq:boundfwo}.
This is clearly seen in Fig.~\ref{fig:scan_fixed_dmm5}.
Points which can explain the BAU for larger mixings, i.e. without having a slow flavour $\alpha$ until $T_{\rm EW}$, necessarily show an exponential fine-tuned behaviour similar to the orange benchmark point shown in Fig~\ref{fig:bp_comparison}.
However, the numerical scan finds less points showing this fine tuned behaviour than in the case of $\Delta M/M = 10^{-10}$. This is because the  overshooting of the asymmetry at earlier times is larger (and needs therefore to be more strongly washed out) for larger $\Delta M/M$. 

{\it Non-degenerate HNLs with  $\Delta M/M = 10^{-2}$ }

\begin{figure}[!t]
\centering
\begin{tabular}{cc}
\hspace{-0.5cm} \includegraphics[width=0.5\textwidth]{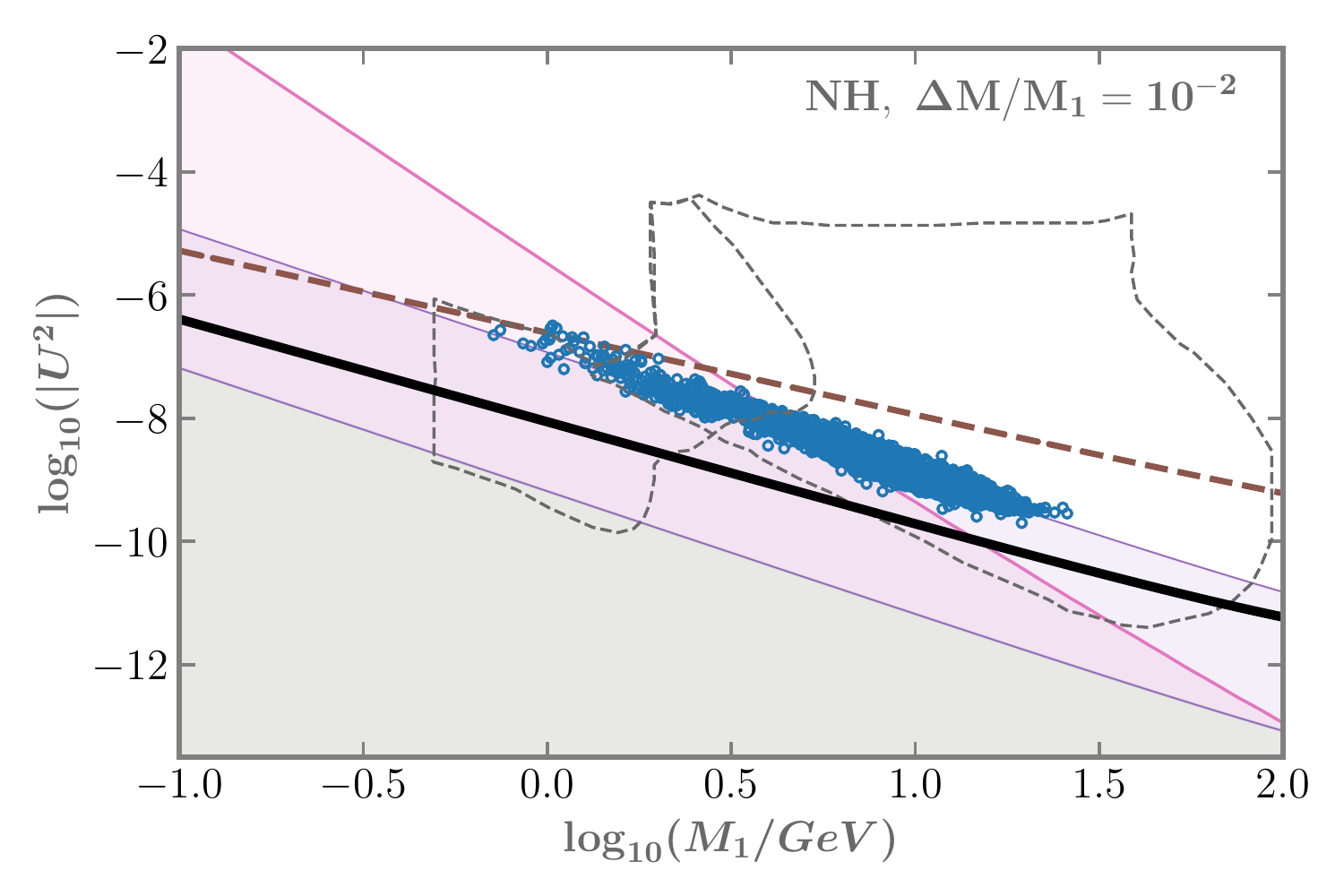} &
\hspace{-0.55cm}  \includegraphics[width=0.5\textwidth]{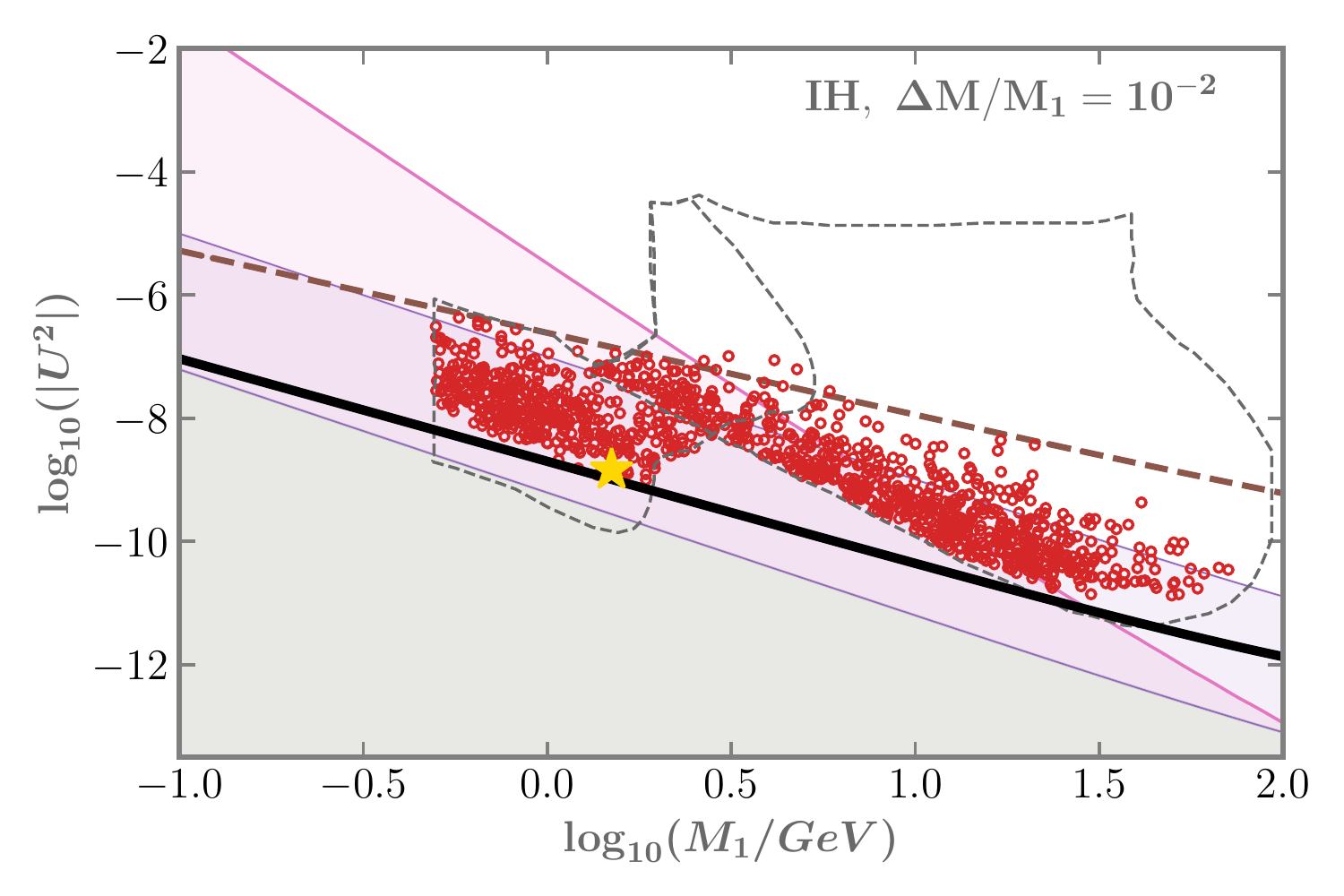}   
\end{tabular}
\vspace{-0.4cm}
\caption{ 
Result of the numerical scan for $\Delta M/M = 10^{-2}$ shown in blue (red) for NH (IH).
The lower bound on the mixing imposed by the BAU in the fast oscillation regime is indicated by the black line, while there is an upper bound given by the requirement of having a slow flavour $\alpha$ at $x_{\rm EW}$. Color coding for the shaded regions as in Fig.~\ref{fig:regimes}.}
\label{fig:scan_fixed_dmm2}
\end{figure}

\begin{figure}[!t]
\centering
\begin{tabular}{cc}
\hspace{-0.5cm}  \includegraphics[width=0.5\textwidth]{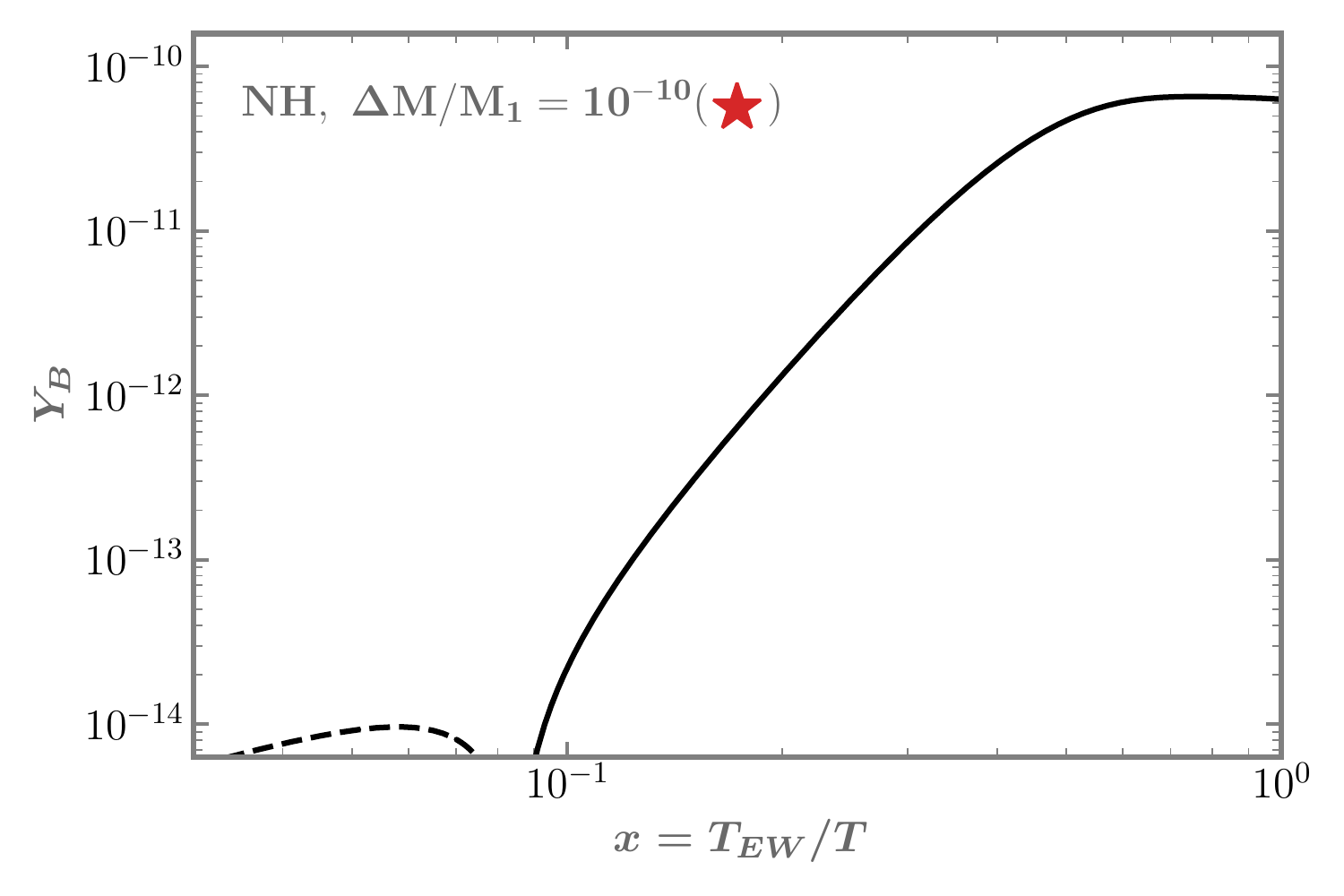} &
\hspace{-0.55cm}  \includegraphics[width=0.5\textwidth]{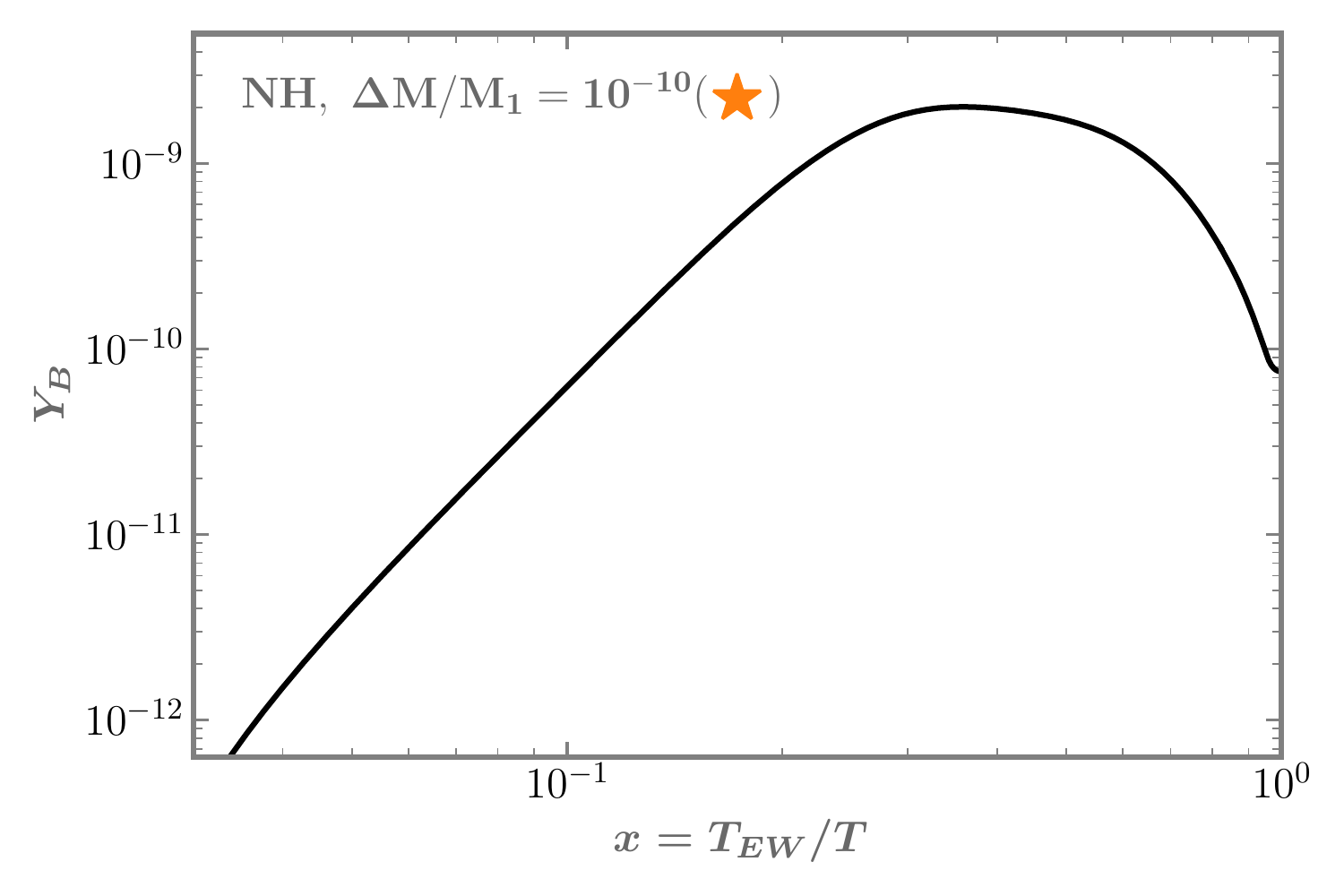} \\
\hspace{-0.5cm}  \includegraphics[width=0.5\textwidth]{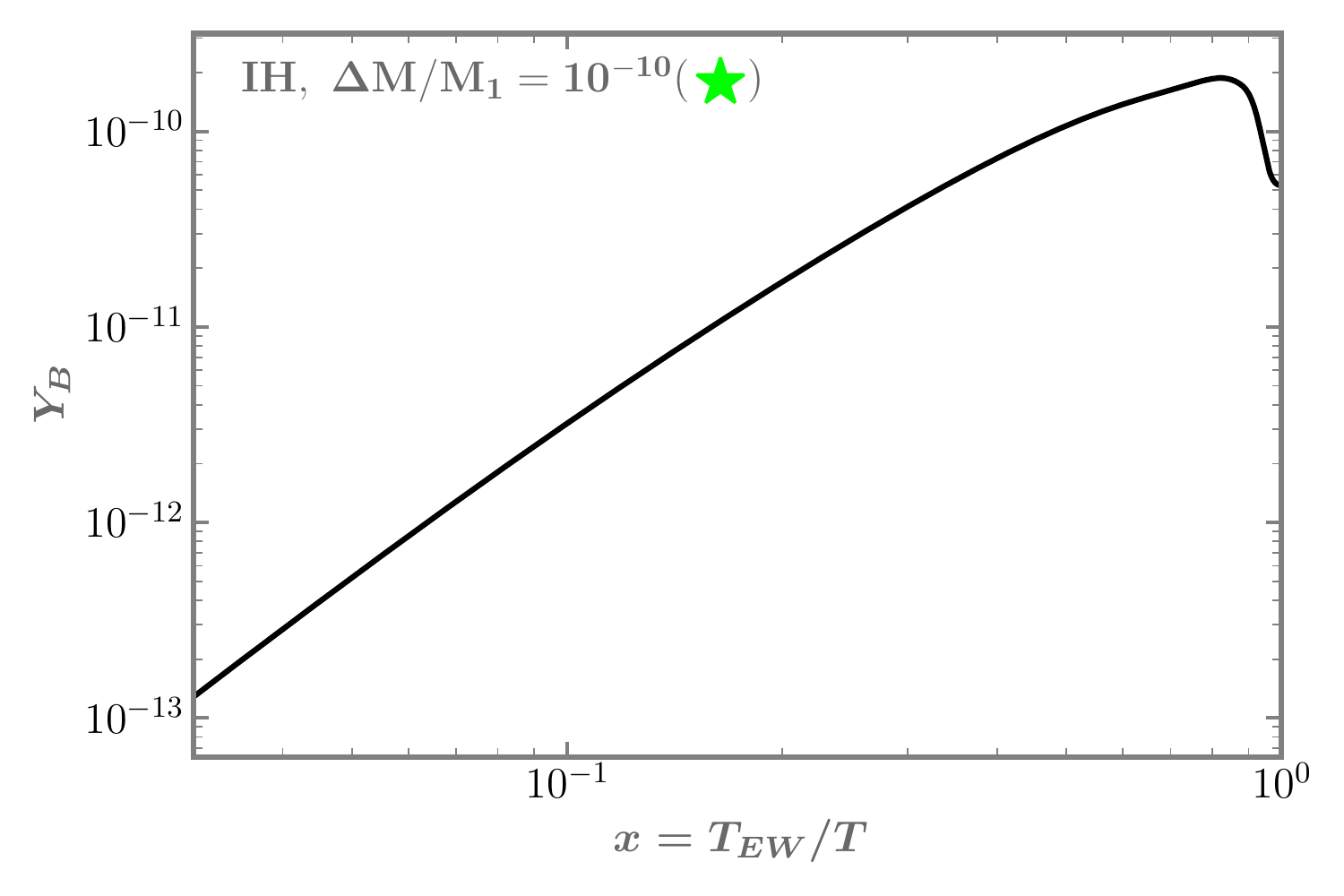} &
\hspace{-0.55cm}  \includegraphics[width=0.5\textwidth]{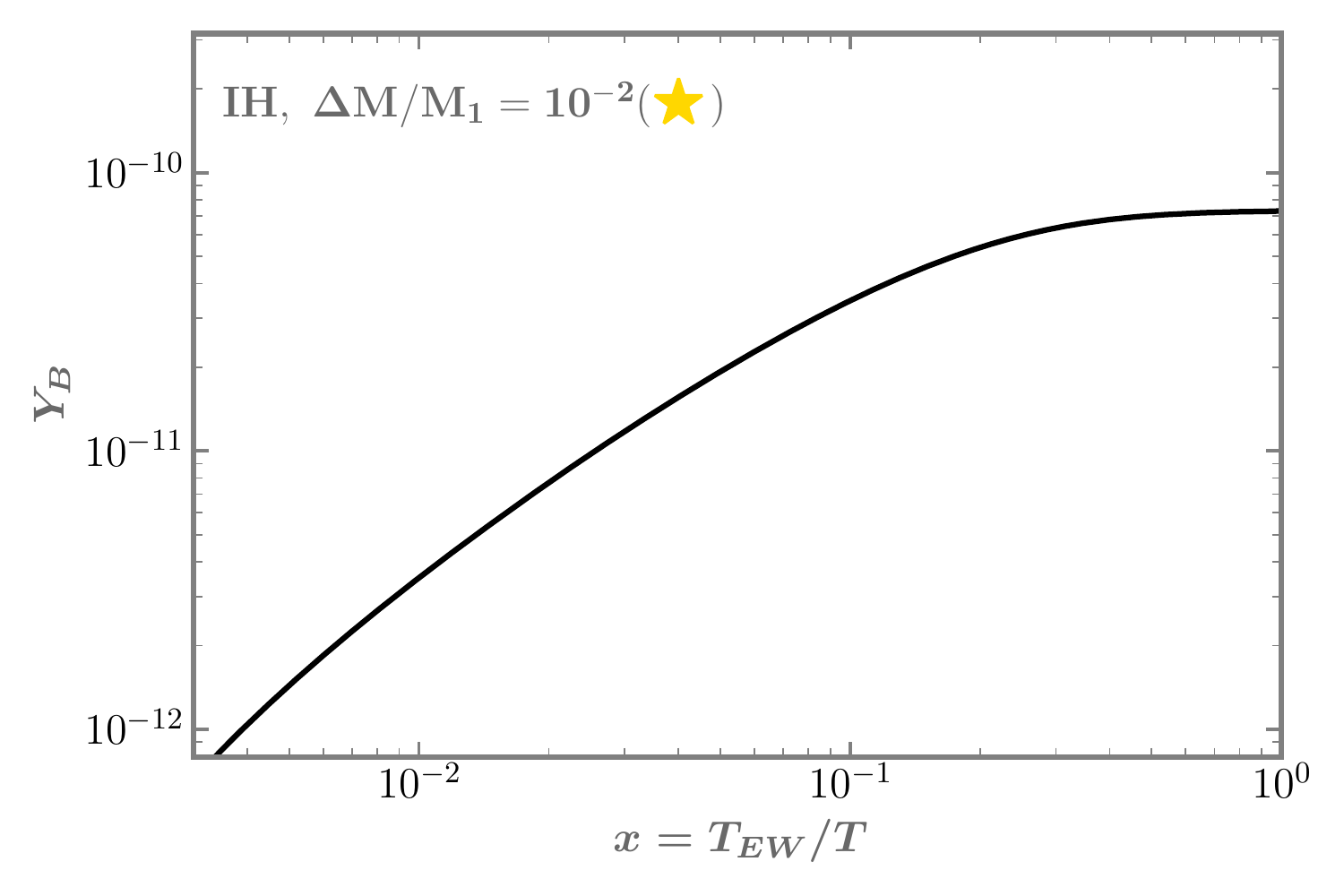} \\
\end{tabular}
\vspace{-0.4cm}
\caption{ 
Benchmark points from the numerical scan representing different qualitative behaviours of the BAU generation, see main text and Figs.~\ref{fig:scan_fixed_dmm10} to \ref{fig:scan_fixed_dmm2}.
}
\label{fig:bp_comparison}
\end{figure}

In this case, the baryon asymmetry is generated always in the fast oscillation regime.
As we have seen in the previous section, in this regime the BAU imposes a lower bound on the HNL mixing, see eq.~\eqref{eq:U2_bound_fast_osc}, indicated by the solid black line in Fig.~\ref{fig:scan_fixed_dmm2}.
This lower bound on $U^2$ is indeed found in the numerical scan as shown in Fig.~\ref{fig:scan_fixed_dmm2}.
We select a benchmark point saturating the lower bound (yellow star) and the evolution of the corresponding BAU generated is shown in Fig.~\ref{fig:bp_comparison}.
The evolution is characterized by an approximate constant asymmetry at late times, indicating the relevance of a weakly coupled flavour $\alpha$.
Smaller mixings would not reproduce the correct BAU. 
On the other hand, the upper bound on the mixing is again given by the criteria of having a slow flavour $\alpha$ during all the evolution.

{\it Global result for variable $\Delta M/M$}

\begin{figure}[!t]
\centering
\begin{tabular}{cc}
\hspace{-0.5cm} \includegraphics[width=0.5\textwidth]{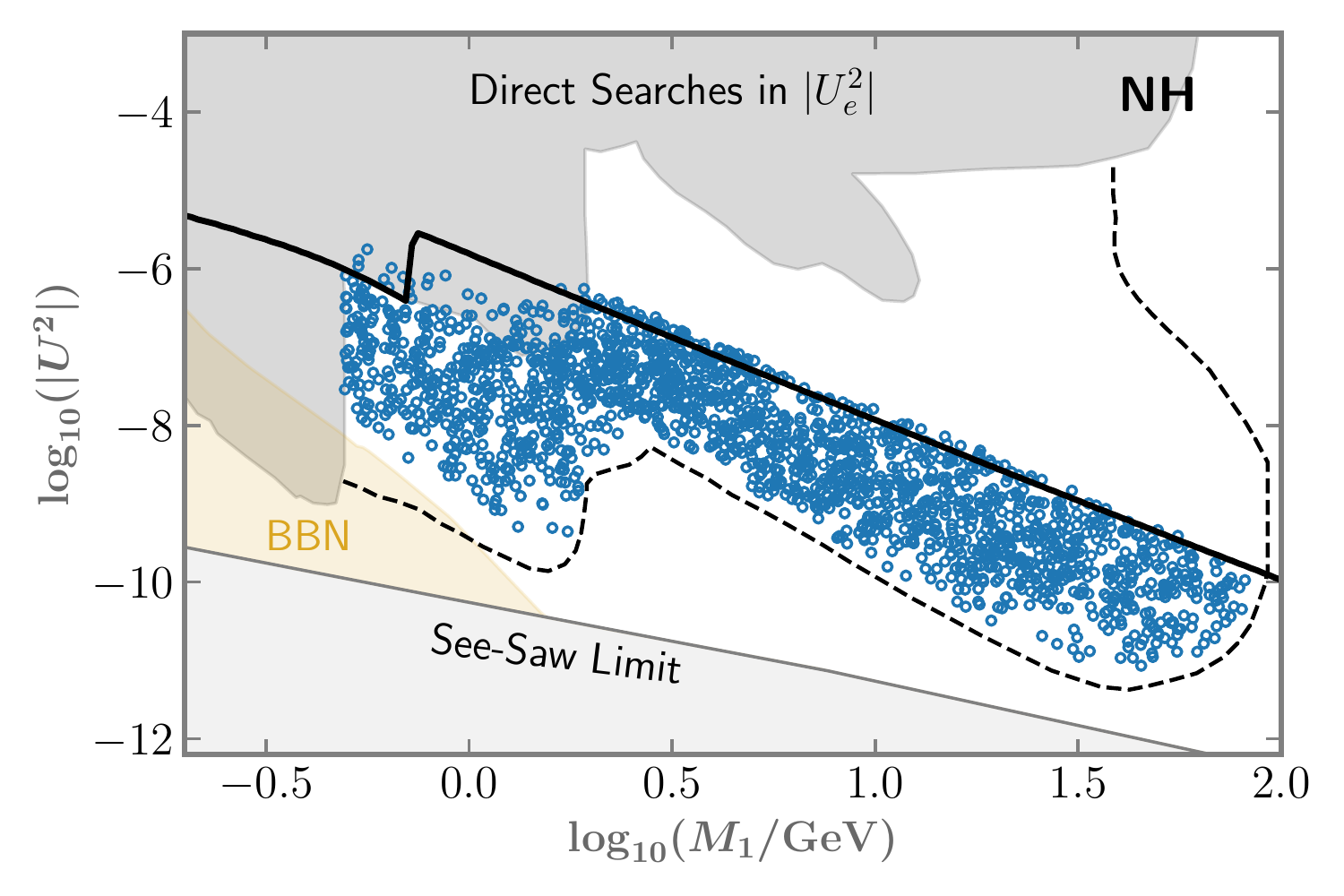} & \hspace{-0.55cm}  \includegraphics[width=0.5\textwidth]{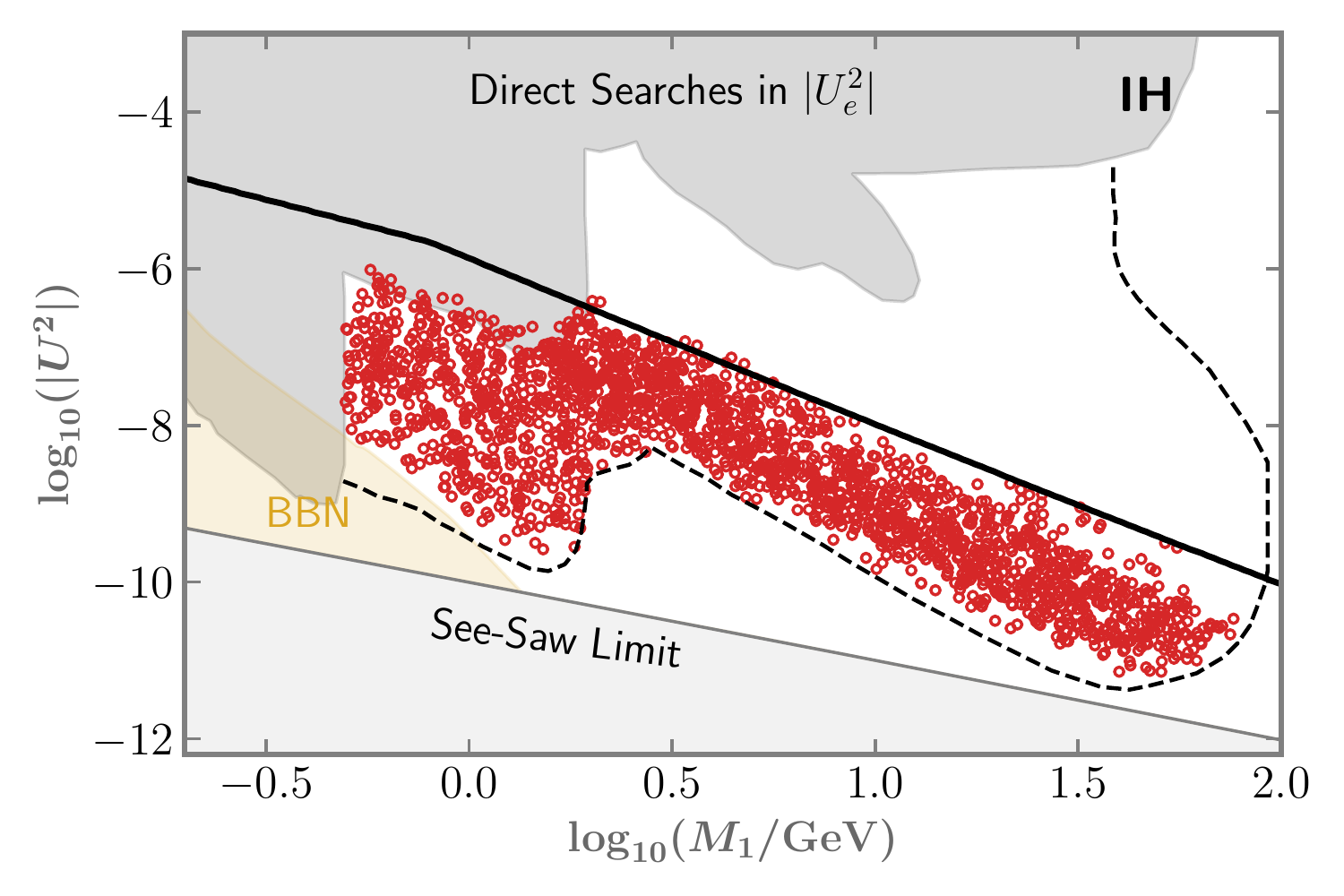}   
\end{tabular}
\vspace{-0.4cm}
\caption{ 
Numerical result of the Bayesian analysis (blue (red) points for NH (IH)) together with the analytical derived upper bound on the HNL mixing (black line).
The grey shaded regions is excluded by direct searches or neutrino masses (seesaw limit), while the yellow one is excluded by big bang nucleosynthesis constraints.
}
\label{fig:scan_general}
\end{figure}

When the mass splitting of the HNLs is not known, different regimes can apply for the same pair of $(U^2,M)$.
Nevertheless, there is an absolute upper bound on the mixing for which the BAU can be reproduced within the model.
We find that the maximal mixing is achieved for the maximum value of $\Delta M/M$ within the overdamped regime.
This is because only within the overdamped regime the mixing is not restricted by the requirement of flavour effects and the asymmetry is linearly proportional to $\Delta M/M$, see eq.~\eqref{eq:Yb_ov_wLNV} and eq.~\eqref{eq:Yb_ov_sLNV}.
The upper bound, however, depends on whether there is a second weak mode at $T_{\rm EW}$ or not. For low masses it is given by eq.~(\ref{eq:U2_ov_wLNV_max_LNC}) in the wLNV regime,
while for larger masses the conservative bound derived from eq.~(\ref{eq:U2_ov_sLNV_max}) in the sLNV regime applies. In Fig.~\ref{fig:scan_general} we show the points of the parameter space leading to the correct BAU found by the bayesian analysis together with the 
analytically derived absolute upper bound. 
We find good agreement between our numerical result and the analytical estimate, as well as with previous numerical results, see for example~\cite{Klaric:2021cpi}.

In appendix~\ref{app:get_dist} we show two dimensional projections of the full numerical scan that reveal some non trivial correlations. In particular, we include the projections on $|U_\alpha|^2$, $\Delta M/M$, $M$ and $\theta$.

\section{Constraints on other observables from the baryon asymmetry}
\label{sec:constraints}

We finally want to discuss the correlation of the baryon asymmetry with other observables. Of particular interest are the flavour 
of the HNL mixings and neutrinoless double-beta decay\footnote{The implications of BAU on charged lepton flavour violating processes, such as $\mu\rightarrow e\gamma$ or $\mu-e$ conversion, has been recently considered in ref.~\cite{Granelli:2022eru}.}. We will also comment on the possible measurement of $\Delta M$ for the extreme degeneracies needed in the overdamped regime. 

\subsection{HNL flavour mixings }

\begin{figure}[!t]
\centering
\includegraphics[width=0.5\textwidth]{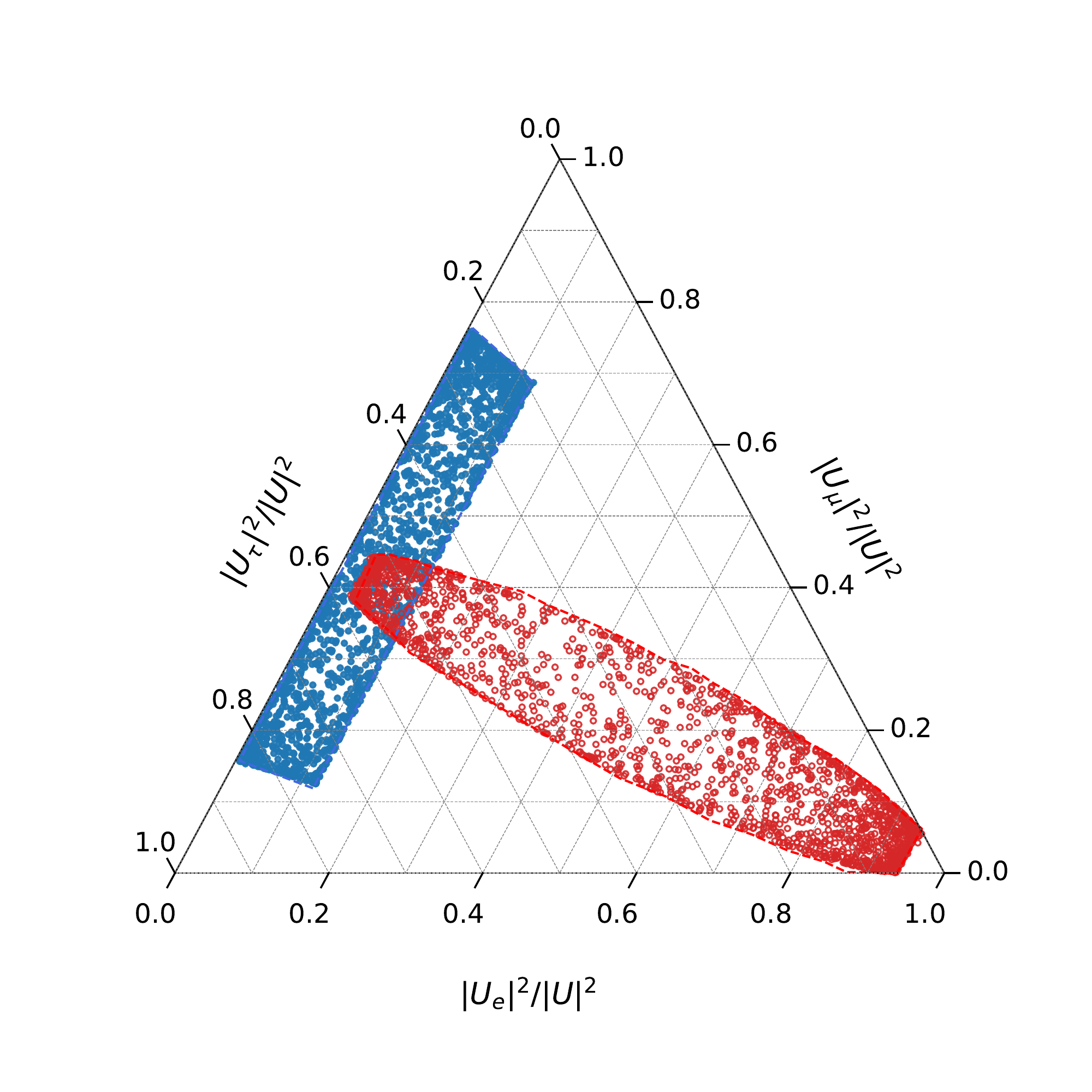} 
\caption{Points of the numerical scan with successful baryogenesis within the sensitivity region of \texttt{SHiP} and \texttt{FCC} for NH (blue) and IH (red).}
\label{fig:ternaryalldm}
\end{figure}

\begin{figure}[!t]
\centering
\begin{tabular}{cc}
\hspace{-0.5cm}
 \includegraphics[width=0.5\textwidth]{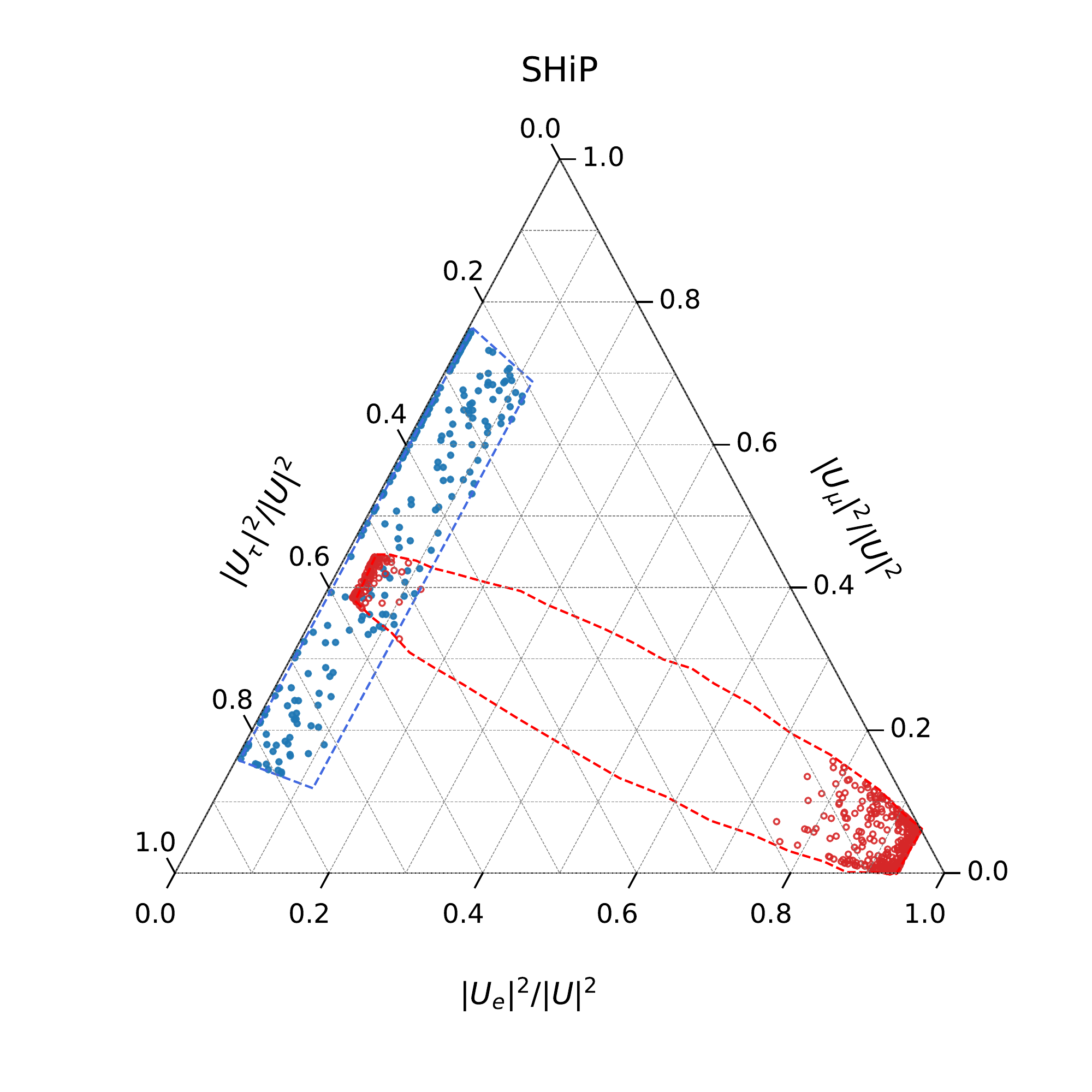}  & \hspace{-0.55cm}  \includegraphics[width=0.5\textwidth]{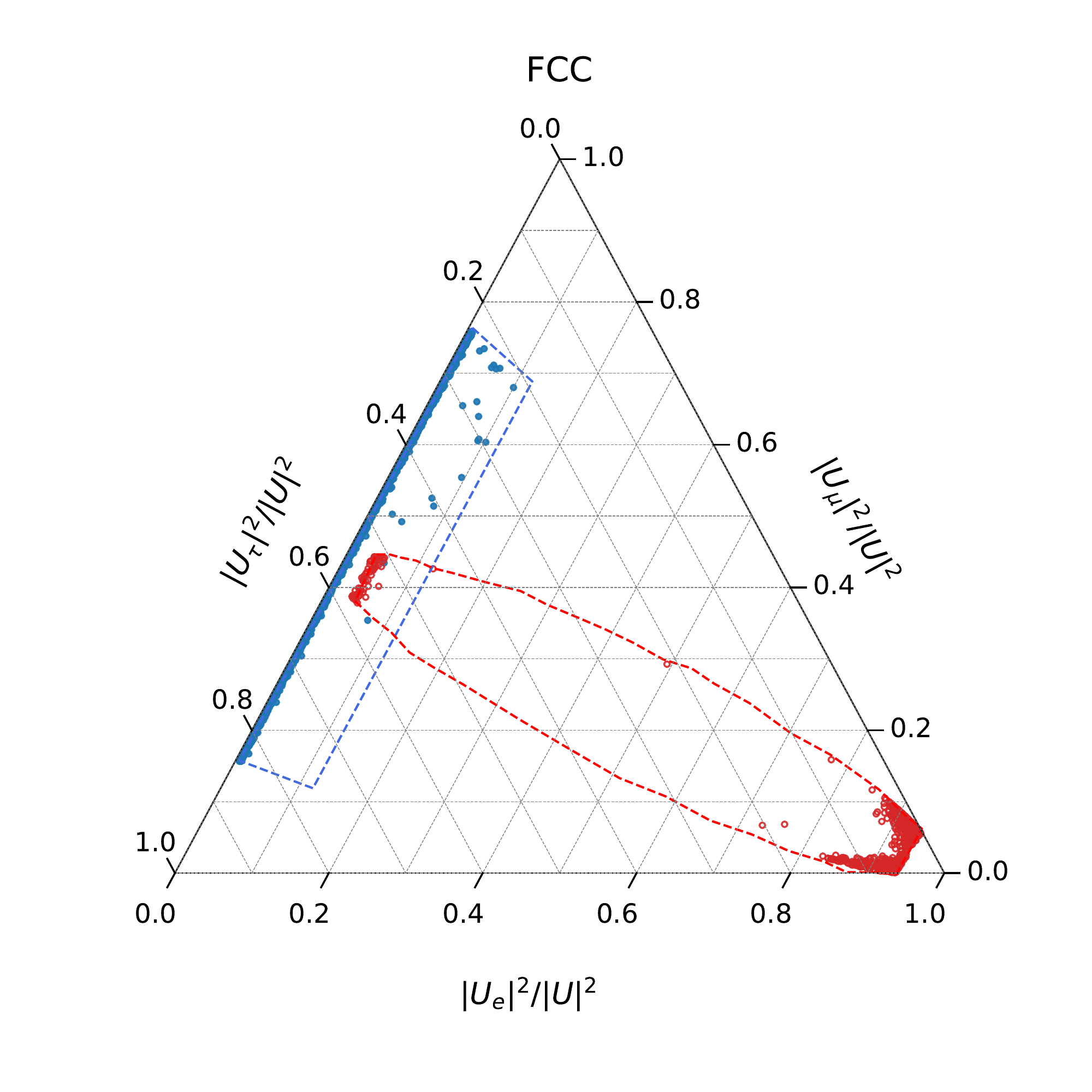} 
\end{tabular}
\vspace{-0.4cm}
\caption{Points fo the numerical scan with successful baryogenesis within the \texttt{SHiP} (left) and \texttt{FCC} (right) regions for fixed $\Delta M/M=10^{-2}$ and NH (blue) or IH (red).  The dashed lines correspond to the regions of Fig.~\ref{fig:ternaryalldm}.}
\label{fig:ternarylargedm}
\end{figure}

It is well known that in the minimal model with two extra singlets, the present constraints on neutrino masses fix to a large extent the ratios
$|U_\alpha|^2/U^2$, where $U_\alpha \equiv \Theta_{\alpha I}$. In fact, those ratios for sufficiently large $U^2$ (or in the approximate LN conserving limit) are completely determined from the light neutrino masses and mixings~\cite{Gavela:2009cd,Hernandez:2016kel}. The unknown CP violating phases in the PMNS matrix lead to some uncertainty in the flavour ratios. This is nicely summarized in a ternary diagram~\cite{Caputo:2017pit}.
The restriction imposed by successful baryogenesis for large mixings on the ternary diagram has been first studied in~\cite{Antusch:2017pkq}.
In Fig.~\ref{fig:ternaryalldm} we show the points on the ternary diagram for NH/IH within the sensitivity region of \texttt{SHiP} and \texttt{FCC}, which successfully explain the baryon asymmetry. 
Since we have not included errors in the oscillation parameters, the only uncertainty is related to the CP phases, $\delta$ and $\phi$, which we assume unconstrained. Explaining the baryon asymmetry does not seem to restrict the region with respect to the one found in ref.~\cite{Caputo:2017pit}.  
However, if we restrict to large values of $\Delta M/M=10^{-2}$ we observe  in Fig.~\ref{fig:ternarylargedm} that the regions significantly shrink. 
These regions can be understood as those that lead to a weak flavour, that is $\epsilon_\alpha \ll 1$ for one or more $\alpha=e, \mu, \tau$.

\begin{figure}[!t]
\centering
\begin{tabular}{cc}
\hspace{-0.5cm}
 \includegraphics[width=0.49\textwidth]{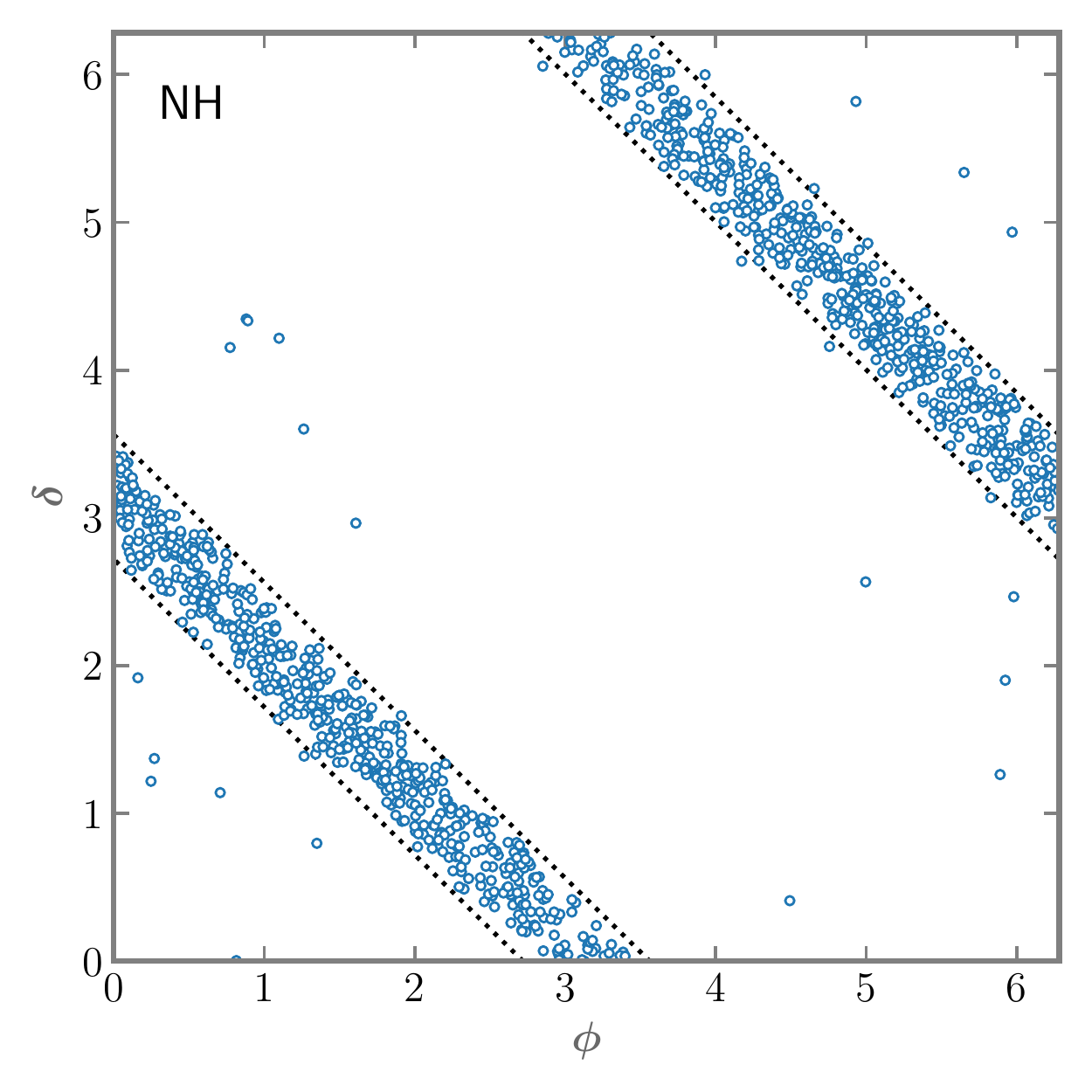}   & \hspace{-0.55cm}   
\includegraphics[width=0.49\textwidth]{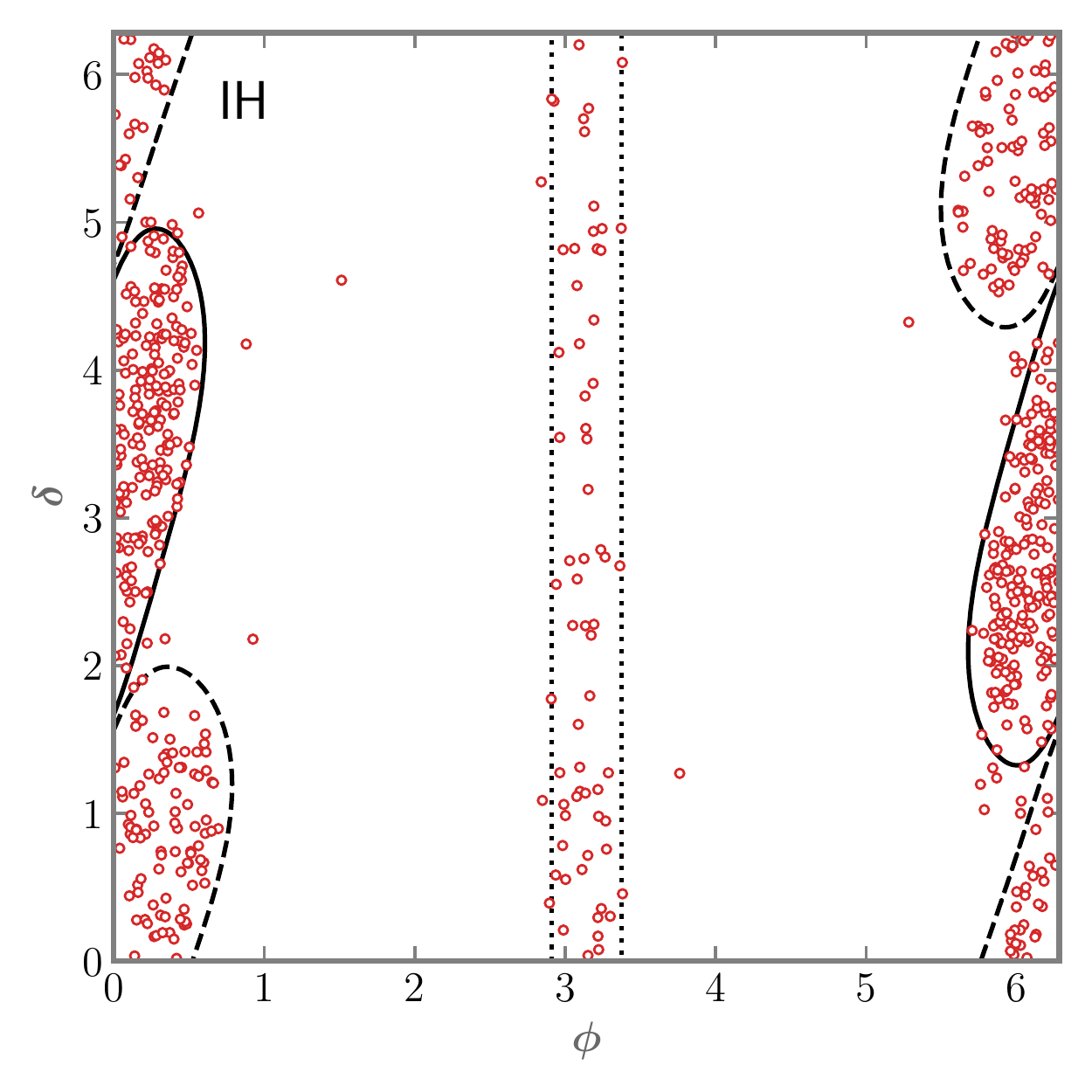} 
\end{tabular}
\vspace{-0.4cm}
\caption{Points from the scan at $\Delta M/M=10^{-2}$ for NH (left) and IH (right).
The black dotted lines enclose the regions where $\epsilon_e\leq 0.01$ (NH), and $\epsilon_e \leq 0.05$ (IH), while the dashed lines enclose the region $\epsilon_\mu \leq 0.03$ and the solid that corresponding to $\epsilon_\tau\leq 0.03$.}
\label{fig:flavourselect}
\end{figure}

As we have seen, for $\Delta M/M=10^{-2}$ the overdamped regime is not possible and flavour effects are necessarily present to explain the baryon asymmetry within the \texttt{SHiP}/\texttt{FCC} regions. These flavour effects are related  to  the minimization of $\epsilon_\alpha$. As we have seen in sec.~\ref{sec:parambounds},  the slow flavour for NH is always $\alpha=e$. The $(\phi,\delta)$ phases leading to a suppressed $\epsilon_e$  are shown in the left panel of  Fig.~\ref{fig:flavourselect}. For IH, the slow mode can be $\alpha=\mu,\tau$ or $\alpha=e$ in the regions shown in the right plot of Fig.~\ref{fig:flavourselect}. 
The points from the scan at fixed $\Delta M/M=10^{–2}$ are superimposed in Fig.~\ref{fig:flavourselect}, demonstrating that beyond the requirement of being in the  weak flavour washout, the baryon asymmetry does not seem to impose further constraints on the PMNS CP phases. This is because the parameter $\theta$ can still be fixed to obtain the correct sign and magnitude of the baryon asymmetry.  

The baryon asymmetry makes therefore  a clear prediction for the  HNL flavoured mixings or the PMNS phases when $\Delta M/M$ is sufficiently large.

\subsection{Neutrinoless double-beta decay}

The amplitude for this process depends on the combination of neutrino parameters $m_{\beta\beta}$, that gets contributions from the light and heavy neutrino sectors

\begin{equation}
m_{\beta\beta}=\left| \sum _{i={\rm light}} U_{ei}^2m_i\;+ \sum _{I={\rm heavy}}\Theta_{eI}^2M_I  \mathcal{M}\left(M_I\right)/\mathcal{M}\left(0\right)\right|\,,
\end{equation}
where $\mathcal{M}\left(M_i\right)$ are the Nuclear Matrix Elements (NME) as a function of the mass of the neutrino mediating the process, as defined in~\cite{Blennow:2010th}. In our analysis we will consider the NMEs computed in~\cite{Blennow:2010th}. Recently, it has been found that a new short-distance mechanism associated to the exchange of hard virtual neutrinos can lead to an apreciably different result for the NMEs and a even modify the dependence on the mass of the exchanged neutrino~\cite{Cirigliano:2018hja,Cirigliano:2020dmx,Jokiniemi:2021qqv,Dekens:2020ttz,Tuo:2022hft}.\footnote{We thank J. de Vries for pointing out this effect.} However, these new effects are currently under study and will thus not be considered here.

In order to illustrate the main dependence of $m_{\beta\beta}$ on the neutrino parameters, using eq.~(\ref{eq:Hlmixing}) together with eqs.~(\ref{eq:Yno_PMNS})-(\ref{eq:Yio_PMNS}) and (\ref{eq:W}), the following approximated expression\footnote{The approximation implies a scaling of the NMEs as $\mathcal{M}\left(M_I\right)\propto 1/M_I^2$. For $M_I\lesssim 3\,{\rm GeV}$ the deviation with respect to the nuclear computation~\cite{Blennow:2010th} is larger than $1\%$.} can be derived~\cite{Blennow:2010th,Ibarra:2011xn,Lopez-Pavon:2015cga,Hernandez:2016kel} for the symmetry protected scenario considered here:

{\it Normal Hierarchy}
\begin{eqnarray}
\label{eq:mbbapprox_IH}
 m_{\beta\beta}^{NH}&=&\left| \sqrt{\Delta m^2_{\rm atm}}\left( c_{12}^2c_{13}^2r-e^{-2i(\delta+\phi)}s_{13}^2\right)\right.\nonumber\\
 &-& \left. 2e^{i\theta}U^2\Delta M f(A) \left(\frac{0.9 {\rm GeV}}{M}\right)^2  
 \left( rs_{12}^2 +2\sqrt{r}s_{12}s_{13}e^{-i(\delta+\phi)}+s_{13}^2e^{-2i(\delta+\phi)} \right) \right|\,. \,\,\,\,\,\,\,
\end{eqnarray}

{\it Inverted Hierarchy}
\begin{eqnarray}
\label{eq:mbbapprox_NH}
m_{\beta\beta}^{IH}&=&\left| \sqrt{\Delta m^2_{\rm atm}}c_{13}^2\left( c_{12}^2-s_{12}^2e^{2i\phi}+\mathcal{O}\left(r^2\right)\right)\right. \nonumber \\
 &-& \left. e^{i\theta}U^2\Delta M f(A) \left(\frac{0.9 {\rm GeV}}{M}\right)^2\left( c_{12}+s_{12}^{i\phi} \right)^2\left(1+\mathcal{O}\left(r^2\right) \right) \right|\,.
\end{eqnarray}
The function $f(A)$ depends on the nucleus under consideration: for $^{48}{\rm Ca}$, $^{76}{\rm Ge}$, $^{82}{\rm Se}$, $^{130}{\rm Te}$ and $^{136}{\rm Xe}$, $f(A) \approx$ $0.035$, $0.028$, $0.028$, $0.033$ and $0.032$, respectively~\cite{Blennow:2010th,Ibarra:2011xn}. The above approximated formulae match with the ones derived in~\cite{Hernandez:2016kel}\footnote{A typo has been noted in the IH expression in ref.~\cite{Hernandez:2016kel}: the factor $1-2e^{i \delta}s_{23}\theta_{13}$ should be removed.} using the mapping to the Casas-Ibarra parameterization presented in appendix~\ref{sec:CasasIbarra}.

\begin{figure}[!t]
\centering
\begin{tabular}{cc}
\hspace{-0.5cm}
 \includegraphics[width=0.5\textwidth]{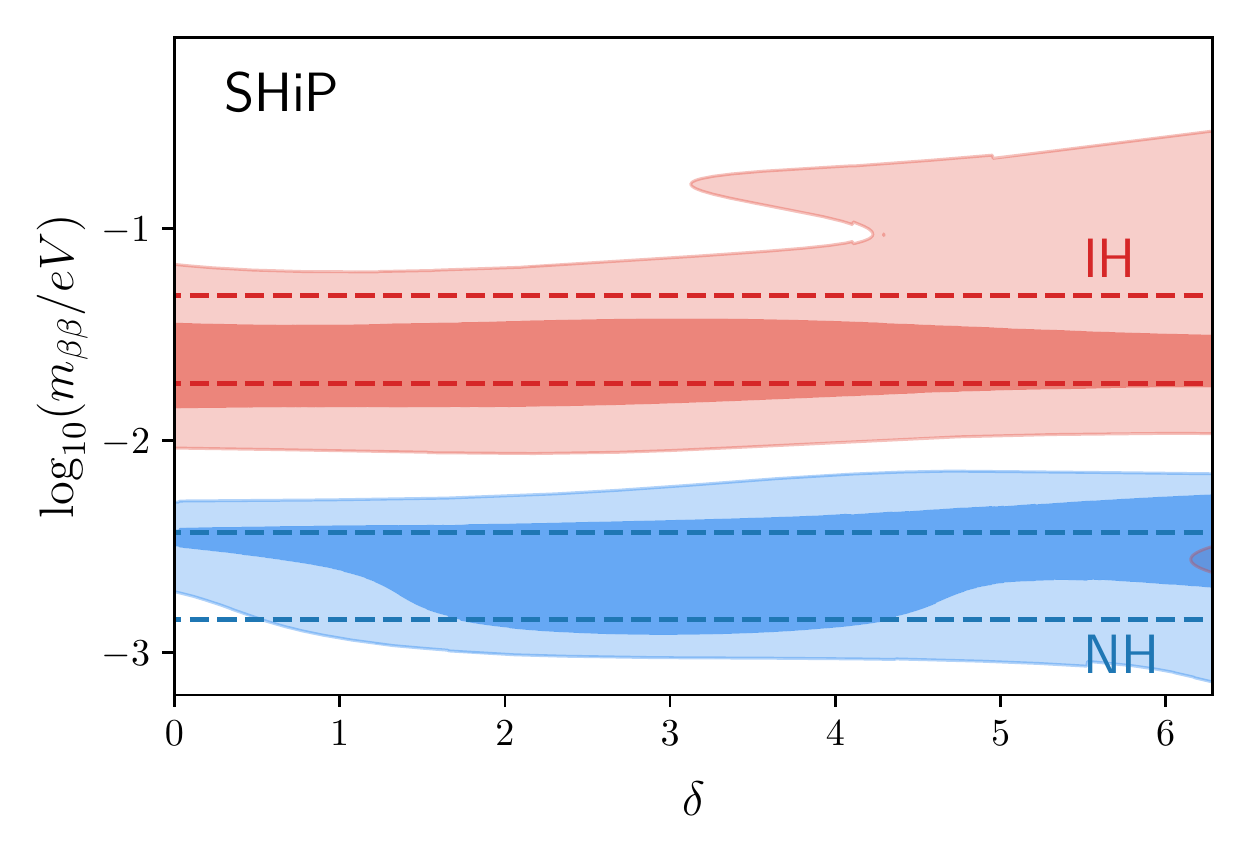}  & \hspace{-0.55cm}  \includegraphics[width=0.5\textwidth]{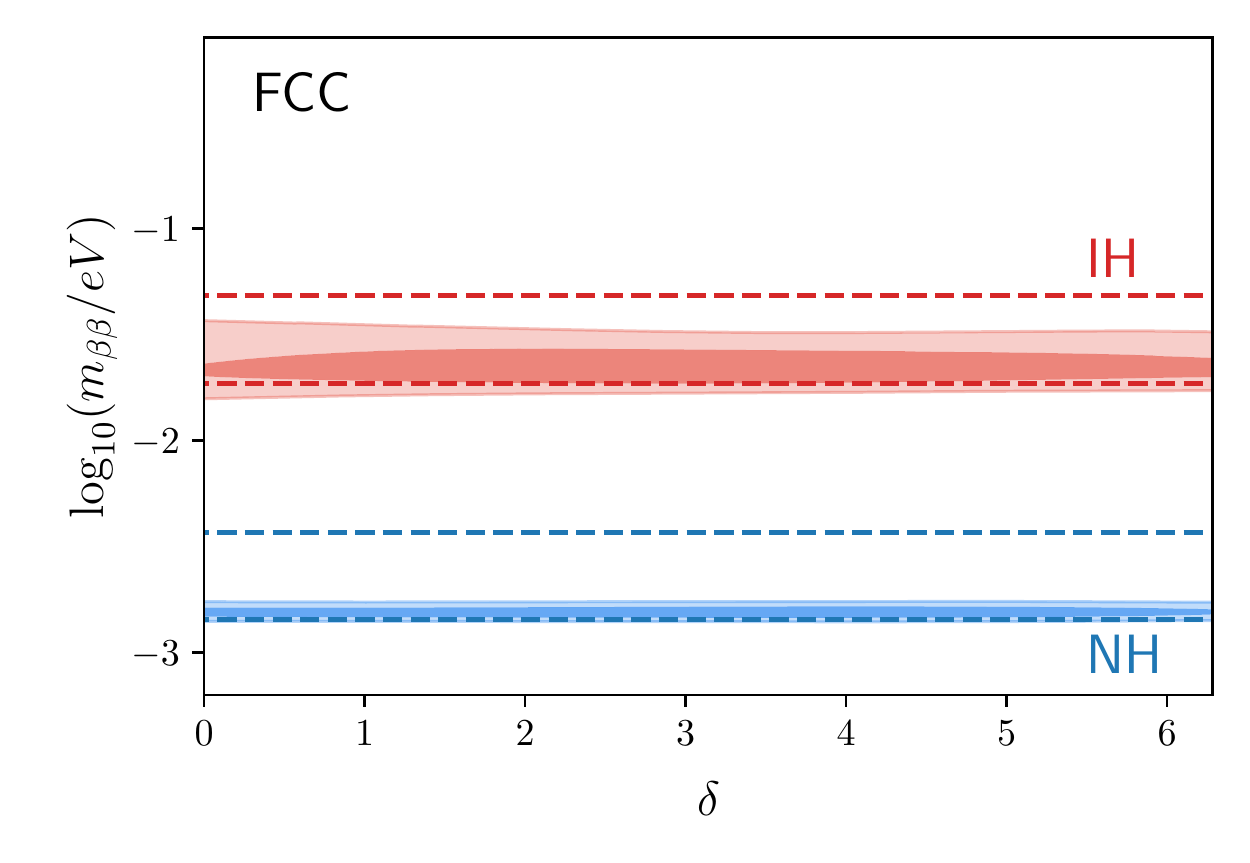} 
\end{tabular}
\vspace{-0.4cm}
\caption{1 and 2 $\sigma$ regions from the numerical scan  on the plane $(\delta,m_{\beta\beta})$ at $\Delta M/M=10^{-2}$ for NH (blue) and IH (red). 
The standard light neutrino contribution is contained in the dashed bands. The left plot corresponds to the \texttt{SHiP} range and the right plot to the \texttt{FCC} one.  }
\label{fig:mbb}
\end{figure}

There are two important implications of successful baryogenesis on the prediction of neutrinoless double-beta decay. First, there can be a sizable non-standard contribution from the heavy states if $M$ is not too large and $\Delta M/M$ not too small. From the above equations, it is clear that the interference of the light and heavy contributions depends on the parameter $\theta$, which is completely unconstrained otherwise, as already shown in~\cite{Hernandez:2016kel}. For $M\gg 100$ MeV, the matrix element associated to the heavy contribution is suppressed, and can be neglected above a few GeV~\cite{Lopez-Pavon:2012yda}. Therefore we expect to find a non-standard contribution only in the range of \texttt{SHiP} and for large enough $\Delta M/M$. On the left plot of Fig.~\ref{fig:mbb}, we show the 1 and 2 $\sigma$ regions from the numerical scan\footnote{Note that due to the imposed constraint on the parameter space while performing the Bayesian analysis the interpretation of the posterior probabilities has to be taken with care.} on the plane $(\delta,m_{\beta\beta})$ for both hierarchies and $\Delta M/M=10^{-2}$ in the range of \texttt{SHiP}. 
The dashed lines correspond to the standard range of the light neutrino contribution to $m_{\beta\beta}$.
Indeed we  observe a significant deviation of the standard expectation for both hierarchies, which furthermore depends on the Dirac CP phase, $\delta$. The presently preferred range of $\delta\geq \pi$~\cite{Esteban:2020cvm,deSalas:2020pgw} seems to be also the region where the HNLs effects on $m_{\beta\beta}$ are more relevant. This dependence on $\delta$ is the result of a non trivial interplay among the CP phases which play a role in the baryon asymmetry, the $U_\alpha$ flavor structure shown in Fig.~\ref{fig:ternarylargedm} (see also Fig.~\ref{fig:flavourselect}), and neutrinoless double-beta decay. 

A second effect is the restriction of the standard light neutrino contribution. Ignoring the uncertainties of nuclear matrix elements, we do not have an accurate prediction of $m_{\beta\beta}$, because it depends on the PMNS CP phases~\cite{Feruglio:2002af}. We have seen that the BAU restricts these phases for large $\Delta M/M$ in order to ensure that at least one flavour remains weak, and this involves the PMNS CP phases. We therefore expect that the prediction for  $m_{\beta\beta}$ will also be restricted by this requirement. 
On the  right plot of Fig.~\ref{fig:mbb}, we show  the 1 and 2 $\sigma$ regions for \texttt{FCC}.  We observe indeed a reduction of the standard regions, which is very significant for NH. Unfortunately it seems to select the smallest range of $m_{\beta\beta}$. This behaviour is easy to understand analytically just considering the dependence on $\delta$ and $\phi$ of the light neutrino contribution given in eq.~(\ref{eq:mbbapprox_NH}) and the flavor selection shown by Fig.~\ref{fig:flavourselect}.

\subsection{HNL mass splitting}

A key parameter regarding the testability of low scale leptogenesis is the HNL mass splitting. This can be kinematically measured for large $\Delta M/M$ depending on the experimental resolution. In the previous sections we have studied the predictions from the baryon asymmetry generation on the flavor structure of HNL mixing, the PMNS CP-phases and the neutrinoless double-beta decay rate, considering a potentially measurable value of $\Delta M$ ($\Delta M/M=10^{-2}$). However, for small values of $\Delta M/M$, a kinematical measurement is essentially hopeless. 

Interestingly, sensitivity to small $\Delta M/M$ can be achieved in future colliders or beam dump experiments via the measurement of coherent HNL oscillations~\cite{Boyanovsky:2014una,Cvetic:2015ura,Anamiati:2016uxp,Antusch:2017ebe,Cvetic:2018elt,Tastet:2019nqj} or the correlation among the HNL decay products~\cite{Dib:2016wge,Arbelaez:2017zqq,Balantekin:2018ukw,Hernandez:2018cgc}. Both effects are essentially driven by the ratio $\Delta M/\Gamma$, where $\Gamma$ is the total HNL decay width, and their experimental observation requires roughly $\Delta M \simeq \Gamma$. We have checked that this condition is not fulfilled in the testable region of the parameter space compatible with successful leptogenesis for values of $\Delta M/M$ larger than $10^{-14}$. Note, that for smaller values of the mass splitting corrections from the Higgs mechanism induced after electroweak symmetry breaking should be included, which are of the order of the light neutrino masses. 
This region can be phenomenologically motivated, for instance, in the $\nu$MSM model~\cite{Asaka:2005pn,Asaka:2005an} in which a third HNL at the keV scale, almost decoupled, may be a Dark Matter candidate. This extremely degenerate case will be considered elsewhere.


\section{Conclusion}
\label{sec:conclu}

We have presented a detailed study of the baryon asymmetry in the context of the minimal type-I seesaw model, with two extra singlet fermions (HNL) with masses in the $0.1-100~\rm{GeV}$ range, that can also explain the light neutrino masses. This scenario has received considerable attention in previous literature, since it can be tested in future experiments such as \texttt{SHiP} or \texttt{FCC}. We have focussed precisely in the region of parameter space accesible to these experiments, which requires relatively large HNL mixings, and studied the constraints imposed by the requirement of successfully reproducing the observed baryon asymmetry. As a first step, we have developed an accurate analytical approximation to the baryon asymmetry, exploiting the approximate lepton number symmetry that must be satisfied to achieve large enough HNL mixings, significantly above the naive seesaw expectation, $U^2\gg m_\nu/M$. This is often called an inverse or linear seesaw scenario and involves almost degenerate HNLs and expansion parameters that permit a perturbative solution of the kinetic equations based on the adiabatic approximation. 
The validity of the approximation has been confirmed by confronting it with the full numerical solutions of the kinetic equations. 

These analytical results have allowed us to map all the washout regimes, where the necessary out-of-equilibrium condition is satisfied by at least one mode. The slow modes have been identified as the oscillation mode in the overdamped regime, a weakly coupled flavour in the presence of flavour hierarchies or the mode associated to the approximate lepton number symmetry.  The regions corresponding to the different regimes are displayed in 
Fig.~\ref{fig:regimes} for two fixed values of the $\Delta M/M$ on the plane of HNL mass and mixing.
Interestingly the complex parameter dependencies of the baryon asymmetry are encoded in CP invariants, that can be easily derived from first principles and can be expressed in terms of measurable parameters: light neutrino masses and mixings, HNL masses and mixings and very importantly CP phases. 
 We have used these non-trivial relations to derive robust bounds on the HNL mixings (upper or lower bounds) depending on the regime, see eqs.~(\ref{eq:U2_ov_wLNV_max_fixed_DMM}), (\ref{eq:U2_ov_sLNV_max_fixed_DMM}), (\ref{eq:boundfwo}), (\ref{eq:U2_bound_fast_osc}), and on the HNL mass degeneracy in eq.~(\ref{eq:dmbound}). Furthermore, strong correlations among CP violating phases for successful baryon asymmetry have been shown to exist in certain regions of parameter space, in particular in regions that are far from the upper/lower bounds. Interestingly, in some regions of parameter space CP phases should be correlated to suppress the angular dependence of the CP invariants, as in Fig.~\ref{fig:f_contours}. Also, for moderate HNL degeneracies, 
 flavour effects are mandatory,  restricting the PMNS CP phases according to Fig.~\ref{fig:flavourselect}. This restriction has interesting observable consequences in the flavour of the HNL mixings, as shown in Fig.~\ref{fig:ternarylargedm}, and in neutrinoless double-beta decay, see Fig.~\ref{fig:mbb}. 
  
 The methods developed in this work will be useful to derive robust bounds in the significantly more complex parameter space of non-minimal models with more than two fermion singlets.


\begin{acknowledgments}

We thank J. de Vries, C. Hagedorn, M. Laine, J. Men\'endez, J. Salvado, J.L. Tastet and I. Timiryasov for useful discussions and/or clarifications. 
This work was partially supported by the European Union Horizon 2020 research and innovation programme under the Marie Sklodowska-Curie grant agreement No 860881-HIDDeN,
by the Spanish Ministerio de Ciencia e Innovacion project PID2020-113644GB-I00 and  by Generalitat Valenciana through the ``plan GenT" program (CIDEGENT/2018/019)
and grant PROMETEO/2019/083.
We gratefully acknowledges the computer resources at Artemisa, funded by the European Union ERDF and Comunitat Valenciana as well as the technical support provided by the Instituto de Fisica Corpuscular, IFIC (CSIC-UV).
The work of SS received the support of a fellowship from ``la Caixa" Foundation (ID 100010434) with fellowship code LCF/BQ/DI19/11730034.

\end{acknowledgments}


\newpage
\appendix


\section{Appendix: CP phases}
\label{sec:app}

In this appendix we will show how all the CP phases can be absorbed in the Yukawa couplings leaving $M_R$ as a real symmetric matrix. Further, we will also demonstatrate that in the symmetry protected scenario under consideration, it can be assummed $\mu_1=\mu_2$ in all generality. Finally, since the CP invariants presented in sec.~\ref{sec:cpinv} are given in the basis in which the Majorana mass term is diagonal (with real and positive entries), we will provide the connection between this basis and the one given by eq.~(\ref{eq:LNVparam2N}) diagonalizing $M_R$.

First of all, notice that there is no CP violation in the symmetric limit ($y'_\alpha=\mu_1=\mu_2=0$) since in such a case all the phases in eq.~(\ref{eq:LNVparam2N}) can be trivially reabsorbed with a rephasing of the $N_i$ and $L^\alpha$ fields. If the symmetry is broken, in principle $M_R$ is a complex symmetric matrix which contains three phases. 
Two of them can be easily removed performing $N_i$ field redefinitions. However, a priori there is a non trivial phase contained in $M_R$ in the general case. It is easy to show that we can start from the following basis:
\be
\tilde{Y}=\begin{pmatrix}
\tilde{Y}_{e 1} & \tilde{Y}_{e 2} \\
\tilde{Y}_{\mu 1} & \tilde{Y}_{\mu 2} \\
\tilde{Y}_{\tau 1} & \tilde{Y}_{\tau 2}
\end{pmatrix},
\;\;\;\;
\tilde{M}_R=\begin{pmatrix}
\tilde{\mu}_1 e^{i\alpha_\mu} & \Lambda \\
\Lambda  & \tilde{\mu}_2 e^{i\alpha_\mu} 
\end{pmatrix}\,,
\ee
where $\tilde{\mu}_1,\tilde{\mu}_2, \Lambda\Subset\,\mathbb{R}^+$ and $0\leq\alpha_\mu\leq2\pi$. Rotating to the basis in which $\tilde{M}_R$ is real and diagonal and expanding over the small LNV parameters we find
\bea
Y&=& \tilde{Y}O,\;\;
\rm{diag}(M_1,M_2)=O^T\tilde{M}_RO\,,
\\
O&=&\frac{1}{\sqrt{2}}\left\{\begin{pmatrix}
1& 1 \\
-1 & 1 
\end{pmatrix} +\frac{(\tilde{\mu}_2-\tilde{\mu}_1)e^{i\alpha_\mu}}{4\Lambda}\begin{pmatrix}
1& 1 \\
-1 & 1 
\end{pmatrix} \right\}\rm{diag}(i\,e^{-i\alpha_1/2},e^{-i\alpha_2/2})\,, 
\eea
where $\alpha_{2(1)}=\rm{Arg}\left\{1\pm(\tilde{\mu}_1+\tilde{\mu}_2)e^{i\alpha_\mu}/2\Lambda\right\}$ and $M_{2(1)}=\Lambda\pm(\tilde{\mu}_1+\tilde{\mu}_2)\cos\alpha_\mu/2$. Here we are neglecting higher order terms in $\tilde{\mu_i}/\Lambda$ and $\tilde{Y}_{\alpha 2}$. Finally, expanding also $e^{-i\alpha_i/2}$, we obtain
\bea
Y_{\alpha 1}&=&\frac{i}{\sqrt{2}}\left(\tilde{Y}_{\alpha 1}-\tilde{Y}_{\alpha 2}^{new}\right)\,,\\
Y_{\alpha 2}&=&\frac{1}{\sqrt{2}}\left(\tilde{Y}_{\alpha 1}+\tilde{Y}_{\alpha 2}^{new}\right)\,,\\
\rm{diag}(M_1,M_2)&=&\rm{diag}(\Lambda-\mu_2,\Lambda+\mu_2)\,,
\eea
where we are neglecting the $\mathcal{O}(\tilde{\mu}^2/\Lambda^2)$ and $\mathcal{O}(\tilde{Y}_{\alpha 2} \tilde{\mu}/\Lambda)$ higher order terms, and
\bea
\tilde{Y}_{\alpha 2}^{new}&\equiv&\tilde{Y}_{\alpha 2}-\frac{(\tilde{\mu}_2-\tilde{\mu}_1)e^{i\alpha_\mu}}{4\Lambda}\tilde{Y}_{\alpha 1}-\frac{\tilde{\mu}_1+\tilde{\mu}_2}{4\Lambda}i\sin\alpha_\mu \tilde{Y}_{\alpha 1},\\
\mu_2 &\equiv& (\tilde{\mu}_1+\tilde{\mu}_2)\cos\alpha_\mu/2=\Delta M/2\,. 
\eea
Now, we can perform the following rotation of the $N_i$ fields 
\be
O=\frac{1}{\sqrt{2}}\begin{pmatrix}
-1& 1 \\
1 & 1 
\end{pmatrix} \rm{diag}(i,1)\,,
\ee
to go back to an initial basis in which there are no phases contained in the Majorana mass term and their diagonal matrix elements are equal:
\be
Y= \begin{pmatrix}
y_{e} e^{i \beta_e} & y'_e e^{i \beta_e'}\\
y_{\mu} e^{i \beta_\mu} & y'_{\mu}  e^{i \beta_\mu'}\\\
y_{\tau} e^{i \beta_\tau} & y'_{\tau} e^{i \beta_\tau'}\ 
\end{pmatrix},\;\;
\;\;
M_R=\begin{pmatrix}
\mu_2 & \Lambda \\
\Lambda  & \mu_2 
\end{pmatrix}\,,
\ee
where $\mu_2$ and $\Lambda$ are real and positive parameters. Diagonalizing $M_R$ and rotating to the basis in which it is diagonal, we obtain in the $N_i$ mass basis
\bea
Y=\begin{pmatrix}
y_{e} e^{i \beta_e} & y'_e e^{i \beta_e'}\\
y_{\mu} e^{i \beta_\mu} & y'_{\mu}  e^{i \beta_\mu'}\\\
y_{\tau} e^{i \beta_\tau} & y'_{\tau} e^{i \beta_\tau'}\ 
\end{pmatrix}W,\;\;\;\;{\rm diag}(M_1,M_2) = W^T M_R W\,,
\eea
with 
\be
\label{eq:W}
W=\frac{1}{\sqrt{2}}\begin{pmatrix}
1 & 1\\
-1  & 1
\end{pmatrix}
{\rm diag}(i, 1)\,,
\ee
and
\bea
M_{2(1)}=\Lambda\pm\Delta M/2, \;\;\;\;\Delta M = M_2-M_1=2\mu_2\,.
\eea


\section{Appendix: Mapping to the Casas-Ibarra parameterization}
\label{sec:CasasIbarra}

The Casas-Ibarra parameterization~\cite{Casas:2001sr} is a perturbative parameterization of the Yukawa couplings, based on the seesaw expansion, which implements the light neutrino mass and mixing constraints. Therefore, it should also be able to describe the symmetry protected scenario considered in this paper. Indeed, our results can be mapped to the Casas-Ibarra parameterization in the large HNL mixing regime explored here. In the Casas-Ibarra language, this limit corresponds to a large imaginary part of the complex angle $z$ appearing in the Casas-Ibarra matrix $R$. There is some arbitrariness in the concrete definition of this matrix and we will, thus, follow the prescription given by eq.~$(2.5)$ in~\cite{Hernandez:2016kel}.

We have checked that we recover the expressions for the weak washout CP invariant, the neutrinoless double-beta decay heavy contribution and the HNL mixings obtained in~\cite{Hernandez:2016kel},
performing the following mapping between parameterizations\footnote{The rephasing in the Majorana phase is required in order to recover positive light neutrino masses}: 

{\it Normal Hierarchy}
\bea
\phi^{\rm} \rightarrow \phi +\pi/2, \;\;\; \theta \rightarrow 2\,{\rm Re}\left[z\right],\;\;\; U^2\approx \frac{y^2v^2}{2 M_1^2}\rightarrow \frac{e^{2{\rm Im}\left[z\right]}\sqrt{\Delta m^2_{\rm atm}}}{4M_1}\,,
\eea
where $z$ is the complex Casas-Ibarra angle. 

Recall that the parameters $y$ and $y'$ can be related to the active heavy mixing $U^2$ and the neutrino masses as
\bea
y y' \approx \frac{M_1+M_2}{4v^2}\left(\sqrt{\Delta m^2_{\rm atm}}+\sqrt{\Delta m^2_{\rm sol}}\right),\;\;\;y^2 \approx \frac{2M_1^2U^2}{v^2}\,,
\eea
and, equivalently, $y'/y$ (which in the large mixing limit corresponds to $e^{-2{\rm Im}\left[z\right]}$) is given by
\be
y'/y=\frac{M_1+M_2}{8M_1^2U^2}\left(\sqrt{\Delta m^2_{\rm atm}}+\sqrt{\Delta m^2_{\rm sol}}\right)\,.
\ee

{\it Inverted Hierarchy}
\bea
\phi^{\rm} = \phi -\pi/2,\;\;\;\theta =2\,{\rm Re}\left[z\right],\;\;\; U^2\approx \frac{y^2v^2}{2 M_1^2} \rightarrow \frac{e^{2{\rm Im}\left[z\right]}\sqrt{\Delta m^2_{\rm atm}}}{2M_1}\,.
\eea
Notice the minus sign in front of $\pi/2$, to be compared with the normal hierarchy case. 

The parameters $y$ and $y'$ are related to the active heavy mixing $U^2$ and the neutrino masses as
\bea
yy' \approx \frac{M_1+M_2}{4v^2}\left(\sqrt{\Delta m^2_{\rm atm}}+\sqrt{\Delta m^2_{\rm atm}-\Delta m^2_{\rm sol}}\right),\;\;\;\;y^2 \approx \frac{2M_1^2U^2}{v^2}\,,
\eea
and
\be
y'/y=\frac{M_1+M_2}{8M_1^2U^2}\left(\sqrt{\Delta m^2_{\rm atm}}+\sqrt{\Delta m^2_{\rm atm}-\Delta m^2_{\rm sol}}\right)\,.
\ee


\section{Appendix: Lepton number conserving limit}
\label{app:lnc_limit}

Analytical approximations of the baryon asymmetry found so far in the literature were obtained in the lepton number conserving limit ($\Gamma_M \to 0$), see for example~\cite{Drewes:2017zyw, Drewes:2016gmt} and references therein.
Our analytical estimates from eqs.~\eqref{eq:overdamped_wLNV_sol} to \eqref{eq:lnv_osc_sol} contain both, LNC and LNV contribution simultaneously. 
This appendix is devoted to the LNC limit of our found analytical expressions.
Due to the clear separation of both contributions, as already expected on general grounds from the CP invariants of section~\ref{subsec:CP_invariants_general}, the LNC limit can be obtained trivially. 

Closed form analytical expressions were only obtained in the fast oscillating regime ($\Gamma_{\rm{osc}} (T_{\rm{osc}})  \gg \Gamma(T_{\rm{osc}})$)~\cite{Drewes:2017zyw}.
In this regime the asymmetry is generically \textit{suppressed} by a factor of 
\be
\eta \simeq \frac{\gamma_1 \kappa}{2\gamma_0 + \gamma_1 \kappa} \simeq 4\,,
\ee
compared to the scenario with strong LNV rates, see eq.~\eqref{eq:lnc_osc_sol}.
Hence, the parameter space for successful explanation of the BAU expands to slightly larger mass splittings when including LNV rates.
This clarifies the numerical enhancement of the asymmetry via LNV rates found previously in the literature, see e.g. ref.~\cite{Antusch:2017pkq}.
The reason is simply given by the competing weak modes and their different time evolution.
The same conclusion also applies for the intermediate regime, see eq.~\eqref{eq:intermediate_fw_sol}, for which, however, no analytical approximation existed in the literature so far.

\begin{figure}[!t]
\centering
\begin{tabular}{cc}
\hspace{-0.5cm}  \includegraphics[width=0.5\textwidth]{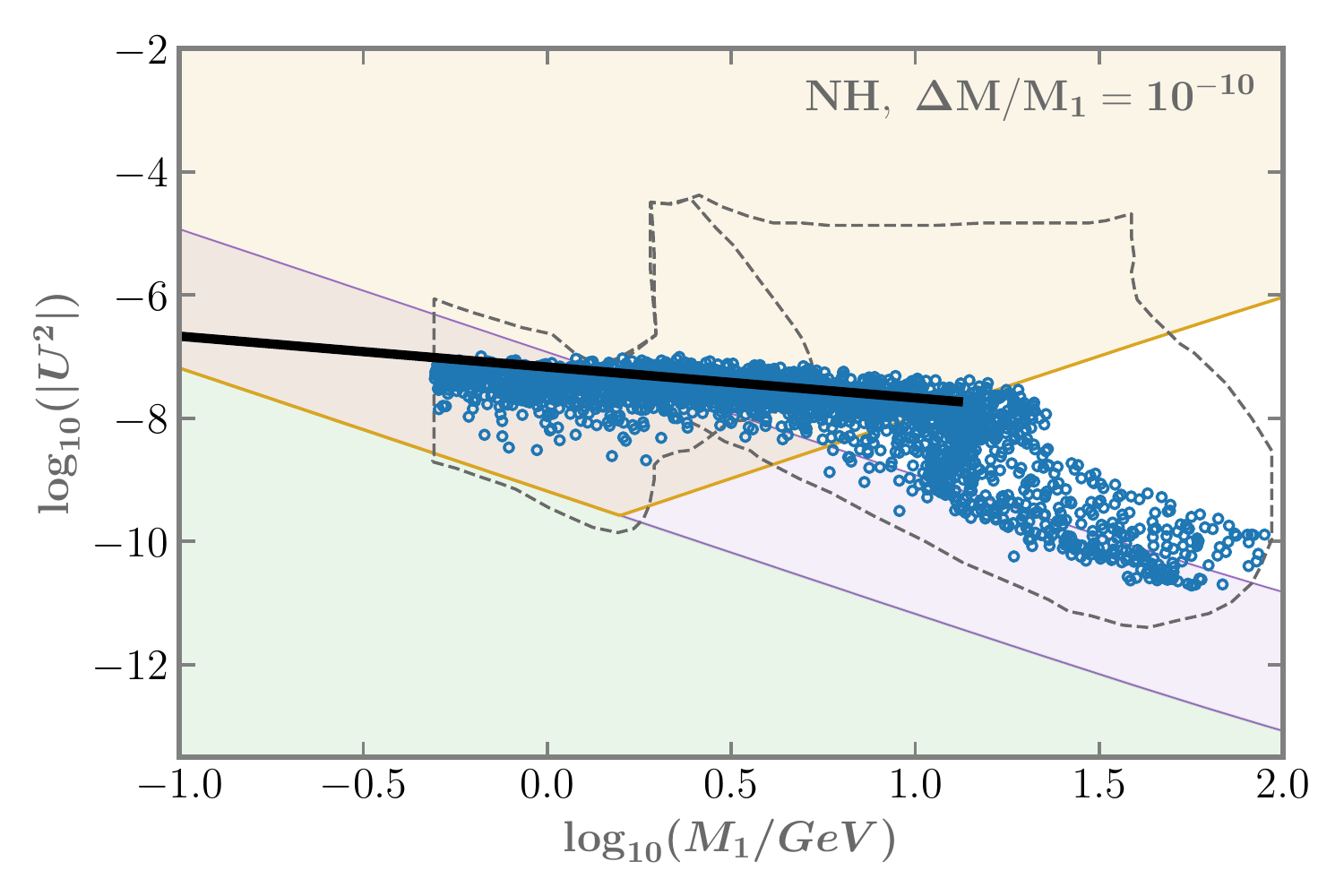} &
\hspace{-0.55cm}  \includegraphics[width=0.5\textwidth]{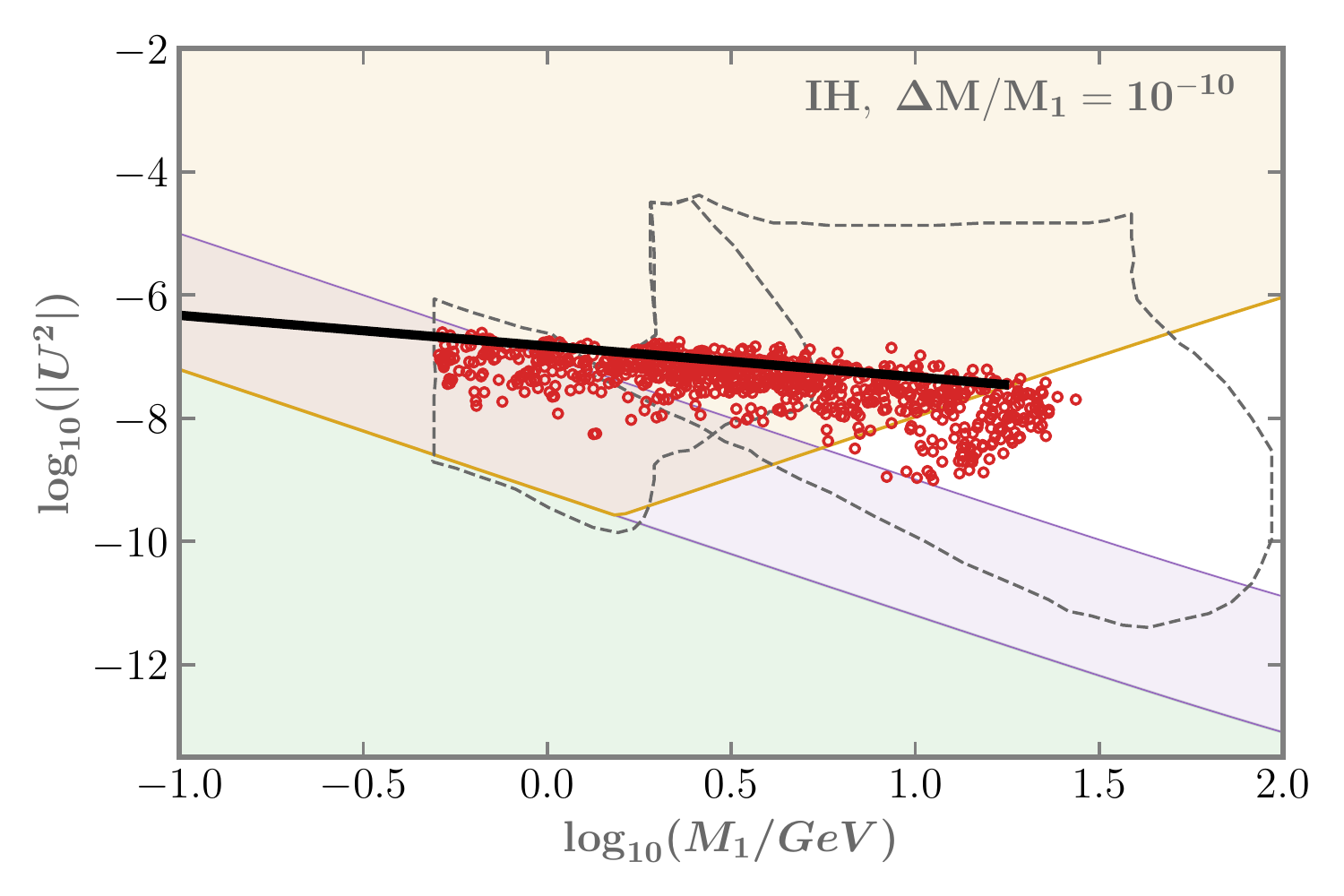} 
\end{tabular}
\vspace{-0.4cm}
\caption{ 
Result of the numerical scan for $\Delta M/M = 10^{-10}$ within the LNC limit shown in blue (red) for NH (IH).
The black lines represent the analytical upper bound on the mixing of eq.~\eqref{eq:U2_ov_LNC_max_fixed_DMM}.
Color coding as in figure~\ref{fig:regimes}.
}
\label{fig:scan_dmm10_lnc}
\end{figure}

Within the overdamped regime ($\Gamma_{\rm osc}^{\rm slow} (T_{\rm EW}) \ll H_u(T_{\rm EW})$) semi-analytical expressions in the LNC limit were found in~\cite{Drewes:2016gmt}.
We find that the impact of LNV rates in this regime is more accentuated than in the fast oscillation and intermediate regime.
This is because LNC and LNV contributions to the asymmetry not only differ dramatically in their time evolution, but also enter with opposite sign.
When neglecting LNV plasma interactions the dynamics of the BAU generation is only coupled to one weak mode, i.e. the overdamped oscillation mode.
This is because in the LNC scenario we have $\Gamma_{M}^{\rm slow} \to 0$ and hence the LNV weak mode decouples completely from the BAU generation.
The asymmetry hence grows in the whole overdamped regime with $x^2$, i.e.
\be
 \left(\sum_\alpha\mu_{B/3-L_\alpha}\right)^{\rm ov-LNC} \simeq {\kappa x^2 \over   6\gamma_0+\kappa \gamma_1}{\gamma_0^2\over \gamma_0^2+ 4 \omega^2} \frac{c_H M_P^*}{T_{EW}^3} \Delta^{\rm ov}_{\rm LNC}\,.
\ee 
Expressing the CP invariant in terms of physical parameters, eq.~\eqref{eq:ovlncinvnh} (eq.~\eqref{eq:CP_invariants_IH_leading_order}) for NH (IH), and using the instantaneous sphaleron freeze-out approximation we can formulate the asymmetry as
\be
\left( Y_B \right)_{\rm ov}^{\rm LNC} \simeq - 2\times 10^{-1}  \frac{\Delta M}{M} \frac{1\,\text{GeV}}{M}  \left(\frac{10^{-7}}{U^2}\right)^2 f_{\rm LNC}^{\rm H}\,.
\ee
The angular function $f_{\rm LNC}^{\rm H}$ is defined in eq.~\eqref{eq:f_lnc_lnv_H} (eq.~\eqref{eq:f_lnc_ih}) for NH (IH).
Maximizing this function leads to an upper bound on the HNL mixing compatible with the BAU
\be
\left( U^2 \right)_{\rm ov}^{\rm LNC} \lesssim 6\, (15) \times 10^{-3} \sqrt{\frac{\Delta M}{M}} \sqrt{\frac{1\,\rm GeV}{M}} \,\,\,\,\, \rm{NH\,(IH)}\,.
\label{eq:U2_ov_LNC_max_fixed_DMM}
\ee   
In figure~\ref{fig:scan_dmm10_lnc} we compare this bound with a numerical analysis within the LNC limit for an exemplary mass splitting of $\Delta M/M = 10^{-10}$.
\begin{figure}[!t]
\centering
\begin{tabular}{cc}
\hspace{-0.5cm}  \includegraphics[width=0.5\textwidth]{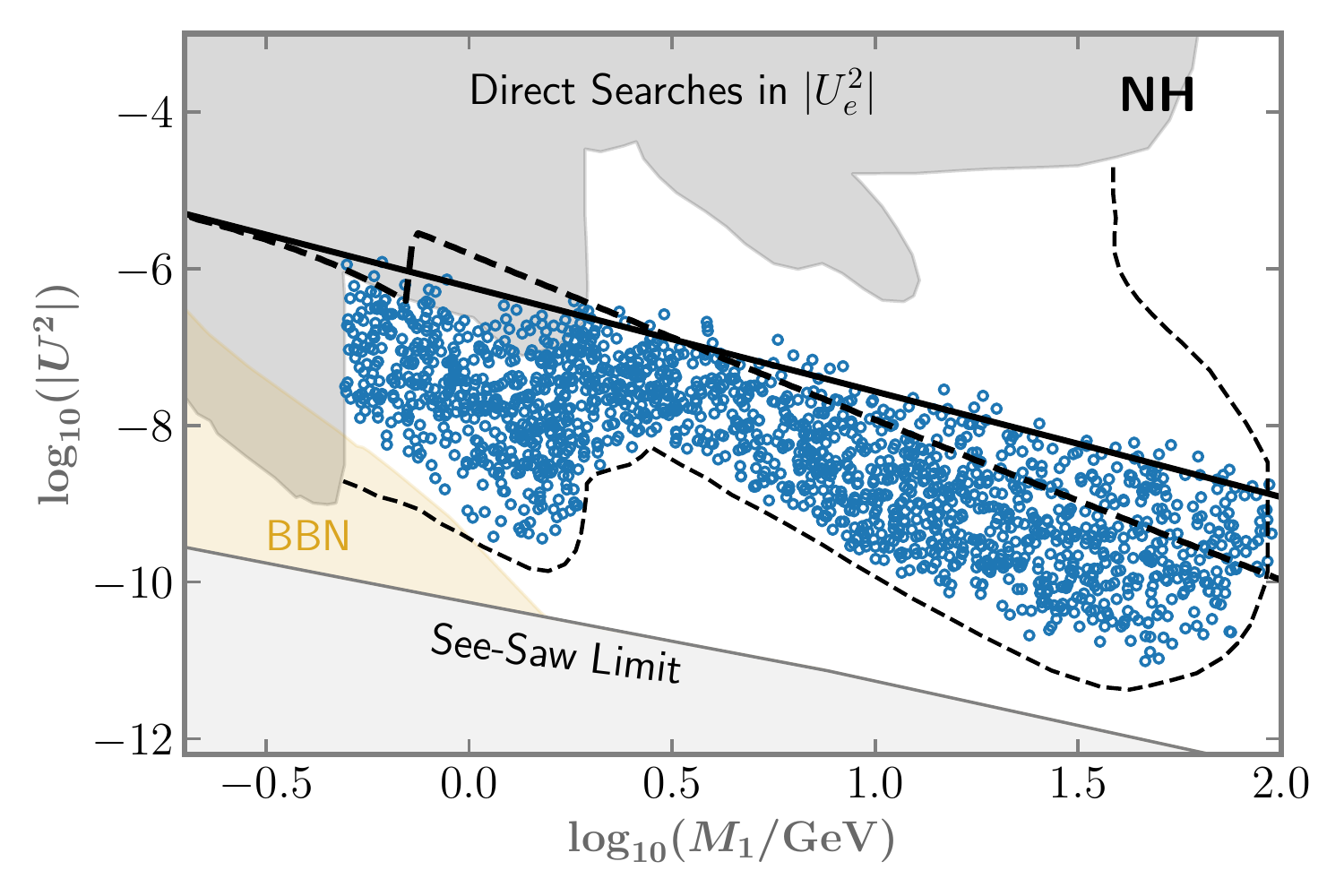} &
\hspace{-0.55cm}  \includegraphics[width=0.5\textwidth]{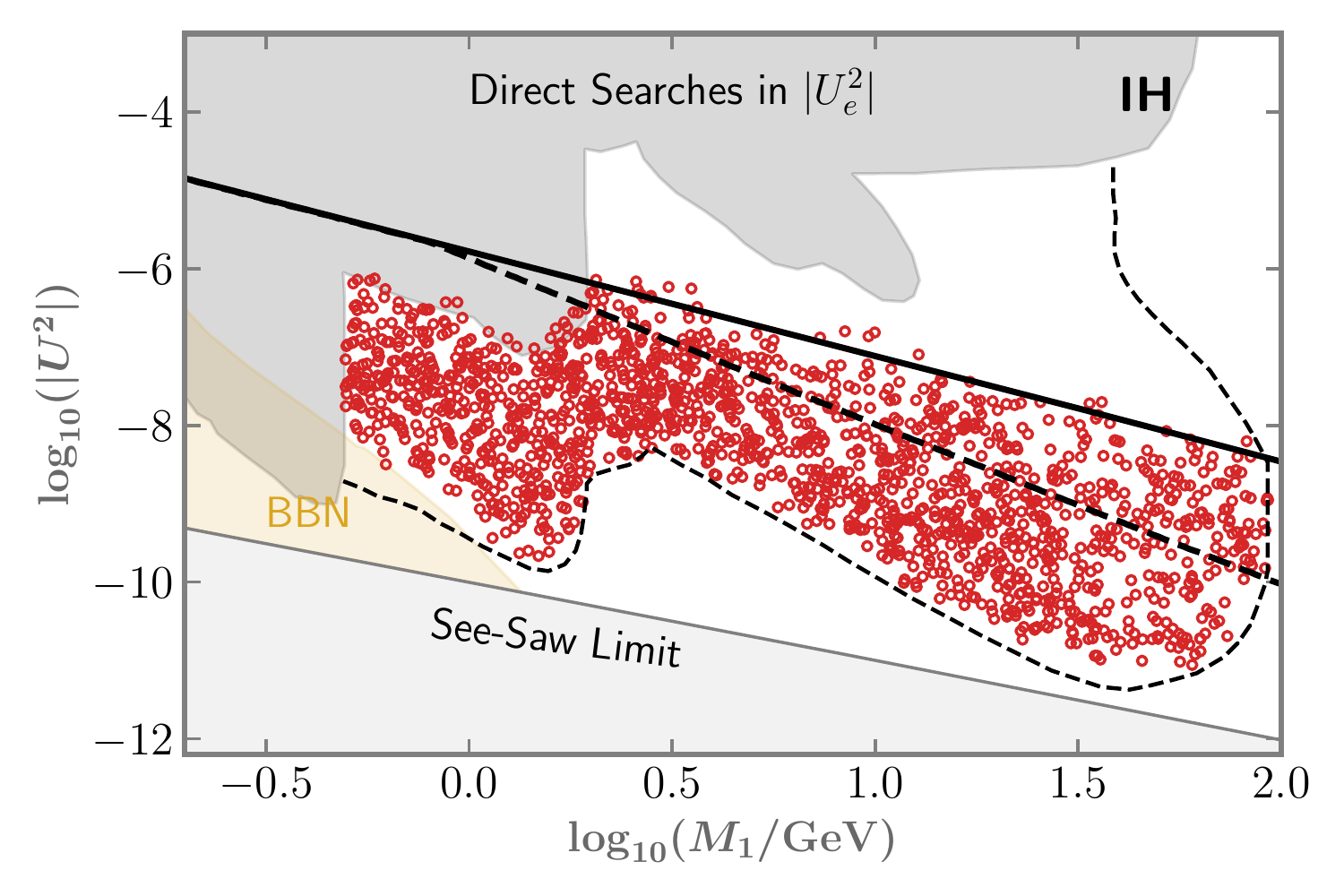} 
\end{tabular}
\vspace{-0.4cm}
\caption{ 
Numerical result of the Bayesian analysis (blue (red) points for NH (IH)) together with the analytical derived upper bound on the HNL mixing (black line).
In dashed we show for comparison the corresponding analytical upper bound when LNV rates are included.
The grey region is excluded by direct searches and the yellow one by big bang nucleosynthesis constraints.
}
\label{fig:scan_general_lnc}
\end{figure}

Having derived eq.~\eqref{eq:U2_ov_LNC_max_fixed_DMM} the general upper bound for variable HNL degeneracy is found by saturating the overdamped condition of eq.~\eqref{eq:dmovmboundov} which leads to
\be
U^2 \lesssim 5\,(17) \times 10^{-7} \left(\frac{1\,\rm{GeV}}{M}\right)^{4/3}  \,\,\,\,\, \rm{NH\,(IH)}\,.
\label{eq:U2_ov_LNC_max}
\ee
The numerical result of a bayesian analysis with variable $\Delta M/M$ within the LNC limit is shown in figure~\ref{fig:scan_general_lnc}.
The priors are the same as given in table~\ref{tab:priors}.
In dashed we show for comparison the upper bound on the mixing when LNV rates are included.


\section{Appendix: Triangle Plots}
\label{app:get_dist}

\begin{figure}[!t]
\centering
\includegraphics[width=0.99\textwidth]{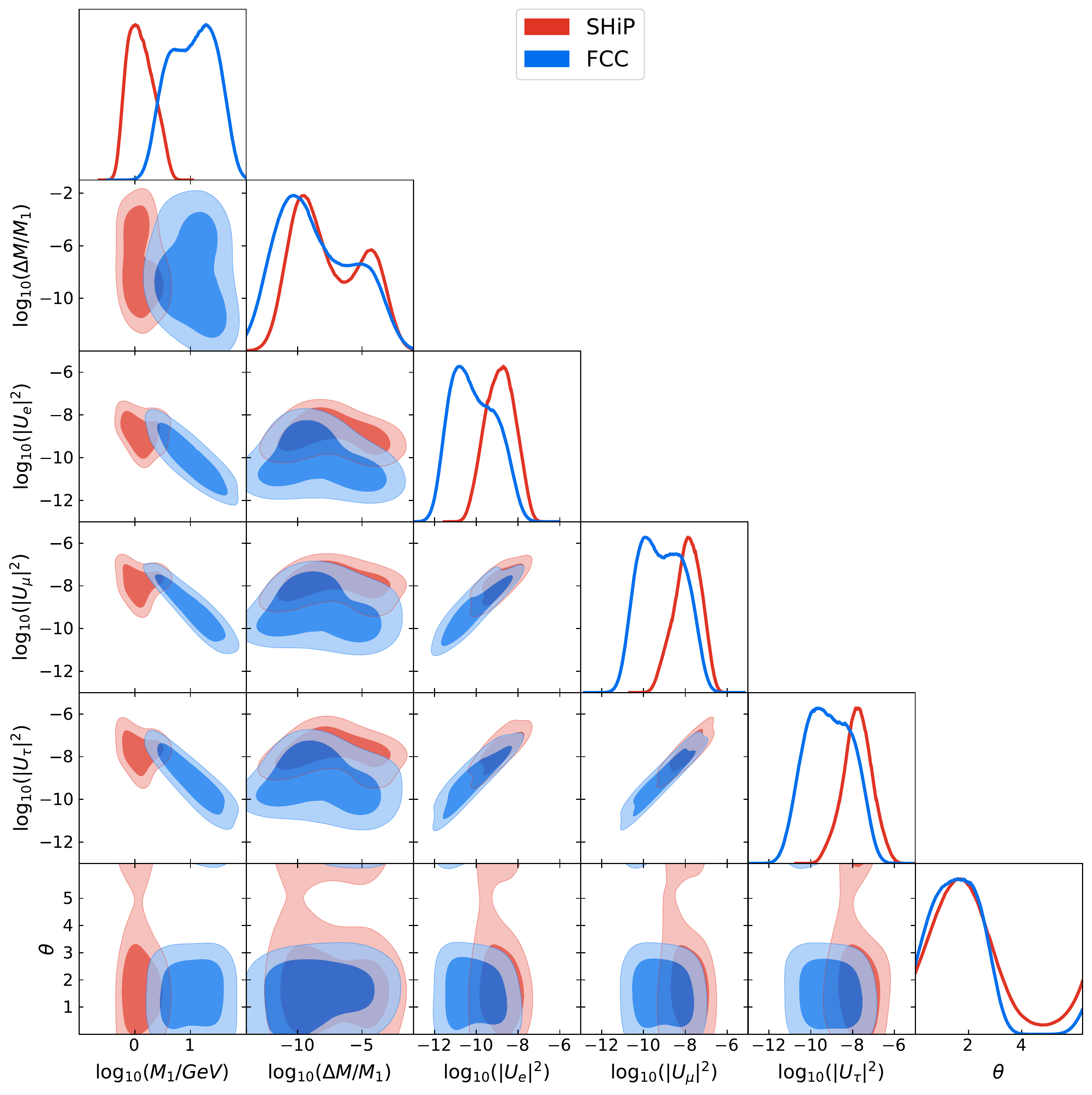} 
\caption{ 
Two-dimensional $1$ and $2$ $\sigma$ posterior probability projections of the global numerical scan for NH, with priors given in table~\ref{tab:priors}. The scan is performed separately for \texttt{SHiP} (red) and \texttt{FCC} (blue).}
\label{fig:get_dist_NH}
\end{figure}

\begin{figure}[!t]
\centering
\includegraphics[width=0.99\textwidth]{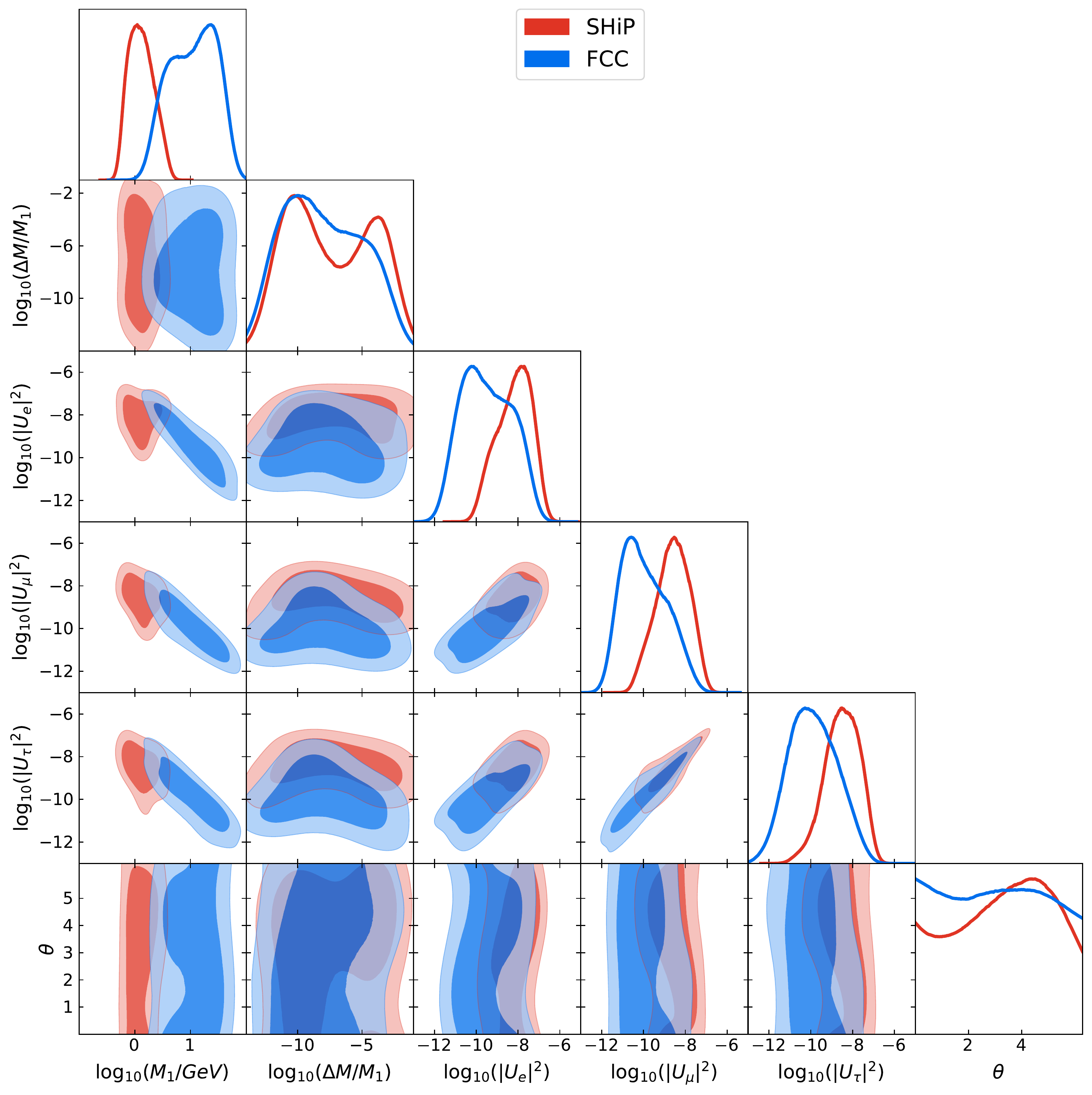} 
\caption{ 
Same as in Fig.~\ref{fig:get_dist_NH} but for IH.
}
\label{fig:get_dist_IH}
\end{figure}

In addition to the $U^2$ vs. $M$ projections shown in Fig.~\ref{fig:scan_general}, 
here we show other two-dimensional posterior probability projections of the global numerical scan for NH (IH) in Fig.~\ref{fig:get_dist_NH} (Fig.~\ref{fig:get_dist_IH}). 

In particular, we include the non trivial projections on $|U_\alpha|^2$, $\Delta M/M$, $M$ and $\theta$. 
We do not show the correlation with the PMNS CP-phases since there is no restriction on those phases when we consider the full prior range of $\Delta M/M$ as given in table~\ref{tab:priors}. 
For the same reason, we do not include $m_{\beta\beta}$ since the preferred range does not differ from the standard active neutrino contribution.

The selection of $\theta < \pi$ in NH arises from the sign of the baryon asymmetry.  At leading order the CP invariants of eqs.~\eqref{eq:ovlncinvnh}, \eqref{eq:intlncinvnh} and \eqref{eq:osclnvnh} only depend on $s_\theta$. Note that only for $M_1\lesssim 1$ GeV this preference is relaxed.
This can be understood due to the dominance of LNV interactions in the overdamped regime for larger values of $M_1$, see eq.~\eqref{eq:Yb_ov_wLNV}. 
For IH this preference disappears because  the PMNS phases can always be adjusted to yield the correct sign of the baryon asymmetry.

We use the Monte Carlo analysis software \href{https://github.com/cmbant/getdist}{\tt GetDist}~\cite{Lewis:2019xzd} to extract the $1$ and $2$ $\sigma$ posterior probabilities. 
We take the sensitivity reach of \texttt{SHiP} and \texttt{FCC} in our posterior probability projections into account via higher order multiplicative bias corrections.
Therefore, any interpretation of the posterior probabilities has to be done with respect to the constrained parameter space and has to be taken with care.


\newpage
~\\
\newpage
\bibliography{biblio}

\end{document}